\documentclass[iop,apj,tighten,twocolappendix,numberedappendix]{emulateapj}
\usepackage{apjfonts}
\usepackage{graphicx}
\usepackage{amsmath,amssymb}
\usepackage{color}
\usepackage{bm}
\usepackage[breaklinks,colorlinks,citecolor=blue,linkcolor=red]{hyperref}
\usepackage[all]{hypcap}
\usepackage{enumitem}

\def\LaTeX{L\kern-.36em\raise.3ex\hbox{a}\kern-.15em
    T\kern-.1667em\lower.7ex\hbox{E}\kern-.125emX}

\newcommand{\vmax}{v_\mathrm{max}}
\newcommand{\vbar}{\bar{v}}
\newcommand{\wbar}{\bar{w}}
\newcommand{\Msmbh}{M_{\rm SMBH}}
\newcommand{\Msun}{M_{\odot}}
\newcommand{\Gpc}{\mathrm{Gpc}}
\newcommand{\yr}{\mathrm{yr}}
\newcommand{\SgrA}{\mathrm{SgrA}^*}
\renewcommand{\inf}{\mathrm{inf}}
\newcommand{\rhop}{\rho_{\mathrm{p}0}}

\begin{document}

\title[Distributions of EBBHs]{Eccentric Black Hole Gravitational-Wave Capture Sources in Galactic Nuclei:\\Distribution of Binary Parameters}

\author{L\'aszl\'o Gond\'an\altaffilmark{1}, Bence Kocsis\altaffilmark{1}, P\'eter Raffai\altaffilmark{1,2}, and Zsolt Frei\altaffilmark{1,2}}
  \affil{$^1$E\"otv\"os University, Institute of Physics, P\'azm\'any P. s. 1/A, Budapest, 1117, Hungary}
  \affil{$^2$MTA-ELTE Extragalactic Astrophysics Research Group}

\label{firstpage}

\begin{abstract}
 Mergers of binary black holes on eccentric orbits are among the targets for second-generation ground-based gravitational-wave detectors. These sources may commonly form in galactic nuclei due to gravitational-wave emission during close flyby events of single objects. We determine the distributions of initial orbital parameters for a population of these gravitational-wave sources. Our results show that the initial dimensionless pericenter distance systematically decreases with the binary component masses and the mass of the central supermassive black hole, and its distribution depends sensitively on the highest possible black hole mass in the nuclear star cluster. For a multi-mass black hole population with masses between $5 \, \Msun$ and $80 \, \Msun$, we find that between $\sim 43 - 69\%$ ($68 - 94\%$) of $30 \, \Msun - 30 \, \Msun$ ($10 \, \Msun - 10 \, \Msun$) sources have an eccentricity greater than $0.1$ when the gravitational-wave signal reaches $10 \, \mathrm{Hz}$, but less than $\sim 10\%$ of the sources with binary component masses less than $30 \, \Msun$ remain eccentric at this level near the last stable orbit (LSO). The eccentricity at LSO is typically between $0.005-0.05$ for the lower-mass BHs, and $0.1- 0.2$ for the highest-mass BHs. Thus, due to the limited low-frequency sensitivity, the six currently known quasi-circular LIGO/Virgo sources could still be compatible with this originally highly eccentric source population. However, at the design sensitivity of these instruments, the measurement of the eccentricity and mass distribution of merger events may be a useful diagnostic to identify the fraction of GW sources formed in this channel.
\end{abstract}


\section{Introduction}
\label{sec:Intro}

 The Advanced Laser Interferometer Gravitational-wave Observatory\footnote{\url{http://www.ligo.caltech.edu/}} (aLIGO; \citealt{Aasietal2015}) and Advanced Virgo\footnote{\url{http://www.ego-gw.it/}} (AdV; \citealt{Acerneseetal2015} have made the first six detections of gravitational waves (GWs) from approximately circular inspiraling binaries \citep{Abbottetal2016c,Abbottetal2016b,Abbottetal2017,LIGOColl2017,Abbottetal2017_2,Abbottetal2017_3} during their first two observing runs, and opened a new window through which the universe can be observed. Two additional GW detectors are planned to join the network of aLIGO and AdV: (i) the Japanese KAGRA\footnote{\url{http://gwcenter.icrr.u-tokyo.ac.jp/en/}} is under construction, with baseline operations beginning in $2018$ \citep{Somiya2012}; while (ii) the proposed LIGO-India\footnote{\url{http://www.gw-indigo.org/}} is expected to become operational in $2022$ \citep{Iyeretal2011,Abbottetal2016a}. These instruments are expected to  continue to make detections of GWs and to answer fundamental questions about their astrophysical sources.	

 One of the most anticipated type of events to be detected with ground-based GW detectors is the circular inspiral and coalescence of compact binaries consisting of neutron stars (NSs) and/or black holes (BHs). The current detections constrain the merger rate density of BH--BH mergers in the universe to $12-213 \, \Gpc^{-3}\yr^{-1}$ \citep{Abbottetal2017}. At a typical detection range of $2 \, \Gpc$, for aLIGO's design sensitivity, this corresponds to a detection rate between $400-7.000 \, \yr^{-1}$. There are various theoretical studies to explain these rates; see \citet{Abadieetal2010} for a partial list of historical compact binary coalescence rate predictions, and \citealt{Dominiketal2013,Kinugawaetal2014,Abbottetal2016g,Abbottetal2016f,Abbottetal2016h,Abbottetal2016d,Abbottetal2016e,Belczynskietal2016,Rodriguezetal2016a,Sasakietal2016,Bartosetal2017,McKernanetal2017,Stoneetal2017,Hoangetal2017} and references therein for recent rate estimates. Because GW emission tends to circularize the orbit as it shrinks the binary separation \citep{Peters1964}, many of the GW sources are expected to have small orbital eccentricity by the time they enter the frequency bands of ground-based GW detectors. For instance, although the eccentricity of the Hulse--Taylor pulsar is currently $0.6171$ \citep{HulseTaylor1975}, it will have an eccentricity of $\sim 10^{-4}$ when it enters the aLIGO band (see Section \ref{subsec:DistIniParam}). However, not all binary sources are circular.

 There are multiple reasons to expect a non-negligible eccentricity for some binary sources emitting GWs in the aLIGO band. One mechanism leading to eccentric mergers is the Kozai--Lidov oscillation in hierarchical triple systems \citep{Wen2003}, in which the binary is perturbed by a distant third object. Indeed, high-mass stars, the progenitors of BHs, are commonly found in triples in the galactic field (i.e. more than $25\%$ of massive stars are in triples; see \citealt{Sanaetal2014}). Further, BH binaries can capture triple companions via dynamical encounters in dense stellar systems such as globular clusters (GCs; \citealt{Wen2003,Thompson2011,Aarseth2012,Antoninietal2014,Antoninietal2016,Breiviketal2016,Rodriguezetal2016a,Rodriguezetal2016b}) and in isolated triples in the field \citep{SilsbeeTremaine2017}. BH binaries in orbit around a supermassive BH (SMBH) in galactic nuclei (GNs) also represent hierarchical triples \citep{AntoniniPerets2012,VanLandinghametal2016,PetrovichAntonini2017,Hoangetal2017,RandallXianyu2018,RandallXianyu2017}. In these systems, eccentricity may be increased to a value close to unity during Kozai--Lidov oscillations while the semi-major axis is fixed, and GW emission may then quickly reduce the binary separation and lead to a merger before the orbits can fully circularize. A related channel to form eccentric mergers is through dynamical multibody interactions in dense stellar systems, which can drive the low-eccentricity inner binary to high eccentricity \citep{Gultekinetal2006,OLearyetal2006,Kushniretal2013,AntoniniRasio2016,Rodriguezetal2017}. If primordial BHs comprise dark matter halos, they may form BH binaries in single--single interactions through GW emission and retain some of their initial eccentricity as they enter the frequency bands of ground-based GW detectors \citep{Cholisetal2016}. Finally, binary BHs forming in clusters through binary--single interactions may also produce LIGO/VIRGO sources with non-negligible eccentricities \citep{Samsingetal2014,SamsingRamirezRuiz2017,Samsing2017,Samsingetal2018}.

 The expected merger rate densities in the Kozai--Lidov channel are uncertain (as are all other rate estimates), but expectations are $1-1.5 \,{\rm Gpc}^{-3} \, {\rm yr}^{-1}$ for BH binaries forming in nuclear star clusters through multibody interactions \citep{AntoniniRasio2016} and $0.14-6.1 \, {\rm Gpc}^{-3} \, {\rm yr}^{-1}$ in isolated triple systems \citep{SilsbeeTremaine2017}. A merger rate density of order $1-5\, {\rm Gpc}^{-3} \, {\rm yr}^{-1}$ is expected for BH binaries forming via the Kozai--Lidov mechanism in GCs \citep{Antoninietal2014,Antoninietal2016,Rodriguezetal2016a,Rodriguezetal2016b} and GNs \citep{AntoniniPerets2012,Hoangetal2017}, and non-spherical nuclear star clusters may produce BH--BH binary merger rates of up to $15\, {\rm Gpc}^{-3} \, {\rm yr}^{-1}$ \citep{PetrovichAntonini2017}. Smaller-sized GNs with intermediate-mass BHs \citep{VanLandinghametal2016} and binary BHs forming in clusters through binary--single interaction \citep{Samsingetal2014,SamsingRamirezRuiz2017,Samsing2017,Samsingetal2017} may produce higher rates. An attractive feature of these merger channels is that they offer a solution to the so-called ``final AU problem,'' i.e. on how the BHs or their progenitors shrink to a separation below which GW emission can drive the binary to merge within a Hubble time \citep{Stoneetal2017}.

 Here, we focus on another channel, which also naturally solves the final AU problem and leads to eccentric mergers. These are close fly-bys between single objects in dense stellar systems, which form binaries due to GW emission. If the velocity dispersion is high in the host environment, as in GNs, then the binary pericenter distance must be small for a binary to form in this way. This implies that the source will form in the LIGO frequency band and/or remain eccentric in the LIGO band until the merger. Despite the required small impact parameter, the merger rate in this channel may be high due to the extremely high number density of BHs in the mass-segregated cusps of GNs.\footnote{Formation rates for these scattering events are proportional to the square of the BH number density.} \citet{OLearyetal2009} showed that the expected aLIGO detection rate at design sensitivity for eccentric mergers forming through GW capture in GNs is higher than $\sim 100 \rm \, yr^{-1}$ if the BH mass function extends to masses above $25 \, \Msun$ \citep[see also][]{KocsisLevin2012}. Since that publication, such heavy BHs have been observed in several LIGO/VIRGO detections \citep{Abbottetal2016f,Abbottetal2017,Abbottetal2017_2}. This also means that the aLIGO detection rate of highly eccentric binaries may be dominated by sources in GNs that form through GW capture.

 In this paper, we determine the region of initial orbital parameters including eccentricity, pericenter distance, and masses where eccentric binary black holes (EBBHs) may form, the probability density function (PDF) of EBBH mergers as a function of these parameters, and calculate the expected PDF of initial orbital parameters. However, the initial eccentricity may be difficult to detect for eccentric binaries that form outside of the aLIGO frequency band. In \citet{Gondanetal2017}, we have shown that the GW detector network will be capable of measuring the final eccentricity of the source to much higher accuracy at the LSO. Thus, we also determine the distribution of eccentricity at the LSO. Additionally, we determine the distribution of eccentricity at the time the GW signal enters the aLIGO band.

 We also calculate the merger rate distribution as a function of total mass and mass ratio in a single GN in this channel. Recently, \citet{OLearyetal2016} showed that, in GCs, independent of the BH mass function, the likelihood of merger is proportional to $M^4$, where $M$ is the total mass of the binary and comparable component masses are most common. Furthermore, \citet{Kocsisetal2017} found that a different universal relation holds for the likelihood of primordial BH binaries formed in the early universe. Such universal results may be valuable to statistically disentangle different source populations and to determine the BH mass function therein.

 In a companion paper, \citet{Gondanetal2017}, we calculate the expected parameter measurement errors of eccentric mergers and show that the initial orbital eccentricity can be measured to sub-percent accuracy with the aLIGO--AdV--KAGRA GW detector network. Thus, the predicted merger rate distributions as functions of source parameters (particularly eccentricity) may be used to confirm or rule out this formation mechanism.

 GW signals of EBBHs are very different from those of standard quasi-circular inspirals. At early times, these GW signals in the time domain consist of repeated bursts where the separation between successive bursts shrinks, and their amplitudes change as a function of time, and the waveform transforms into an eccentric quasi-periodic chirp signal before the merger. Eccentricity leads to multiple orbital frequency harmonics in the waveform, pericenter precession causes each harmonic to split into a frequency triplet, and eccentricity also changes the evolution of GW phase. This makes them rich in features and very unique among other signals. Furthermore, these GW signals are more luminous in the aLIGO band than circular sources, which makes them detectable at longer distances \citep{OLearyetal2009}.

 Recent studies \citep{BrownZimmerman2010,HuertaBrown2013,Huertaetal2014,Huertaetal2017} have shown that methods based on circular binary templates could miss eccentric signals if the orbital eccentricity exceeds $\sim 0.1$ at the time these GW signals enter the aLIGO band. As we will show in this paper, a significant fraction of EBBHs forming through GW capture in GN hosts have $e > 0.1$ when their GW signals reach the aLIGO band. In such cases, search methods using circular binary templates will be ineffective in finding GW signals of these EBBHs.

 So far, three search methods have been developed to date to find the signals of stellar-mass eccentric BH binaries in data streams of GW detectors \citep{Taietal2014,Coughlinetal2015,Tiwarietal2016}. All three methods achieve substantially better sensitivity for eccentric BH binary signals than existing localized burst searches or chirp-like template-based search methods. However, the complex nature of EBBH waveforms makes it necessary to develop more efficient search algorithms to detect these GW signals for moderately small signal-to-noise ratios. Such algorithms would greatly benefit from the detailed prior knowledge of source parameter distributions in the parameter space, particularly the likely range of parameter values, which must be covered with the search pipeline.

 Once a source is detected, different algorithms are used to recover its physical parameters. For compact binary coalescence sources on circular inspiral orbits, several algorithms have been developed for this purpose, e.g. \textsc{BAYESTAR} \citep{SingerPrice2016}, \textsc{LALInference} \citep{Veitchetal2015}, \textsc{gstlal} \citep{Cannonetal2012,Priviteraetal2014}, \textsc{Coherent WaveBurst} \citep{Klimenkoetal2016}, \textsc{BayesWave} \citep{CornishLittenberg2015}, and \textsc{LALInferenceBurst} \citep{Veitchetal2015}. The development of algorithms recovering the parameters of compact binaries on eccentric orbits are currently underway. These algorithms will play an important role in the astrophysical interpretation of eccentric sources.

 The paper is organized as follows. In Section \ref{GN}, we introduce the adopted models for the GN host environment and stellar populations properties. In Section \ref{sec:Evolution}, we summarize the phases of GW capture-induced binary evolution. In Section \ref{sec:NumMethod}, we describe Monte Carlo (MC) calculations of the parameter distribution of merger rates. In Section \ref{sec:PIOP}, we present an analytical derivation of the rates, which captures the leading-order behavior. In Section \ref{sec:NumRes}, we present our results in two parts. First, we present the distributions of orbital parameters of EBBHs at different stages of their time evolution, then we determine the merger rate distributions and the corresponding aLIGO detection rate distributions. We also identify characteristics of GN host environments to which the results are sensitive. Finally, we summarize the results of the paper and draw conclusions in Section \ref{sec:Summary}. Details of our analytic calculations can be found in Appendices \ref{sec:CalRmin}--\ref{sec:aLIGOeventrate}.

We use $G=1=c$ units in this paper.

\section{Galactic Nuclei and Relaxed Stellar Populations}
\label{GN}

Here, we describe the characteristics of GN host environments (Section \ref{subsec:GNModels}) and stellar populations (Section \ref{subsec:PropStellPop}) that we adopt in this study.

\subsection{Galactic Nuclei}
 \label{subsec:GNModels}

 GNs are assumed to be relaxed systems of spherically symmetric stellar populations (e.g. white dwarfs or WDs, main sequence stars or MSs, NSs, and BHs) gravitationally bound to a central SMBH. The relaxation of multi-mass stellar populations around an SMBH has been thoroughly investigated \citep{BahcallWolf1977,Freitagetal2006,HopmanAlexander2006b,AlexanderHopman2009,Keshetetal2009,OLearyetal2009,BarOr2013,AharonPerets2016,BarOrAlexander2016,Alexander2017,Baumgardtetal2017}; these studies have found that, within the GN's radius of influence, objects segregate to the central regions and form a power-law number density profile, $n(r) \propto r^{-\alpha}$, in which the $\alpha$ exponent is higher for more massive objects. \footnote{For SMBHs more massive than a few million solar masses, the GN cluster does not have time to reach an equilibrium distribution because the two-body relaxation timescale in that case is larger than a Hubble time \citep{Merritt2010,AntoniniMerritt2012,GualandrisMerritt2012,Antonini2014,DosopoulouAntonini2017}. This might lead to  number density profiles different from that obtained for relaxed cusps.} The radius of influence is given as
 \begin{equation} \label{eq:rmax}
   r_\mathrm{max} = G\Msmbh/\sigma_*^2 \, ,
 \end{equation}
 where $\Msmbh$ is the mass of the SMBH, $\sigma_*$ is the velocity dispersion of the underlying stellar populations in the nucleus near the SMBH, and we use the $\Msmbh$--$\sigma_*$ relationship
 \citep{Tremaineetal2002,Walcheretal2005,MisgeldHilker2011,Norrisetal2014}:
\begin{equation}\label{eq:Msmbhsigma}
  \Msmbh \simeq 1.3 \times 10^8 \,\Msun \left( \frac{ \sigma_* }{ 200\ \mathrm{km}\ \mathrm{s}
  ^{-1}} \right) ^4 
 \end{equation}
 to estimate $\sigma$ in Equation (\ref{eq:rmax}). We choose the lower limit of the SMBH mass range to be $10^5 \, \Msun$ \citep{Barthetal2005,GreeneHo2006}. We conservatively set the upper limit of the SMBH mass to $10^7 \, \Msun$, corresponding to relaxed stellar populations; see Appendix \ref{subsec:CritI} for details.

 For spherically symmetric and relaxed GNs, we use the \citet{BahcallWolf1976} one-body phase space distribution $f(\mathbf{r}, \mathbf{v})$ generalized for a multi-mass system \citep{OLearyetal2009} to derive the distribution of the magnitude of relative velocity between single objects (Appendices \ref{App:AppGN} and \ref{App:HandRelvel}). For objects with mass $m$, this has the form
 \begin{equation}  \label{eq:frv}
   f_m(\mathbf{r},\mathbf{v}) = C_m E(r,v)^{p_m}
 \end{equation}
 if $r_{\rm min} \leqslant r \leqslant r_\mathrm{max}$ and $ E(r,v) > 0$, $C_m$ is a normalization constant,
 \begin{equation}  \label{eq:Erv}
    E(r,v) = \frac{ \Msmbh}{ r } - \frac{ v^2 }{ 2 }
 \end{equation}
 is the Keplerian binding energy in the field of the SMBH, and the $p_m$ exponent depends on mass due to mass segregation. For the light stellar components, such as MSs, WDs, and NSs, $p_m\approx 0$, and for the heavier components such as BHs,
\begin{equation}   \label{eq:pm}
   p_m = p_0 \frac{ m }{ m_\mathrm{BH,max}}\,,
\end{equation}
 where $m_\mathrm{BH,max}$ is the highest possible BH mass in the population and $p_0 \approx 0.5 - 0.6$ based on Fokker--Planck simulations \citep{OLearyetal2009}. We consider the standard value of $p_0$ to be $0.5$. Equation (\ref{eq:Erv}) is valid outside the loss cone \citep{ShapiroLightman1976,SyerUlmer1999}. We assume it to be valid between $r_{\rm min} \leqslant r \leqslant r_\mathrm{max}$ if $v$ is less than the escape velocity at radius $r$, where $r_{\rm min}$ is an inner radius where the density cusp exhibits a cutoff. We calculate $r_{\min}$ by requiring that (i) the number density profile of these populations reaches steady state and forms a power-law density cusp within the age of the galaxy; and (ii) the inspiral timescale into the SMBH is larger than the relaxation time (Appendix \ref{sec:CalRmin}).

 Because the binding energy is positive for objects bound to the SMBH, $E(r,v) >0$, the velocity at distance $r$ from the SMBH must be less than the escape velocity
 \begin{equation}\label{eq:vmax}
 v_\mathrm{max}(r)=\sqrt{\frac{2G  \, M_{\rm SMBH} }{r}} \, .
\end{equation}

 The 3D number density distribution of mass $m$ objects at radius $r$ may be obtained from Equation (\ref{eq:frv}) as
 \begin{equation}  \label{eq:n(r)}
  n_m(r) = \int f_m(\mathbf{r},\mathbf{v}) d^3 v = n_{\rm inf} \left( \frac{r}{r_{\rm max}}
  \right)^{-\alpha_{m}} \, ,
 \end{equation}
 where
\begin{equation}  \label{eq:alpham}
    \alpha_{m} = \frac{ 3 }{ 2 } + p_m \, ,
\end{equation}
 and $n_{\rm inf}$ is the number density of objects at the radius of influence. For MSs of $m = 1 \, \Msun$, this follows from the $M$--$\sigma$ relation (Equation \ref{eq:Msmbhsigma}), assuming that the enclosed stellar mass within the radius of influence $r_{\max}$ is $2M_{\rm SMBH}$ \citep[e.g.][]{OLearyetal2009}
\begin{equation} \label{eq:nrmasMS}
  n_{\inf,\rm MS} \equiv n_\mathrm{MS} (r_\mathrm{max}) \ \simeq  1.38 \times 10^5 \mathrm{pc}^{-3}
  \sqrt{ \frac{ 10^6 \Msun }{ \Msmbh } } \, .
\end{equation}

 We define the normalization constants $n_{\inf}$ and the exponent of the number density distribution $\alpha_m$ separately for the different stellar species in Section \ref{subsec:PropStellPop}.

\subsection{Stellar Populations in Galactic Nuclei}
\label{subsec:PropStellPop}

 The initial mass functions (IMF), which extend from the brown dwarf boundary $\sim 0.1 \, \Msun$ to $\sim 100 \, \Msun$ (e.g. the \citep{Salpeter1955} IMF and its subsequent refinements, the \citep{MillerScalo1979} and \citep{Kroupa2001} IMFs), result in evolved populations that naturally separate into two mass scales: the $\sim 1 \, \Msun$ scale of low-mass MSs, WDs, and NSs; and the $\sim 10 \, \Msun$ scale of stellar-mass BHs, Wolf--Rayet, O-, and B-stars. In case of a GN, the shape of the present-day mass function (PMF) varies  significantly above the scale of $\sim 10 \, \Msun$ \citep{AlexanderHopman2009}. The typical PMF in GNs is not well-understood due to the fact that star formation deep in the potential well of a SMBH can be very different from that of the field. \citet{Bartkoetal2010} find evidence that the stellar disks in the GN of the Galaxy are extremely top-heavy. \citet{Luetal2013} infer that the mass function of these young stars is proportional to $m^{-1.7 \pm 0.2}$, which implies that the number of massive stars is currently higher than
 that for a standard Salpeter mass function.

 We follow \citet{AlexanderHopman2009} and assume single-mass MS, WD, and NS populations in GNs with component
 masses $1 \, \Msun$, $0.6 \, \Msun$, and $1.4 \, \Msun$, respectively.

 We normalize the number density distribution for WD, MS, and NS populations in Equations (\ref{eq:n(r)}) as
 \begin{equation}  \label{eq:nNS}
  n_{\inf,i} = n_{\inf, \rm MS} C_{i} \, ,
 \end{equation}
 where $i$ labels $\{\rm WD, MS, NS\}$, and set the number density exponents to
 \begin{equation}
    \alpha_{\rm MS}=\alpha_{\rm WD}=1.4 {\rm ~~and~~} \alpha_{\rm NS}=1.5\,
 \end{equation}
 \citep{HopmanAlexander2006b}. Here, $C_{i}$ represents the number fraction ratios of MSs, WDs, and NSs as $1:0.1:0.01$
 for continuous star-forming populations \citep{Alexander2005}, implying that
 \begin{equation}
  C_{\rm MS}=1\,,\quad C_{\rm WD}=0.1\,,\quad C_{\rm NS} = 0.01 \frac{3 - \alpha_{\rm NS}}{3 - \alpha_{\rm MS}}\,.
 \end{equation}

 We carry out calculations for three types of BH PMF models: a simple single-component mass distribution, a power-law multi-mass distribution, and a multi-mass distribution given by a population synthesis method, as follows.
 \begin{enumerate}
 \item In the single-mass BH population model, we assume that all BHs have $m_{\rm BH}=10 \, \Msun$ following \citet{Morris1993}, \citet{MiraldaEscudeGould2000}, and \citet{AlexanderHopman2009} for a $10 \, \mathrm{Gyr}$ old coeval population, implying that the number density exponent in equation  (\ref{eq:nrmasMS}) is $\alpha_ \mathrm{BH,s} = 2$ and to determine $C_\mathrm{BH,s}$ we account for BHs segregated into the nucleus. For the Milky Way, as many as $\sim 20,000$ BHs with $10 \, \Msun$ each are expected to have segregated into the  nucleus \citep{Morris1993,MiraldaEscudeGould2000,Freitagetal2006,HopmanAlexander2006b}. The number of BHs segregated into the GN is proportional to the mass of the SMBH through the infall rate of BHs (see Equation (9) in \citet{MiraldaEscudeGould2000}), therefore it can be approximated as
 \begin{equation}  \label{eq:encBH}
   N_{\rm BH} \approx 20,000 \times \frac{M_{\rm SMBH}}{M_{\SgrA}} \, .
 \end{equation}
 Here, $M_{\SgrA} = 4.3 \times 10^6 \, \Msun$ is the mass of the SMBH, Sgr A$^{*}$, in the Milky Way-size nucleus \citep{Gillessenetal2009}\footnote{We choose $M_{\rm SMBH}=M_{\SgrA}$ in our fiducial numerical calculations, but also explore other values.}. The number density distribution of BHs in this single-mass model is of the form
\begin{equation}
  n_\mathrm{BH,s} = C_\mathrm{BH,s} n_\mathrm{inf,MS} \left( \frac{ r }{ r_\mathrm{max} } \right)
  ^{- \alpha_\mathrm{BH,s}} \, ,
\end{equation}
 where $C_\mathrm{BH,s}$ can be determined using Equation (\ref{eq:nNS}) together with Equation (\ref{eq:encBH}), which gives
\begin{equation}
  C_\mathrm{BH,s} = 0.023 \frac{ 3 - \alpha_\mathrm{BH,s} }{ 3 - \alpha_\mathrm{MS} } \, .
\end{equation}

 \item In this multi-mass BH population model, we assume a power-law PMF $m_{\rm BH}^{-\beta}$ following \citet{AlexanderHopman2009} for $10 \, \mathrm{Gyr}$ of continuous star formation. The probability distribution function of BH mass is then
 \begin{equation}\label{e:f(m)}
   \mathcal{F}(m_{\rm BH})= \frac{ (1-\beta) \, m_{\rm BH}^{-\beta} }{ m_{\rm BH,max}
   ^{1-\beta} - m_{\rm BH,min}^{1-\beta} } \, .
 \end{equation}
 We consider $\beta$ in the range $[1,3]$ and arbitrarily choose its standard value to be $\beta \equiv 2.35$, although the exponent for BHs need not be related to the Salpeter IMF of massive stars \citep{Salpeter1955}. This range is consistent with the currently announced GW detections \citep{Abbottetal2017}. We set $m_{\rm BH,min} = 5 \, \Msun$, based on the observations of X-ray binaries  \citep{Bailynetal1998,Ozeletal2010,Farretal2011,Belczynskietal2012}, and we set the fiducial value for the highest possible BH mass to be $m_\mathrm{BH,max} = 30 \, \Msun$, but also explore cases between $10 \, \Msun$ and $80 \, \Msun$ \citep{Belczynskietal2010}. Taking into account the mass distribution of BHs, the number density normalization at the radius of influence in Equation (\ref{eq:n(r)}) for BHs in a multi-mass BH population is
\begin{equation}  \label{eq:nBHmulti2}
  n_\mathrm{BH,m} (m_{\rm BH}) = n_{\rm MS} \, \mathcal{F}(m_{\rm BH}) \, C_{\mathrm{BH},m}  \, ,
\end{equation}
 where $C_\mathrm{BH,m}$ is set numerically by the total number of BHs inside the GN as
\begin{align}
 N_{\rm BH} & = \int_{m_{\rm BH,min}}^{m_{\rm BH,max}} dm_{\rm BH} \, n_\mathrm{BH,m} \, (m_{\rm BH})
 \nonumber \\
 & \times \int_0^{r_{\max}} 4 \pi r^2 \left(\frac{r}{r_{\max}}\right)^{- \alpha_\mathrm{BH,m}} \, dr
  \, .
\end{align}
 Here, $\alpha_\mathrm{BH,m}$ is given by combining Equations (\ref{eq:pm}) and (\ref{eq:alpham}).

 \item To check the dependence of the result for a qualitatively different mass distribution, we also carried out the calculations for a PMF obtained in a population synthesis study, \citet{Belczynskietal2014_2}, for single low-metallicity stars (see Figure 4 therein). In this case, the PMF ranges between $\sim 3 M_{\odot}$ and $\sim 26 M_{\odot}$ and has three peaks in its distribution, near $7 M_{\odot}$, $14 M_{\odot}$, and $24 M_{\odot}$. 
 \end{enumerate}

 We show in Appendix \ref{subsec:CritII} that the region in single GN where EBBHs are expected to form through GW capture is weakly sensitive to the assumptions on the mass distribution of stellar populations.

\section{Phases of Binary Evolution}
 \label{sec:Evolution}

 In this section, we introduce the GW capture binary formation mechanism and calculate resulting distribution of initial orbital parameters. Next, we examine the evolution through the eccentric inspiral and the long-term interaction between the formed binary and a third object in simulations.

 In the following, we denote the masses of BHs forming the binary as $m_A$ and $m_B$, the symmetric mass ratio of the binary as $\eta = (m_A  m_B) /(m_A+m_B)^2$, the total mass as $M_\mathrm{tot} = m_A+m_B$, and the mass ratio as $q = m_A / m_B.$ The reduced mass and symmetric mass ratio satisfy $\mu = \eta M_\mathrm{tot}$ and $\eta=q/(q+1)^2$. Similar to \citet{OLearyetal2009}, we define the dimensionless pericenter distance as $\rho_\mathrm{p} = r_\mathrm{p} / M_\mathrm{tot}$, where $r_\mathrm{p}$ is the pericenter distance changing with each orbit.

\subsection{Formation of Binaries through GW Capture and the Eccentric Inspiral}
\label{subsec:FormBHB}

 Two BHs form a binary if they undergo a close encounter and they emit enough energy, in form of GWs, to become bound. Due to the relativistic nature of such events and the low velocity dispersion compared to $c = 1$ in GN hosts, the encounters are almost always nearly parabolic \citep{QuinlanShapiro1987,Lee1993}. Thus, the initial orbit of the formed binary is typically highly eccentric with $e \sim 1$. In this limit, the amount of energy emitted in GWs during the encounter is
\begin{equation}\label{eq:dEGW}
 \delta E_\mathrm{GW} =- \frac{ 85 \pi \eta^2 M_\mathrm{tot}^{9/2} }{ 12
 \sqrt{2}  r_\mathrm{p0}^{7/2} }
\end{equation}
 \citep{PetersMathews1963,Turner1977}, where $r_\mathrm{p0}$ is the distance of closest approach (we use
 $0$ in the lower index to denote initial values of parameters throughout the paper, i.e. their values at
 the time of EBBH formation):
\begin{center}
\begin{equation}\label{eq:rhop0}
 r_\mathrm{p0}  =  \left(\sqrt{\frac{ 1 }{ b^2 } + \frac{M_\mathrm{tot}
 ^2}{ b^4 w^4 } } + \frac{M_\mathrm{tot}}{b^2 w^2} \right)^{-1}
\end{equation}
\end{center}
 \citep{OLearyetal2009}, where $b$ is the impact parameter of the encounter, and $w$ is the magnitude of relative velocity between the BHs forming the EBBH.

 The initial properties of the EBBH are determined by the encountering system's final energy, $E_{\rm fin}$, and final angular momentum, $L_\mathrm{fin}$, after the first encounter, where
\begin{equation}\label{eq:Efin}
 E_{\rm fin} = E_{\rm kin} + \delta E _\mathrm{GW} \, ,
\end{equation}
\begin{equation}\label{eq:L0}
 L_{\rm fin} = M_\mathrm{tot} b \eta w + \delta L  \, ,
\end{equation}
 where $E_{\rm kin} = \mu w ^2 / 2$ is the kinetic energy in the center of mass system, and $\delta L$ is the amount of angular momentum lost during an encounter \citep{Peters1964}. \citet{OLearyetal2009} found that $\delta L$ is negligible for nearly all first encounters; therefore, we set $\delta L \approx 0$. If the final energy of the system after the encounter is negative, $E_{\rm fin} < 0$, then the system remains
 bound with an initial semi-major axis of
\begin{equation}\label{eq:a0}
 a_0 = \frac{\eta M_\mathrm{tot}^2 }{2 |E_{\rm fin}|} \, ,
\end{equation}
 and with an initial eccentricity of
\begin{equation} \label{eq:e0}
 e_0  = \sqrt{1 - \frac{2 |E_{\rm fin}| w ^2 b^2}{M_\mathrm{tot}^3 \eta}} \, .
\end{equation}
 Note that the initial pericenter distance of the EBBH system $r_\mathrm{p0}$ can be expressed in terms of $e_0$ and $a_0$ as
\begin{equation} \label{eq:rho0}
 r_\mathrm{p0}= a_0  (1 - e_0) \, .
\end{equation}

 The criterion $E_{\rm fin} < 0$ for the two BHs to form a bound binary sets an upper limit on the impact parameter of the approach \citep{OLearyetal2009}. Additionally, a lower limit on $b$ is set by the fact that we require the BHs to avoid direct coalescence during the first close approach. Therefore, to leading order, a bound EBBH system can form if $b$ satisfies
\begin{equation}\label{eq:impactlim}
  b_\mathrm{min} \equiv  \frac{ 4 M_\mathrm{tot} }{ w } <  b  <
  \left( \frac{340 \pi \eta}{ 3 } \right) ^{1/7} \frac{ M_\mathrm{tot}
  }{ w ^{9/7}} \equiv b_\mathrm{max}  \, .
\end{equation}
 Here, the upper limit corresponds to Equations (\ref{eq:dEGW}) and (\ref{eq:Efin}), and the lower limit is valid in the test-particle limit around a Schwarzschild BH \citep[e.g.,][]{Kocsisetal2006,OLearyetal2009}. The leading-order approximation in the allowed range of $b$ values expressed by Equation (\ref{eq:impactlim}) is in a good match with $2.5$ and $3.5$ order post-Newtonian (PN) simulations for relative velocity $w \lesssim 0.01$ \citep[see Figure 1. in ][]{KocsisLevin2012}). Relativistic corrections modify the capture cross section by less than $10\%$ \citep{KocsisLevin2012}

 Due to the escape velocity $v_{\max}$ at radius $r$ from the SMBH (Equation~\ref{eq:vmax}), the initial relative velocity is bounded by
 \begin{equation}\label{eq:wlim}
  0 \leqslant w \leqslant 2 v_{\max}(r) = \sqrt{8M_{\rm SMBH}/r}\, .
\end{equation}
 The allowed range of initial binary parameters $\rho_{\mathrm{p}0}$ and $e_0$ may be calculated from the bounds on $b$ and $w$ given by Equations (\ref{eq:impactlim}) and (\ref{eq:wlim}) using Equations (\ref{eq:a0}) and (\ref{eq:rho0}).

 After the binary is formed with initial parameters $e_0$ and $\rho_{\rm p0}$, it evolves due to GW radiation reaction. We use the leading-order orbital evolution equation \citep{Peters1964}
\begin{equation}\label{eq:rhoe}
 \rho_\mathrm{p} (e) = \frac{c_0}{M_\mathrm{tot}}\frac{ e^{12/19}}{(1+e)}
 \left( 1+ \frac{ 121 }{ 304 } e^2 \right)^{ \frac{870}{2299} } \, ,
\end{equation}
 where $c_0/M_{\mathrm{tot}}$ may be expressed with $e_0$ and $\rho_\mathrm{p0}$, by solving Equation (\ref{eq:rhoe}) for $\rho_\mathrm{p}=\rho_\mathrm{p0}$ and $e=e_0$.

 This evolution equation is valid when the two BHs are relatively far from each other's horizon, for $\rho_\mathrm{p} \gg 2$, up until the binary reaches the LSO. After that, the evolution is no longer quasi-periodic and the BHs quickly coalesce. In the leading-order approximation for infinite mass ratio and zero spins, the eccentricity of the LSO, $e_\mathrm{LSO}$, can be calculated by numerically solving the following equation \citep{Cutleretal1994}
\begin{equation}\label{eq:ELSO}
  \rho_\mathrm{p} (e_\mathrm{LSO}) =
  \frac{ 6+ 2 e_\mathrm{LSO} }{ 1+e_\mathrm{LSO} } 
\end{equation}
 for $e_\mathrm{LSO}$ using Equation~(\ref{eq:rhoe}). The initial pericenter distance is set uniquely by $e_\mathrm{LSO}$ and $e_0$ according to
\begin{equation}
\rho_{\mathrm{p}0} = \frac{6+ 2e_\mathrm{LSO}}{1+e_0}
\frac{ e_0^{12/19}}{e_\mathrm{LSO}^{12/19}}
 \left[\frac{ 1+ (121/304) e_0^2}{ 1+ (121/304) e_{\mathrm{LSO}}^2} \right]^{ \frac{870}{2299} } \, .
\end{equation}
 In the limit $e_0\approx 1$, which we find to be the relevant case (Section \ref{subsec:DistIniParam}),
\begin{equation}  \label{eq:rhopeLSO}
\rho_{\mathrm{p}0} \approx  \frac{3+ e_\mathrm{LSO}}{e_\mathrm{LSO}^{12/19}}
  \left( 1+ \frac{121}{304} e_{\mathrm{LSO}}^2 \right)^{ -\frac{870}{2299} }\,.
\end{equation}
 We invert Equation (\ref{eq:rhopeLSO}) numerically to obtain $e_{\rm LSO}(\rho_{\mathrm{p}0})$. The result is well-fitted\footnote{We use the method of least squares to fit the prefactor and exponent. The fit error is $0.001$ for both parameters for the $95 \%$ confidence bounds in the range $\rho_{\mathrm{p}0} \in [8,1000]$. Most EBBHs form within this range in GNs; see Section \ref{subsec:DistIniParam}.} by
\begin{equation}  \label{eq:eLSOrhop0}
  e_{\rm LSO} =  6.3096 \, \rho_{\mathrm{p}0}^{-1.599} \, .
\end{equation}
 This shows that $\rho_{\mathrm{p}0}$ generally is a monotonically decreasing function of $e_{\mathrm{LSO}}$, where $\rho_{\mathrm{p}0} \gtrsim 10$ corresponds to $e_{\mathrm{LSO}} \lesssim 0.19$.

\subsection{Interaction with a Third Object}
\label{sec:EIAPD}

 Although the stellar number density is the highest in GNs, \citet{OLearyetal2009} have shown that the binaries, which form by GW capture, are so tight that they typically merge before encountering a third object that would alter their orbital parameters. We conservatively ignore the rare EBBHs in our Monte Carlo (MC) sample that have an encounter with a third object before the merger of the EBBH (Section \ref{subsec:MCCode}), i.e. if the encounter timescale is shorter than the merger timescale,
\begin{equation}  \label{eq:conddisruption}
  t_{\rm merge}>t_\mathrm{enc}\,.
\end{equation}
 The typical timescale for a close encounter between a binary and a single object can be approximated as
 \begin{equation}\label{eq:dEGW4}
    t_\mathrm{enc}  \approx  \frac{ w^3 }{ 12 \pi M_{\rm tot}^2 n_{\rm tot} }
 \end{equation}
 \citep{OLearyetal2009}, where
\begin{equation}  \label{eq:ntot}
  n_\mathrm{tot}= n_\mathrm{WD}(r) + n_\mathrm{MS}(r) + n_\mathrm{NS}(r) + n_\mathrm{BH}(r)
\end{equation}
 is the combined number density of all types of stellar objects at distance $r$ from the SMBH. By substituting Equation (\ref{eq:rhoe}) into Equation (5.7) in \citet{Peters1964}, the time remaining until coalescence is
\begin{equation}\label{eq:tGW}
 t_{\rm merge} (\rho_{\rm p0}, m_A, m_B, e_0) =
 \frac{15 \, c_0^4(e_0,\rho_{\rm p0}) }{ 304 M_\mathrm{tot}^3 \eta} \mathcal{F}(e_\mathrm{LSO},e_0) \, ,
\end{equation}
 where
\begin{equation}\label{eq:FAB}
 \mathcal{F}(A,B) \equiv \int _{A} ^{B} de \, \frac{ e^{29/19} \left(  1+ \frac{121}{304} e^2
 \right)^{ \frac{1181}{2299}} }{ \left( 1 - e ^2 \right)^{3/2} } \, .
\end{equation}
 Note that $\mathcal{F}(e_\mathrm{LSO},e_0) \approx \mathcal{F}(0,e_0)$ within $0.1 \%$ for \mbox{$e_{\rm LSO} \ll e_0 \sim 1$}, and thus $t_{\rm merge}$ can practically be approximated as
\begin{equation}\label{eq:tGW2}
 t_{\rm merge} (\rho_{\rm p0}, m_A, m_B, e_0) \approx \frac{15 \, c_0^4(e_0,\rho_{\rm p0})}{
 304 M_\mathrm{tot}^3 \eta} \mathcal{F}(0,e_0) \, .
\end{equation}

 Because the fraction of binaries that satisfy Equation (\ref{eq:conddisruption}) represents a small fraction of all objects (Section \ref{subsec:DistIniParam}), ignoring this population does not bias the initial binary orbital parameter distributions.

\section{Distributions of Initial Orbital Parameters and EBBH mergers}
\label{sec:NumMethod}

 In this section, we derive the distributions of the masses and the initial orbital parameters of EBBHs in GNs.

 We parameterize the formation of an EBBH through GW capture via four parameters: the two masses of the BHs involved ($m_A$ and $m_B$), the magnitude of the relative velocity between the two BHs ($w$), and the impact parameter of the approach ($b$). Here, we first treat $m_A$ and $m_B$ as being fixed, while $w$ and $b$ as free parameters. We use the phase space distribution function  $f_A(\mathbf{r}_A, \mathbf{v}_A)$ and $f_B (\mathbf{r}_B, \mathbf{v}_B)$, given by Equation (\ref{eq:frv}). The distributions depend on the masses $m_A$ and $m_B$ through the exponents $p_A$ and $p_B$ in Equations (\ref{eq:frv}) and (\ref{eq:pm}) as heavier components form steeper density cusps due to mass segregation.

 The differential rate of EBBH mergers involving BHs $A$ and $B$ can be given in the twelve-dimensional phase space of these two objects as
\begin{align}  \label{eq:infformrate}
 d^{12} \Gamma _{A B} & = \sigma w f_A(\mathbf{r}_A, \mathbf{v}_A) f_B(\mathbf{r}_B, \mathbf{v}_B)
 \nonumber \\
 & \times d^3 r_A \, d^3 r_B \, d^3 v_A \, d^3 v_B \, ,
\end{align}
 where $\sigma\equiv \sigma(w)=\pi b_\mathrm{max}^2(w)-\pi b_\mathrm{min}^2(w)$ is the cross section for two BHs to form an EBBH during their encounter; see Equation (\ref{eq:impactlim}) for $b_\mathrm{min}$ and $b_\mathrm{max}$.

 \subsection{Analytic Estimates}

 The total merger rates may be obtained from integrating Equation (\ref{eq:infformrate}) over phase space. The partial merger rate distribution as a function of parameters (e.g. radius in the GN, mass, initial eccentricity, and pericenter distance) may be obtained by integrating only over the complementary phase space dimensions. We refer the reader to Appendix \ref{sec:AppA} for details, and present the results of the calculations in Section \ref{sec:NumRes}.

\subsection{The Monte Carlo Code}
\label{subsec:MCCode}

 To generate a random sample of EBBHs with component masses $m_A$ and $m_B$, we first generate $\sim 10^3$ random radius values from the galactic center over the radius range \mbox{$r \in [r_{\rm min}^{A,B}, r_\mathrm{max}]$} using Equations (\ref{eq:formrateBHBH_multi}) and (\ref{eq:formrateBHBH_single}). We then randomly draw a pair of $w$ and $b$ for each $r$ from $P_{AB}(w,b)$, using Equations (\ref{APe1}) and (\ref{e:P(w|r)}) in Appendix \ref{sec:AppA}.

 EBBHs can form with $w$ as high as $w \simeq 0.05$ near $r_{\rm min}^{A,B}$. The region of validity of Equation (\ref{eq:impactlim}) for \mbox{$b_{\rm min} \propto w^{-1}$} and \mbox{$b_{\rm max} \propto w^{-9/7}$} is as follows. For high $w$, the condition \mbox{$b_{\rm max}(w) > b_{\rm min}(w)$} may be  violated over
\begin{equation}  \label{eq:wbmax}
   w_{{\rm max},b} = \frac{1}{4^3} \sqrt{ \frac{ 85 \pi \eta }{ 3 } } \, .
\end{equation}
 These systems cannot form an eccentric binary system, but may only suffer a direct head-on collision. We discard all such $(w,b)$ pairs in our MC sample that do not satisfy $w \leqslant w_{{\rm max},b}$. We find numerically that only $\la 0.01 \%$ ($\la 1\% $) of EBBHs form with $w \ga 0.05$ ($ w \ga w_{\mathrm{max},b}$) over the considered ranges of BH mass, SMBH mass, and BH population parameters; thus, direct head-on collisions represent a negligible fraction of sources.

 Furthermore, we keep only those $(w,b)$ pairs that avoid interaction with a third object satisfying Equation (\ref{eq:conddisruption}).

 For each random sample of $(w,b)$, we calculate the corresponding initial pericenter distance and eccentricity of the binary $(\rho_\mathrm{p0},e_0)$, using formulae introduced in Section \ref{subsec:FormBHB}.

 Given that most of these binaries have $e_0 \sim 1$ (Section \ref{subsec:DistIniParam}), we introduce the new variable
\begin{equation}   \label{eq:defeps0}
     \epsilon_0 \equiv 1-e_0^2
\end{equation}
 in order to accurately represent the PDF of $e_0$ in the vicinity of unity; see Section \ref{sec:NumRes}.

\subsection{Bounds on the Initial Orbital Parameters}
\label{sec:PIOP}

\begin{figure}
    \centering
    \includegraphics[width=85mm]{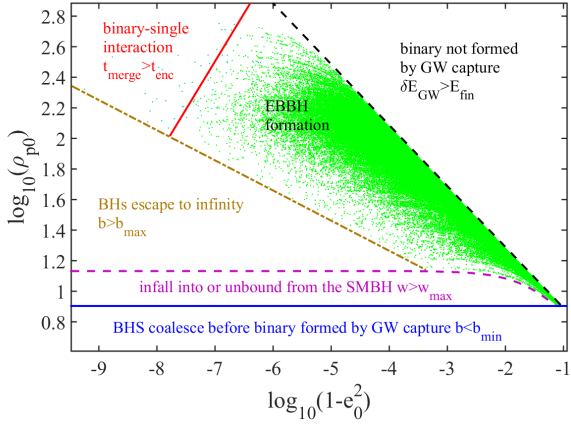}
    \caption{ \label{fig:boundary} A Monte Carlo realization of the EBBH parameters in the space of initial  eccentricity $e_0$ and initial dimensionless pericenter distance \mbox{$\rho_{\rm p0} = r/M_{\rm tot}$} showing the allowed parameter values of EBBHs. The initial eccentricity is typically very close to unity, so we show $\log(1-e_0^2)$ to resolve the deviation from $e_0 = 1$. The allowed region is bounded by curves that represent physical constraints on the binary formation and evolution, as indicated and explained in the text. We assume a Milky Way-sized nucleus with $M_{\rm SgrA*} = 4.3 \times 10^6 \, \Msun$ and show mergers with $m_A = m_B = 10 \, \Msun$ within a BH population with a power-law present-day  mass function (PMF) $dN/dm_{\rm BH} \propto m_{\rm BH}^{-2.35}$, a maximum BH mass of $m_\mathrm{BH,max} = 30 \, \Msun$, and a mass segregation parameter $p_0 = 0.5$ (Equation \ref{eq:pm}). We find that $99 \%$ of EBBHs (represented here by green dots) merge before a dynamical encounter may happen with a third object (i.e. $t_{\rm enc} > t_{\rm merge}$ holds), shown by a solid red line. }
\end{figure}

 Figure \ref{fig:boundary} shows a  $2.5 \times 10^5$ MC sample of EBBHs in the $\epsilon_0$--$\rhop$ plane for a Milky Way-sized nucleus with \mbox{$M_{\rm SMBH} = M_{\rm SgrA*}$} and for EBBHs having $m_A = m_B = 30 \, \Msun$. The distribution is bounded by five curves, where one of the conditions of a stable EBBH formation is violated as labeled. We will next discuss these conditions next and derive analytic approximations for each of them.

 Stable EBBH formation from two single objects bound to the central SMBH requires the following.
\begin{enumerate}[label=(\Roman*),ref=(\Roman*)]
  \item\label{i:I} They must avoid direct collision without forming a stable EBBH, $b>b_\mathrm{min}$.
  
  \item\label{i:II} They must also avoid escape to infinity, $b< b_\mathrm{max}$.
  
  \item\label{i:III} The initial kinetic energy must be positive \mbox{$E_{\rm kin} \geqslant 0$}. The two BHs radiate more GW energy during their encounter than the total initial kinetic energy, \mbox{$\delta E_\mathrm{GW} < E_{\rm kin}$}, which implies that $E_{\rm fin} \geqslant \delta E _\mathrm{GW}$.
  
  \item\label{i:IV} EBBHs should not interact with a third object throughout their evolutions, $t_{\rm merge} < t_\mathrm{enc}$.
  
  \item\label{i:V} BHs bound to the central SMBH have an initial velocity less than the escape velocity $v_{\rm max}(r)$, which sets a bound on the maximum relative speed, $w < w_\mathrm{max}(r) \equiv 2 v_{\rm max}(r)$ (Equation \ref{eq:wlim}). Objects exist in the GN within the radius of influence of the SMBH, but sufficiently far to avoid infall into the SMBH due to GW emission, \mbox{$r_{\rm min}^{AB} \leqslant r \leqslant r_{\rm max}$}. This sets a bound on $w$.
\end{enumerate}

 The above five criteria define sharp boundaries in the $\epsilon_0 - \rho_\mathrm{p0}$ plane as follows.

\begin{enumerate}[label=(\Roman*),ref=(\Roman*)]
    \item[\ref{i:I}] From Equation (\ref{eq:rhop0}), the leading-order approximation for $\rho_
          \mathrm{p0}=r_\mathrm{p0}/M_\mathrm{tot}$ in terms of $w$ is
          \begin{equation}  \label{eq:42}
              \rho_\mathrm{p0} \simeq \frac{ b^2 w^2 }{ 2 M_\mathrm{tot} ^2 } \, .
          \end{equation}
          Using Equation (\ref{eq:42}) and the definition of $b_\mathrm{min}$ given in Equation
          (\ref{eq:impactlim}), the criterion $b > b_\mathrm{min}$ can be translated into the
          constraint $\rho_ \mathrm{p0}> 8$, to leading post-Newtonian order.

    \item[\ref{i:II}] The criterion $b< b_\mathrm{max}$ means that the two encountering BHs must pass each other closely enough to become bound following GW emission, and thus $E_{\rm fin}$ must satisfy $E_{\rm fin}<0$. Using Equation (\ref{eq:Efin}) and the definition of $\epsilon_0$ given in (\ref{eq:defeps0}), this can be translated into a constraint $e_0 < 1$, or equivalently $\epsilon_0 > 0$. Moreover, this criterion must be satisfied for the possible range of $w$ values in the GN, $0 \leqslant w<w_{\rm max}$. Here, $w_\mathrm{max}$ is the highest $w$ occurring in the GN below $w_{{\rm max},b}$. Combining Equations (\ref{eq:Efin}) and (\ref{eq:46}) together with Equation (\ref{eq:dEGW}), $w$ may be expressed in Equation (\ref{eq:dEGW4}) in terms of $\epsilon_0$ and $\rho _\mathrm{p0}$ as
          \begin{equation} \label{eq:w_eps_rhop0}
             w\simeq \left (\frac{85\pi \eta}{6 \sqrt{2} \rho_\mathrm{p0}^{7/2}} -
             \frac{\epsilon_0}{2\rho_\mathrm{p0}}  \right )^{1/2} \, .
          \end{equation}
         Combining Equation (\ref{eq:42}) with Equation (\ref{eq:w_eps_rhop0}), $b$ can be expressed in terms of $\rho_{\rm p0}$ and $\epsilon_0$ as
         \begin{equation} \label{eq:b_termsepsrho}
           b = \sqrt{2 \rho_{\rm p0}} M_{\rm tot} \left (\frac{85\pi \eta}{6 \sqrt{2}
           \rho_\mathrm{p0}^{7/2}} - \frac{\epsilon_0}{2\rho_\mathrm{p0}} \right )^{1/4} \, .
         \end{equation}
         Conversely, solving for $\rho_{\mathrm{p}0}$ and $\epsilon_0$, gives Equation~(\ref{eq:42}) and
         \begin{align}
\epsilon_0 (b,w) &= \frac{ 340 \pi\, \eta M_{\rm tot}^5 }{ 3\, b^5 w^5 } -
  \frac{ b^2  w^4}{ M_{\rm tot}^2 } \, .
\end{align}
   The constraint on $b$ in Equation (\ref{eq:impactlim}) gives a boundary curve in the $\epsilon_0-
   \rho_\mathrm{p0}$ plane.

   \item[\ref{i:III}] Using Equations (\ref{eq:Efin}) and (\ref{eq:42}), $E_{\rm fin}$ can be expressed in terms of $\epsilon_0$ and $\rho_\mathrm{p0}$ as
      \begin{equation}   \label{eq:46}
          E_{\rm fin} \simeq - \frac{ M_\mathrm{tot} \eta \epsilon_0 }{ 4 \rho_\mathrm{p0} } \, .
      \end{equation}
       Finally, using Equation (\ref{eq:dEGW}) together with Equation (\ref{eq:46}), the criterion $\delta E _\mathrm{GW}< E_{\rm fin}$ defines the following upper limit on $\epsilon_0$:
       \begin{equation} \label{eq:47}
          \epsilon_0 < \frac{85\pi \eta}{3\sqrt{2}}\rho_\mathrm{p0}^{-\frac{5}{2}} \, .
       \end{equation}
       Note that the $\delta E_\mathrm{GW} = E_{\rm fin}$ case corresponds to the $w=0$ limit; see Equation (\ref{eq:Efin}).

     \item[\ref{i:IV}] The boundary curve in the $\epsilon_0-\rho_\mathrm{p0}$ plane defined by $t_{\rm merge} = t_\mathrm{enc}$ can only be constructed numerically by expressing all terms in Equations (\ref{eq:dEGW4}) and (\ref{eq:tGW}) as functions of $\epsilon_0$ and $\rho_\mathrm{p0}$. Here, $a_0$ and $c_0$ can be expressed in terms of $\epsilon_0$ and $\rho_\mathrm{p0}$ using Equations (\ref{eq:rho0}) and (\ref{eq:rhoe}). This boundary must be calculated at each radius in the GN because $n_{\rm tot}$ depends on $r$ (Equation \ref{eq:ntot}). However, an upper boundary can be constructed for the whole EBBH population merging in the GN by choosing $r=r_{\rm max}$, which we use to identify that region of the $\epsilon_0-\rho_\mathrm{p0}$ plane in which EBBHs surely interact with a third object before the merger phase of the EBBH evolution.

   \item[\ref{i:V}] BHs A and B should have a relative speed, $w$, lower than $2 v_{\rm max}(r)$ at distance $r$ from the center of the nucleus, i.e. $w<w_\mathrm{max}(r)\equiv 2 v_{\rm max}(r)< w_{\rm max}$. Using Equation (\ref{eq:w_eps_rhop0}), the criterion              $w<w_\mathrm{max}$ can be expressed as
              \begin{equation} \label{eq:wwmax}
               \epsilon_0 > \frac{85\pi \eta}{3 \sqrt{2}} \rho_\mathrm{p0}^{-5/2} - 2 \rho_\mathrm{p0}
               w_\mathrm{max}^2(r_{\rm min}^{A,B})
              \end{equation}
              in the $\epsilon_0$--$\rho_\mathrm{p0}$ plane.
\end{enumerate}

 These bounds on $\epsilon_0$ and $\rho_{\rm p0}$ lead to the following scaling relations. The expressions of $b_{\rm min}$ and $b_{\rm max}$ defined by Equation (\ref{eq:impactlim}) may be expressed in the form $M_{\rm tot} w^h \eta^{\alpha}$, where $h=-9/7$ and $-1$ and $\alpha = 0$ and $1/7$, respectively. Thus, the impact parameter lies in the range
\begin{equation}  \label{eq:bscaling}
  b \propto M_{\rm tot} w^h \eta^{\alpha} \, , \quad -9/7 \leq h \leqslant -1 \, , \quad
  0 \leqslant \alpha \leqslant 1/7 \, .
\end{equation}
 These relations, together with Equation (\ref{eq:42}), describe $\rho_{\rm p0}$. They may be written as
\begin{equation} \label{eq:rhoscaling}
  \rho_{\rm p0} \propto w^{\beta} \eta^{2 \alpha} \, ,  \quad  -4/7 \leqslant \beta \leqslant 0 \, .
\end{equation}
 To get a similar expression for $\epsilon_0$, we first examine $E_{\rm fin}$. Combining Equations (\ref{eq:dEGW}) and (\ref{eq:Efin}) together with Equations (\ref{eq:bscaling}) and (\ref{eq:rhoscaling}), $E_{\rm fin}$ can be given as
\begin{equation}
  E_{\rm fin} = C_1 \eta M_{\rm tot} w^2 - C_2 M_{\rm tot} w^{-7(h + 1)} \eta^{2-7 \alpha} \,,
\end{equation}
 where $C_1$ and $C_2$ are constants. Because $w \ll 1$ for encounters over the considered ranges of BH mass, SMBH mass, and BH population parameters (Section \ref{subsec:MCCode}), therefore $E_{\rm fin}$ is dominated by its second term, and thereby scales as
\begin{equation}  \label{eq:EfinScaling}
  E_{\rm fin} \propto - M_{\rm tot} w^{-7 (h + 1)} \eta^{2 - 7 \alpha} \, .
\end{equation}
 Finally, using Equations (\ref{eq:Efin}), (\ref{eq:bscaling}), and (\ref{eq:EfinScaling}) together yields
\begin{equation} \label{eq:epsscaling}
  \epsilon_0 \propto w^{\gamma} \eta^{\kappa} \, , \quad 0 \leqslant \gamma \leqslant 10/7 \, ,
  \quad 2/7 \leqslant \kappa \leqslant 1 \, .
\end{equation}
 These relations help interpret the results for MC simulations given next.

\section{Results}
\label{sec:NumRes}

 We now present the PDF of initial binary parameters describing EBBH events.

\subsection{Radial Distribution of EBBH Mergers in Galactic Nuclei}
\label{subsec:FormRate}

 Let us now determine the relative encounter rate as a function of the radius from the center of the GN, $P_{AB}(r)$, for objects with masses $m_A$ and $m_B$.

 In Appendix \ref{sec:AppA}, we show that the distribution of merger rates as a function of radius from the center of the GN has the form
\begin{align}  \label{eq:radiusPDF}
  \frac{\partial^3 \Gamma}{\partial r\partial m_A \partial m_B} & = 4 \pi^2 r^2 n_{m_A}(r) \,
  n_{m_B}(r) \, N_{\rm BH}^2\mathcal{F}(m_A) \mathcal{F}(m_B)
  \nonumber \\
  & \times  C_{rA} C_{rB} \left(\zeta_{\rm capt}'(r)- \zeta_{\rm coll}'(r) \right) \, ,
\end{align}
 where $n_m(r)$ and $N_{\rm BH}$ are given by Equations (\ref{eq:n(r)}) and (\ref{eq:encBH}), respectively, $C_{rA}$ and $C_{rB}$ are constants defined by Equation (\ref{eq:Cr}), and $\zeta_{\rm capt}'(r)$ and $\zeta_{\rm coll}'(r)$ are terms that follow from the cross section for GW capture and direct collisions, $\pi b_{\max}^2$ and $\pi b_{\min}^2$, respectively, defined by Equations (\ref{eq:captzeta}) and (\ref{eq:collzeta}) in Appendix \ref{sec:AppA}.

\begin{figure}
    \centering
    \includegraphics[width=80mm]{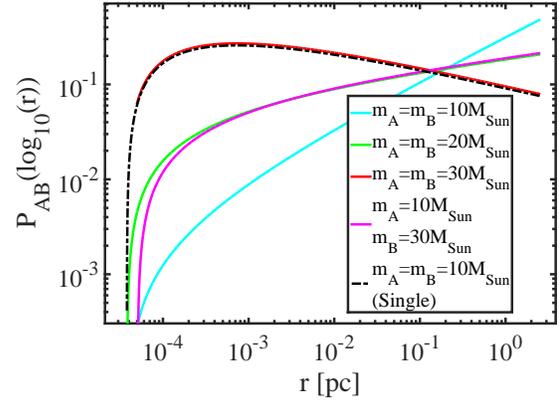}
\caption{ \label{fig:DisPAB} Probability density $\mathrm{log}_{10}(P_\mathrm{AB}(\mathrm{log}_{10}(r)))$ of an EBBH formed by BHs with masses $m_\mathrm{A}$ and $m_\mathrm{B}$ merging at $r$ from the central SMBH of a Milky Way-size nucleus. In this example, the mass of the central SMBH is \mbox{$M_{\rm SgrA*} = 4.3 \times 10^6 \, \Msun$}, and stellar populations surrounding the SMBH are assumed to form spherically symmetric and relaxed populations within the radius of influence of the SMBH, $r_{\rm max}$. We give $P_\mathrm{AB}(r)$, using Equations (\ref{eq:formrateBHBH_multi}) and (\ref{eq:formrateBHBH_single}), for various combinations of $m_\mathrm{A}$ and $m_\mathrm{B}$ values indicated in the figure legend, and for $r\in [r_{\rm min}^{A,B},r_{\rm max}]$ as defined by Equations (\ref{eq_rminAB}) and (\ref{eq:rmax}). Here, $r_{\rm min}^{A,B}$ defines the inner radius at which BHs \emph{A} and \emph{B} can still form an EBBH (see Appendix \ref{sec:CalRmin}). Solid lines correspond to a fiducial multi-mass BH population with a PMF $dN/dm_{\rm BH} \propto m_{\rm BH}^{-2.35}$, $m_\mathrm{BH,max} = 30 \, \Msun$, and $p_0=0.5$ in Equation (\ref{eq:pm}). The dashed-dotted line corresponds to a single-mass BH population of $10 \Msun$. }
\end{figure}

 For a multi-mass BH population, the probability distribution of the EBBH events among events with fixed BH masses $m_A$ and $m_B$ as a function of $r$ may be obtained by normalizing Equation (\ref{eq:radiusPDF}) over the radius range $[r_{\rm min}^{A,B}, r_\mathrm{max}]$, where BHs exist in the GN; see Appendix \ref{sec:CalRmin}. We get
\begin{align}  \label{eq:formrateBHBH_multi}
  P_{AB}(r)  = C_1 \left[\left( \frac{ 340 \pi \eta }{ 3 } \right)^{\frac{2}{7}}
  \frac{r^{-\frac{3}{14} - p_0 \frac{m_A + m_B}{ m_{\rm BH,max} }}} {M_{\rm SMBH}^
  {11/14}} - \frac{ 16 \, r^{-\frac{1}{2} - p_0
 \frac{m_A + m_B}{m_{\rm BH,max}}}}{M_{\rm SMBH}^{1/2}}\right] \, ,
\end{align}
 where $\eta=q/(1+q)^2$, $q=m_A/m_B$, and $C_1$ is a normalization constant set by the requirement $\int_ {r_{\min}}^{r_{\max}} P_{AB}(r) dr = 1$. For a single-mass distribution, $p_0=0.5$, and $m_A = m_B = \mathrm{max}(m_\mathrm{BH}) = 10 \, \Msun$ (i.e. $q=1$, $\eta=1/4$); thus we get
\begin{equation}  \label{eq:formrateBHBH_single}
  P_{AB}(r) = C_1 \left[\left( \frac{85 \pi}{3} \right)^{\frac{2}{7}}
  \frac{r^{-\frac{17}{14}}}{M_{\rm SMBH}^{11/14}} - \frac{16 \,
  r^{-3/2}}{ M_{\rm SMBH}^{1/2}}\right] \, .
\end{equation}

 The distribution $P_{AB} (\mathrm{log}_{10}(r))$ is displayed in Figure \ref{fig:DisPAB} for various component masses. For a single-mass BH population, the profile is identical to that of the most massive BHs in a multi-mass BH population when $p_0=0.5$; see Equation \ref{eq:pm}. We find that $P_{AB}(r)$ is steeper for more massive EBBHs as a function of radius, which is due to the steeper density gradient for higher masses (Equation \ref{eq:formrateBHBH_multi}). The profile of $P_{AB}(r)$ is weakly sensitive to the mass ratio for fixed $M_\mathrm{tot}$, i.e. proportional to $\eta^{2/7}$, which is due to the fact that $P_{AB}(r)$ is dominated by its first term in Equation (\ref{eq:formrateBHBH_multi}). Note that the merger rate distribution on a logarithmic radial scale is $d\Gamma_{AB}/d\ln r = r d\Gamma_{AB}/dr \propto r P_{AB}(r)$. By using Equation (\ref{eq:formrateBHBH_single}) for $P_{AB}(r)$ we find that $P_{AB} (\ln r) \propto r^{-3/14}$ for the most massive BHs in the BH population, showing that the merger rate is approximately uniform on a logarithmic scale between $10^{-4}$ and $3 \, \mathrm{pc}$. In fact, roughly $50 \%$ of the most massive EBBHs are formed in the innermost $4 \times 10^{-3} \, \mathrm{pc}$, and $97 \%$ of the events are within $1.2 \, \mathrm{pc}$. However, the least massive EBBHs with $m_A = m_B = 5 \, \Msun$, $p_0=0.5$, and $m_{A,B} \ll m_{\mathrm{ BH,max}}$ are distributed as $P(\ln r) \propto r^{11/14}$. Thus, only $\sim 2 \%$ of them are within $4 \times 10^{-3} \, \mathrm{pc}$, and only roughly $50 \%$ of them are within $1.2 \, \mathrm{pc}$. These low-mass binaries form preferentially further out, near the radius of influence.

 Figure \ref{fig:DisPAB} shows that $P_{AB}(r)$ drops off quickly near $r_{\rm min}^{A,B}$. Thus, for fixed component-mass EBBHs, the distributions of orbital parameters do not depend significantly on the exact value of $r_{\rm min}^{A,B}$. For EBBHs with fixed component masses, $r_{\rm min}^{A,B}$ is approximately $2 \times$ lower for the \citet{Belczynskietal2014_2} PMF than for the same BH mass range with a power-law PMF (Appendix \ref{sec:CalRmin}). As a consequence, $P_{AB}(r)$ and the distributions of orbital parameters are very similar for the \citet{Belczynskietal2014_2} and for a power-law PMF.

\subsection{Distributions of Initial Orbital Parameters}
\label{subsec:DistIniParam}

 In Appendix \ref{sec:AppA}, we determine the EBBH merger rate distribution as a function of relative velocity and impact parameter. The initial orbital parameters $e_0$ and $\rho_\mathrm{p0}$ may be calculated from these parameters using Equations (\ref{eq:rhop0}) and (\ref{eq:Efin}).

\begin{figure}
    \centering
    \includegraphics[width=85mm]{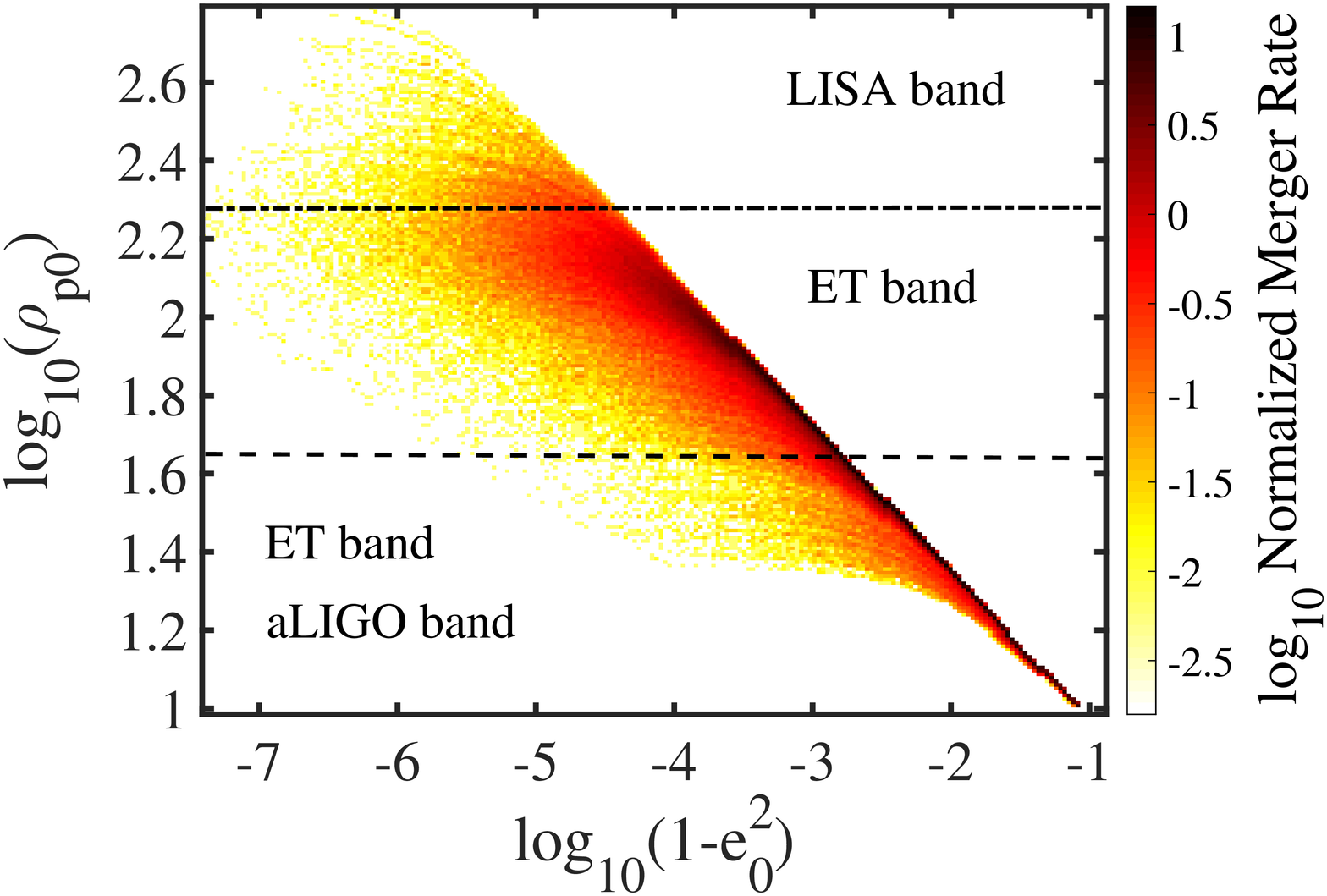}
    \includegraphics[width=85mm]{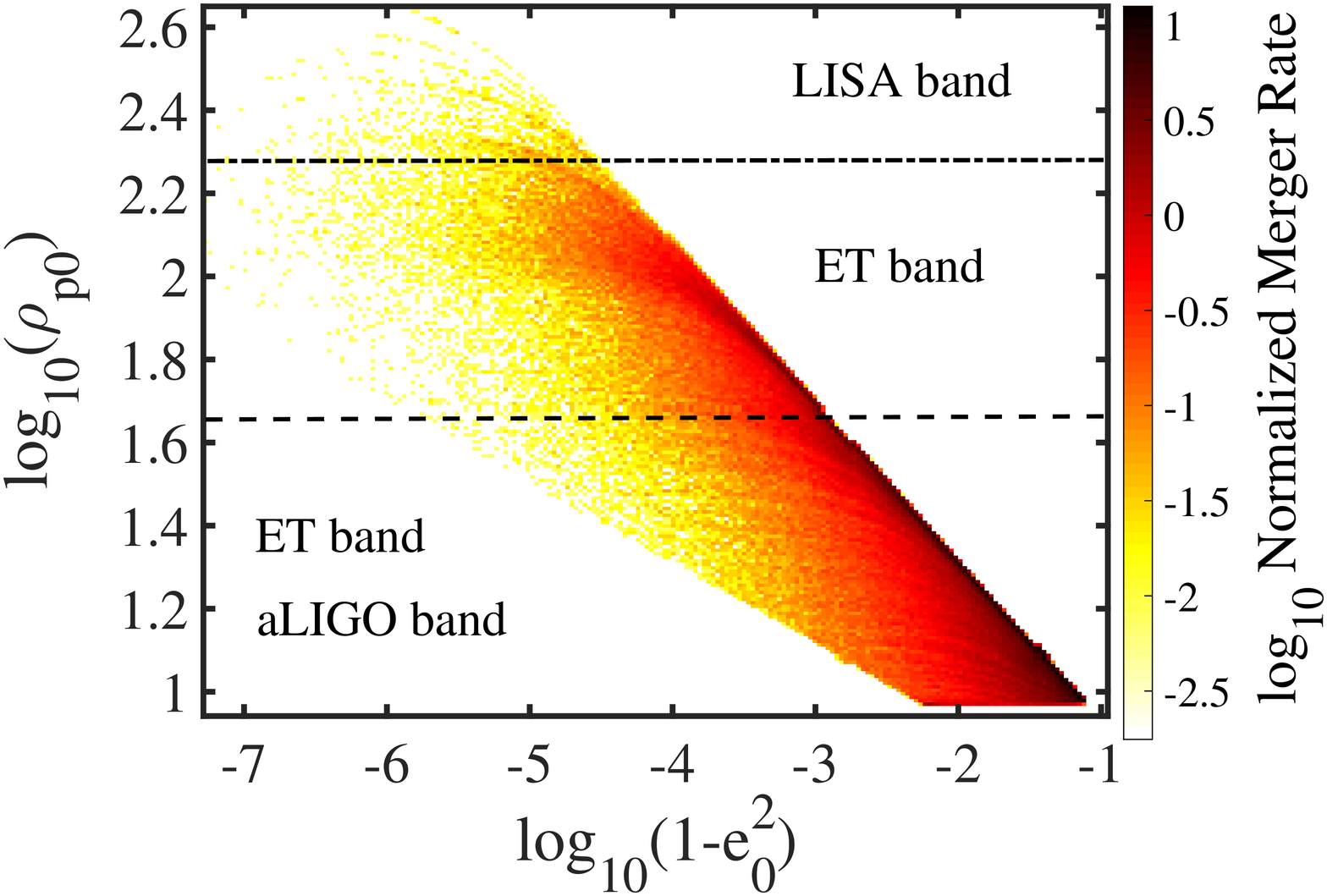}
    \includegraphics[width=85mm]{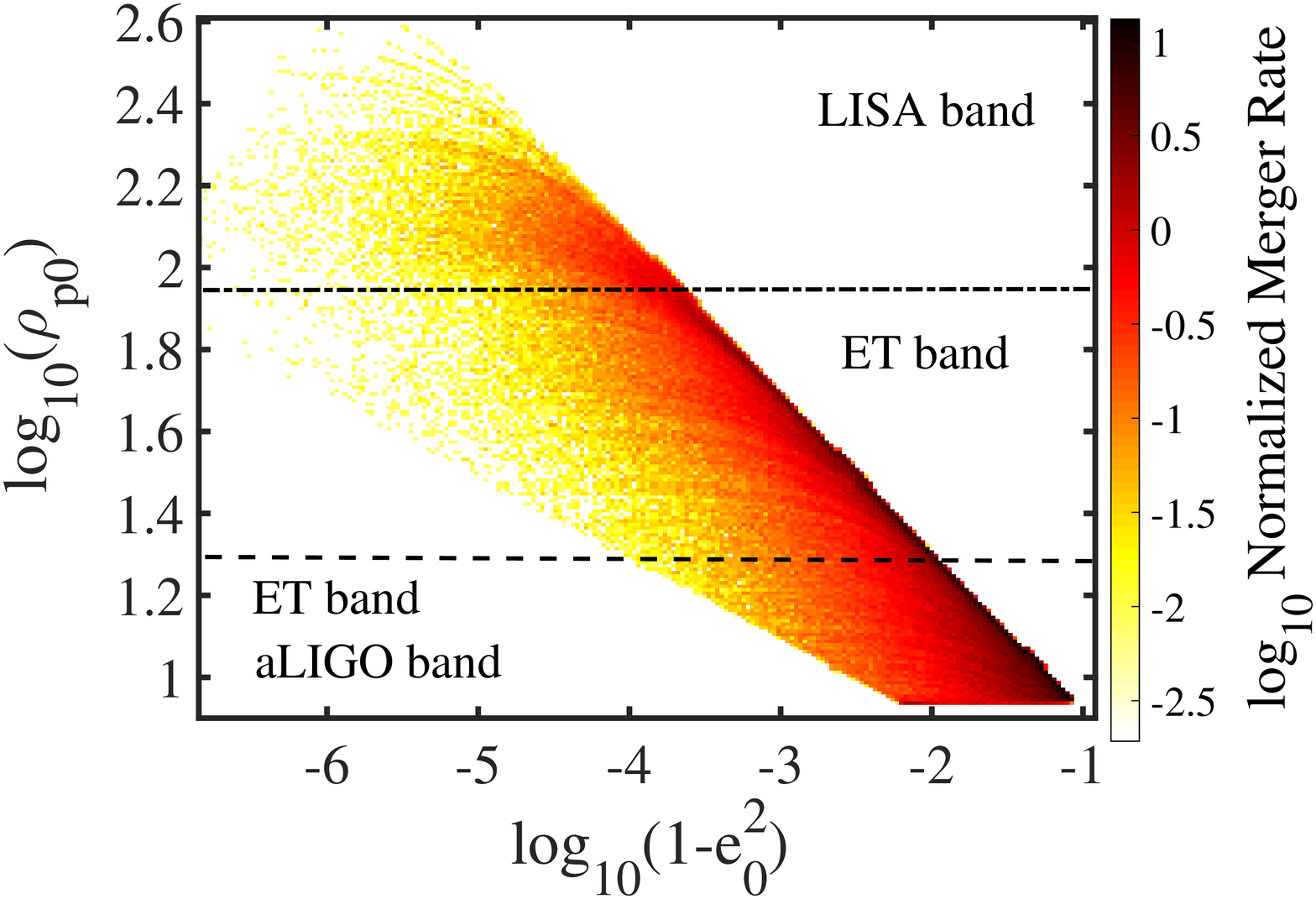}
 \caption{ \label{fig:DistFormPlane10Multi} Two-dimensional histograms of data points generated by the Monte Carlo code. The top and bottom panels correspond to $10 \Msun-10 \Msun$ and $30 \, \Msun-30 \, \Msun$ EBBHs forming in a fiducial multi-mass BH population with a PMF $dN/dm_{\rm BH} \propto m_{\rm BH}^{-2.35}$, $p_0 = 0.5$ and $m_\mathrm{BH,max} = 30 \, \Msun$, and the middle panel corresponds to EBBHs forming in a single-mass BH population. We assumed a Milky Way-size nucleus in all cases. We simulated $10^6$ binaries: those that did not interact with a third object before the merger phase of the EBBH evolution are shown. Histograms were normalized by the total number of the data points. The dashed and dashed-dotted lines correspond to $10 \, \mathrm{Hz}$ and $1 \, \mathrm{Hz}$ characteristic GW frequencies (Equation \ref{eq:rhoaLIGO}), separating the regions detectable by LISA and aLIGO; the intermediate region may be detectable by ET. We find, in panels from top to bottom, that about $(41,80,56) \%$ of EBBHs form within the aLIGO band. The middle and bottom panels are very similar due to the same radial distribution of EBBH merger rate inside the nucleus; see Figure \ref{fig:DisPAB}. From top to bottom panels, respectively, $(1.5, 4.5, 0.2) \%$) of sources form in the LISA band, and $(57.5, 15.5, 43.8) \%$) of sources form in the ET band but outside the aLIGO band.  }
\end{figure}

 Figure \ref{fig:DistFormPlane10Multi} shows the merger rate distribution as a function of the initial eccentricity and dimensionless pericenter distance. Two-dimensional histograms\footnote{Histograms are normalized by the total number of the data points.} are shown in the \mbox{$\mathrm{log}_{10} (\epsilon_0)$--$\mathrm{log}_{10}(\rho_{\rm p0})$} plane for three choices of the component masses in the three panels, respectively: $m_A = m_B = 10 \, \Msun$ and $m_A = m_B = 30 \, \Msun$ within a fiducial multi-mass BH population (i.e. $p_0=0.5$, $\beta=2.35$, and $m_\mathrm{BH,max}=30 \, \Msun$), and for a single-mass BH population with $m_A = m_B = 10 \, \Msun$. Clearly, the distributions are bounded by the boundaries derived in Figure \ref{fig:boundary} in Section \ref{sec:PIOP}. For low-mass EBBHs forming in a multi-mass BH population, we find that a significant fraction of events form relatively close to the $\delta E_\mathrm{GW} = E_{\rm fin}$ boundary line. For massive EBBHs forming in a multi-mass BH population, as well as for EBBHs forming in a single-mass BH population, we find that a significant fraction of events form at low $\rho_{\rm p0}$ close to the $\delta E_\mathrm{GW} = E_{\rm fin}$ boundary line, which corresponds to an approximately parabolic encounter. The merger rate distribution for a single-mass population is similar to the merger rate distribution for the heaviest components in a multi-mass distribution, as shown by the bottom two panels in Figure \ref{fig:DistFormPlane10Multi}.

 The Monte Carlo samples show that EBBHs typically form with high eccentricities ($0.9 \la e_0$), and $\rho_{\rm p0}$ ranges from $\sim 8-9$ up to $\sim 400-900$. The typical range of $\rho_{\mathrm{p}0}$ for GW capture sources can be understood using analytic arguments presented in Appendix \ref{sec:AppA}. Among mergers occuring at a fixed radius $r$, $\rho_{\mathrm{p}0}$ is distributed uniformly for $\rho_{\mathrm{p}0} \leqslant \rho_{\mathrm{p}0, \mathrm{uni}}$
 \begin{equation}\label{eq:rhop0uni}
\rho_{\mathrm{p}0, \mathrm{uni}} = \left(\frac{85\pi\eta}{24\sqrt{2}}\right)^{2/7} v_{\max}^{-4/7}
 =\left(\frac{85\pi \eta}{48\sqrt{2}}\right)^{2/7} \left(\frac{r}{M_{\mathrm{SMBH}}}\right)^{2/7} \, .
 \end{equation}
 Note that here the first term is unity ($0.9952$) for equal mass binaries with $\eta=0.25$. If half of the events are located within $0.01\,\mathrm{pc} \sim 5 \times 10^4 \, M_{\rm SgrA*}$, this implies a characteristic cutoff scale at $\rho_{\mathrm{p}0, \mathrm{uni}} \sim 22$. The cutoff scale $\rho_{\mathrm{p}0, \mathrm{uni}}$ depends on the mass of the binary system, due to their different radial merger distributions shown in Figure \ref{fig:DisPAB}. For the most massive BHs in a Milky Way-size nucleus, roughly $50 \%$ of them are located within $ 4 \times 10^{-3} \,\mathrm{pc} \sim 2000 \, M_{\rm SgrA*}$ which implies a cutoff scale at $\rho_{\mathrm{p}0, \mathrm{uni}} \sim 9$, while roughly $50 \%$ of the least massive EBBHs are located within \mbox{$1.2 \,\mathrm{pc} \sim 5.8 \times 10^6 \, M_{\rm SgrA*}$}, which implies a cutoff scale at $\rho_{\mathrm{p}0, \mathrm{uni}}\sim 85$.

 The characteristic scale of $\rho_{\mathrm{p}0}$ given by Equation (\ref{eq:rhop0uni}) for a given $v_{\max}$ is applicable for an arbitrary host population. The escape velocity is much lower in GCs than in GN with an SMBH. Plugging in $v_{\max}\sim 60\,\mathrm{km} \, \mathrm{s}^{-1}$ ($30\,\mathrm{km} \, \mathrm{s}^{-1}$) gives a cutoff scale of $\rho_{\mathrm{p}0, \mathrm{uni}} = 160$ (230). Thus, a smoking gun signature of GW capture events in high velocity dispersion environments (i.e. GN) is their small characteristic $\rho_{\mathrm{p}0}$ values below $\sim 100$.

 We find that the inner cutoff of $\rho_{\rm p0}$ is insensitive to the component masses of the merging binary, the mass of the SMBH, and the mass distribution of the BH population. This is due to the fact that its leading-order expression gives $\rho_{\rm p0} \geqslant 8$ (Section \ref{sec:PIOP}) independently of system parameters. However, the upper limit of $\rho_{\rm p0}$ is determined by the segment of boundary lines defined by criteria \ref{i:III} and \ref{i:IV} (Figure \ref{fig:boundary}), and it significantly depends on $M_{\rm SMBH}$ and the component masses $m_A$ and $m_B$. Moreover, the results are weakly sensitive to the $p_0$ and $\beta$ parameters of the BH mass distribution, and more sensitive to $m_{\rm BH,max}$.

\begin{figure}
    \centering
    \includegraphics[width=85mm]{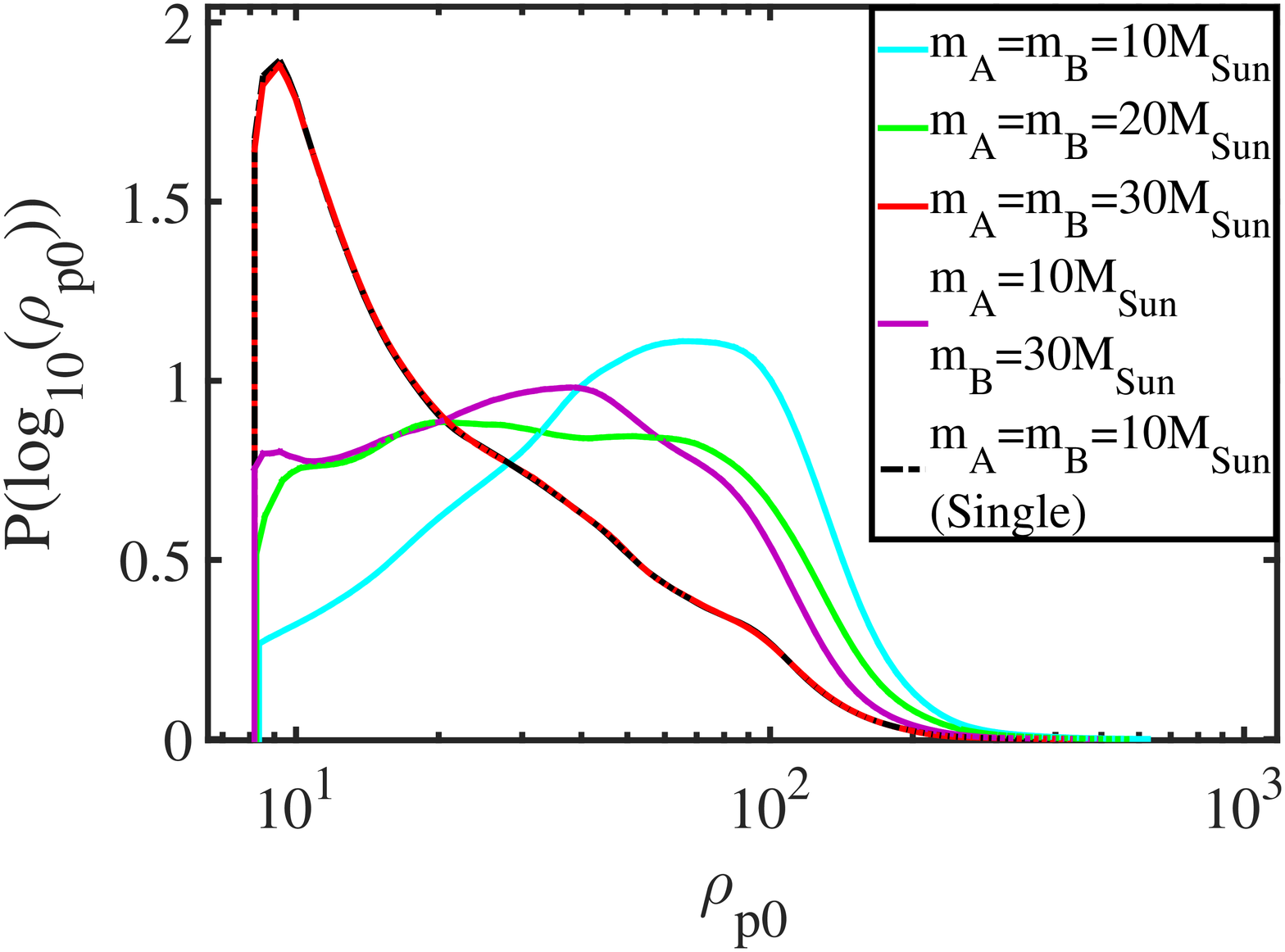}
    \includegraphics[width=85mm]{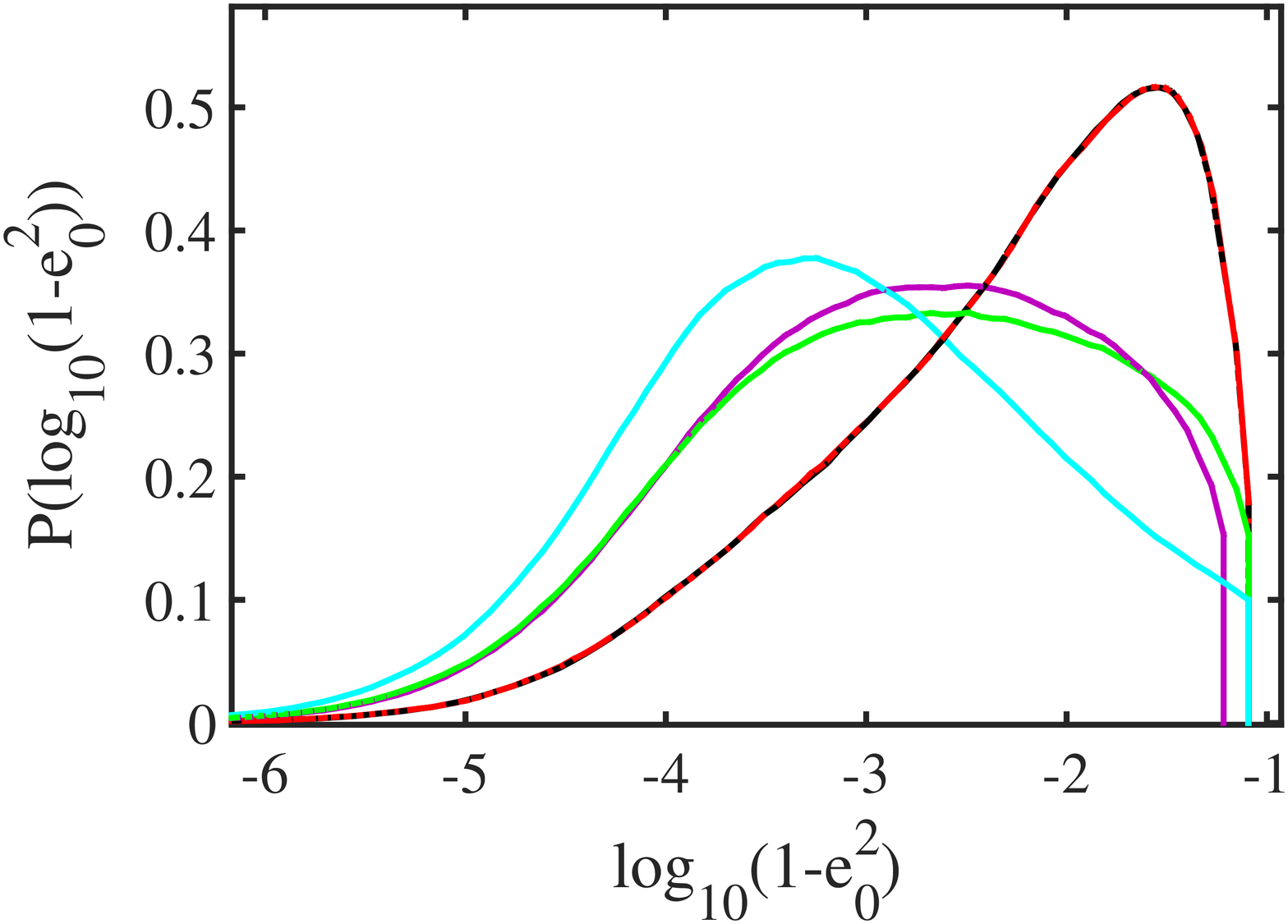}
\caption{ \label{fig:Distrhop0} PDF of $\rho_{\rm p0}$ (top panel) and $1-e_0^2$ (bottom panel) for different component-mass EBBHs forming in a Milky Way-size nucleus. Solid lines correspond to EBBHs forming in a fiducial multi-mass BH population with a PMF $dN/dm_{\rm BH} \propto m_{\rm BH}^{-2.35}$, $m_\mathrm{BH,max} = 30 \, \Msun$, $p_0=0.5$ in Equation (\ref{eq:pm}), and dashed-dotted lines correspond to EBBHs forming in a single-mass BH population. We find that more massive EBBHs have lower $\rho_{\rm p0}$ and $e_0$, and EBBHs forming in a single-mass BH populations have lower $\rho_{\rm p0}$ and $e_0$ than similar EBBHs forming in a multi-mass BH population. Moreover, the PDFs of $\rho_{\rm p0}$ and $e_0$ for EBBHs forming in a single-mass BH population are very similar to those of EBBHs formed by the most massive BHs of a multi-mass BH population, which is due to them having the same PDF profile for the EBBH merger rate inside the nucleus (Figure \ref{fig:DisPAB}). We find that both $\rho_{\rm p0}$ and $\epsilon_0$ are lower for binaries of lower $q$ at fixed $M_{\rm tot}$, and the discrepancy between PDFs of different mass ratio EBBHs increases with $M_{\rm tot}$.  }
\end{figure}

 Figure \ref{fig:Distrhop0} displays the one-dimensional PDF of the initial pericenter distance and eccentricity, respectively, by marginalizing the 2D distribution shown in Figure \ref{fig:DistFormPlane10Multi} over the other parameters for EBBH sources (cf. Figure 4 of \citealt{OLearyetal2009}). Different lines represent the distribution of different component masses for merging binaries. Solid lines show results for a fiducial multi-mass BH population with $dN/dm \propto m^{-2.35}$, $m_\mathrm{BH,max} = 30 \, \Msun$, and $p_0=0.5$.

 The main findings of Figures \ref{fig:Distrhop0}--\ref{fig:Distrho_Param} are summarized as follows.
\begin{enumerate}[label=(\Alph*),ref=(\Alph*)]
\item \label{i:A}
 At their formation, we find that most sources generally have pericenter distances in the range $\rho_{\mathrm{p}0} \sim 10-80$ and eccentricity $e_0\sim 0.9-0.9999$. More massive EBBHs form with systematically lower $\rho_{\rm p0}$ and with less extreme $e_0$ (e.g. $\rho_{\mathrm{p}0}\lesssim 20$ and $e_0\sim 0.9$ for the heaviest EBBHs). This is due to the fact that, in a multi-mass BH population, more massive EBBHs form closer to the center of the nucleus (Figure \ref{fig:DisPAB}), ergo $w$ extends to higher values $2v_{\rm max}(r)$ (Equation \ref{eq:vmax}), and binary formation by GW capture implies $\rho_{\rm p0} \propto w^{\beta}$ with $-4/7 \leqslant \beta \leqslant 0$ and $\epsilon_0 \propto w^{\gamma}$ with $0 \leqslant \gamma \leqslant 10/7$ (Equations \ref{eq:rhoscaling} and \ref{eq:epsscaling}).\footnote{The encounter cross section is typically dominated by gravitational focusing, which implies a uniform $P(\rho_{\mathrm{p}0})$ distribution for isotropic distribution of relative velocity with a fixed magnitude $w$ \citep{Kocsisetal2006,OLearyetal2009} in the range $0 \leqslant \rho_{\mathrm{p}0} \leqslant k \eta^{2/7} w^{-4/7}$, where $k=(85\pi/6\sqrt 2)^{2/7}$. The decrease of $P(\rho_{\mathrm{p}0})$ is due to the distribution of $w$ in the GN.}

\begin{figure}
    \centering
    \includegraphics[width=85mm]{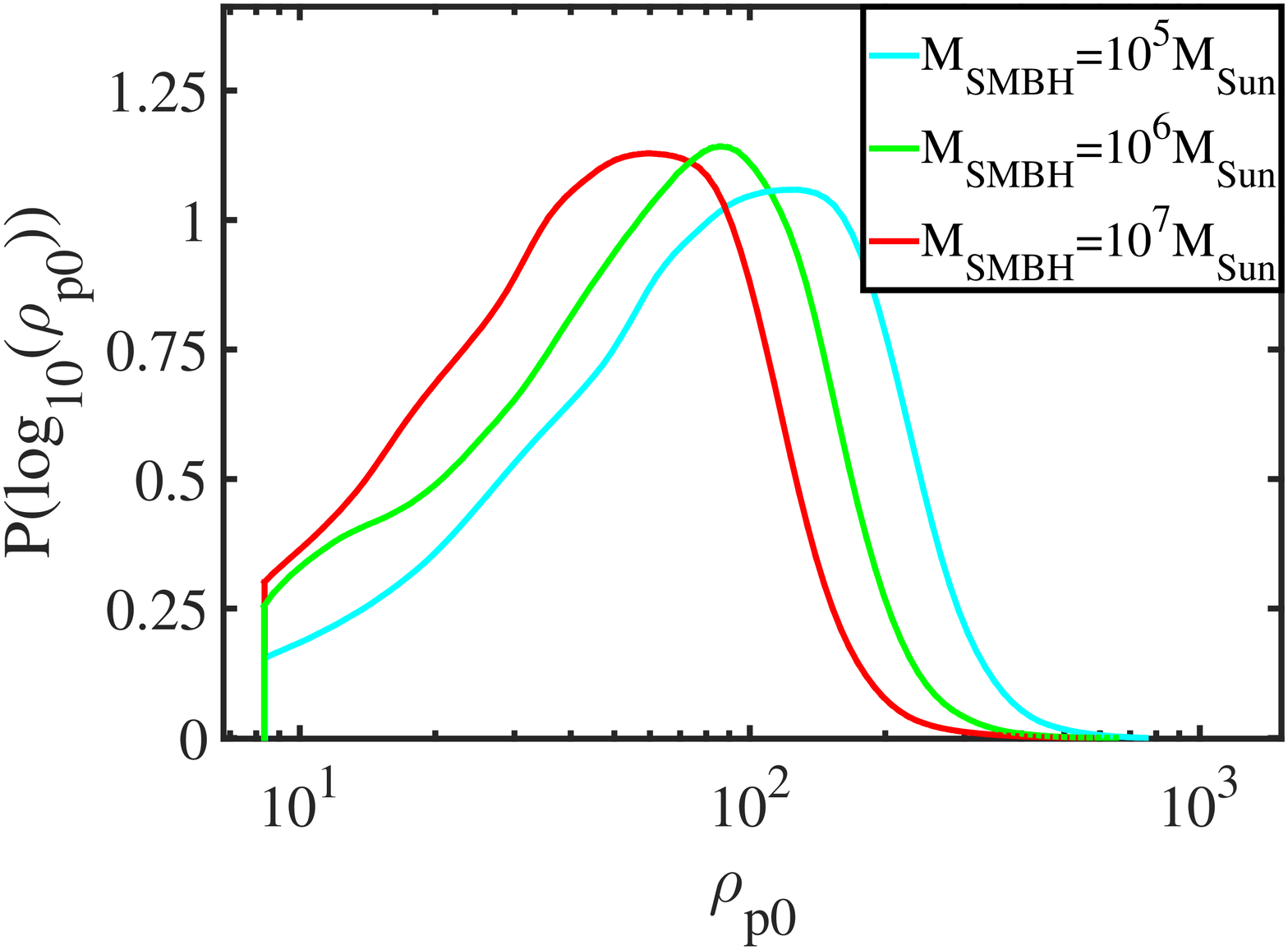}
    \includegraphics[width=85mm]{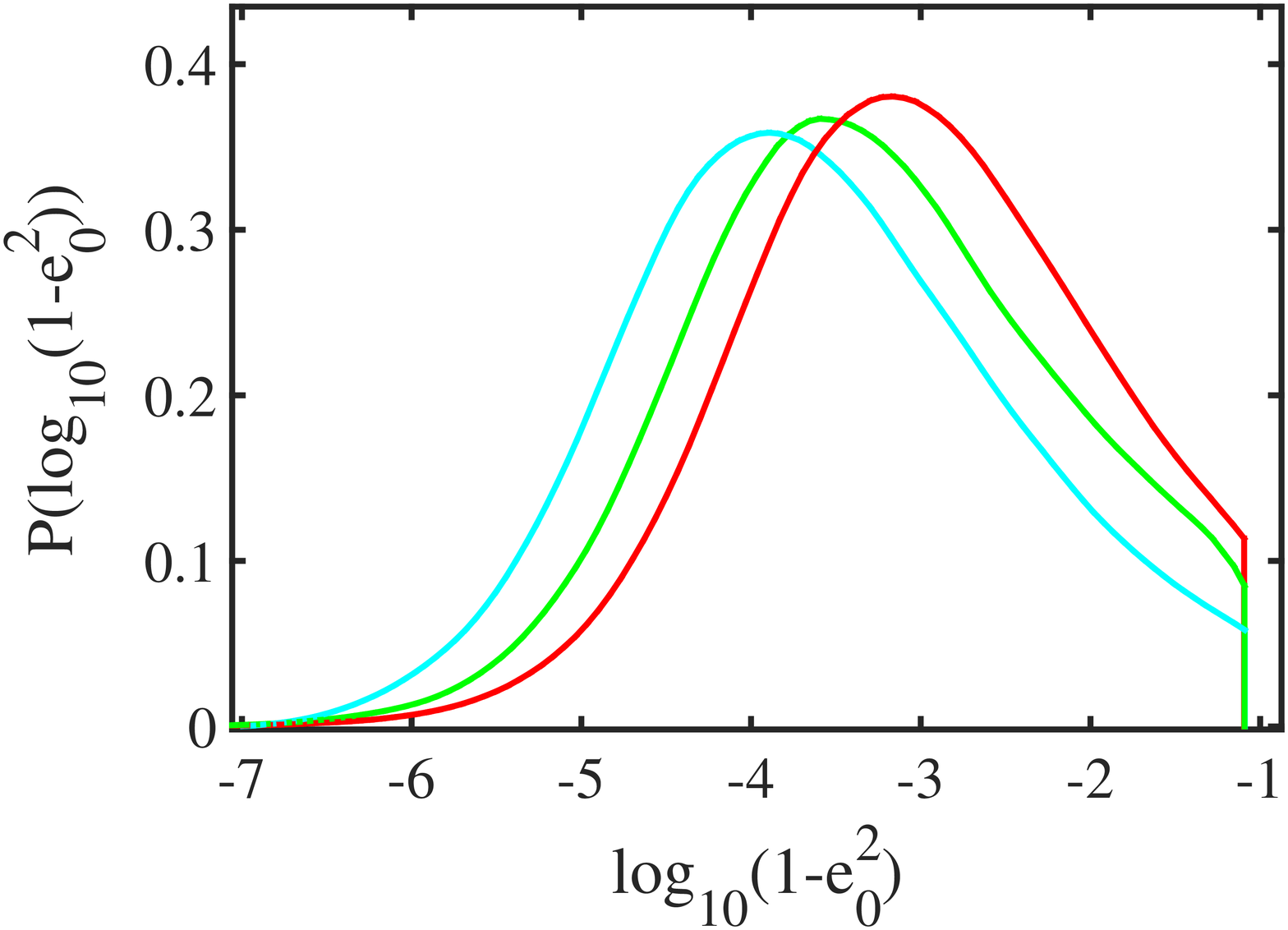}
\caption{ \label{fig:SMBHdeprhopaeps0} An example of how different SMBH masses influence the PDF of $\rho_{\rm p0}$ (top panel) and $1-e_0^2$ (bottom panel). Results are shown for $10 \, \Msun - 10 \, \Msun$ EBBHs forming in a fiducial multi-mass BH population with a PMF $dN/dm_{\rm BH} \propto m_{\rm BH}^{-2.35}$, $m_\mathrm{BH,max} = 30 \, \Msun$, and $p_0=0.5$. }
\end{figure}

\item \label{i:B} The maximum BH mass in the distribution $m_{\rm BH,max}$ significantly influences the PDF of $\rho_{\rm p0}$ and $e_0$ for different BH mass mergers. However, the mass segregation parameter (assuming that it is in the range $0.5<p_0<0.6$) and the slope of the BH mass function (assuming the range $1 < \beta < 3$) do not significantly influence these PDFs. This result is consistent with analytic expectations presented in Section \ref{subsec:FormRate}.

\item \label{i:C} Both $\rho_{\rm p0}$ and $\epsilon_0$ are somewhat lower for EBBHs of lower $q$ at fixed $M_{\rm tot}$, which arise due to the Equations (\ref{eq:rhoscaling}) and (\ref{eq:epsscaling}) for $\rho_{\rm p0}$ and $\epsilon_0$. However, because $\rho_{\mathrm{p}0, \mathrm{uni}}\propto \eta^{2/7}$, this dependence is rather weak. Moreover, the difference between the PDFs of EBBHs with different mass ratios increases with $M_{\rm tot}$ due to the $M_{\rm tot}$ dependence of the minimum radius, $r_{\rm min}^{AB}$, and the corresponding increased value of $w$ for increasing $M_{\rm tot}$.

\begin{figure}
    \centering
    \includegraphics[width=85mm]{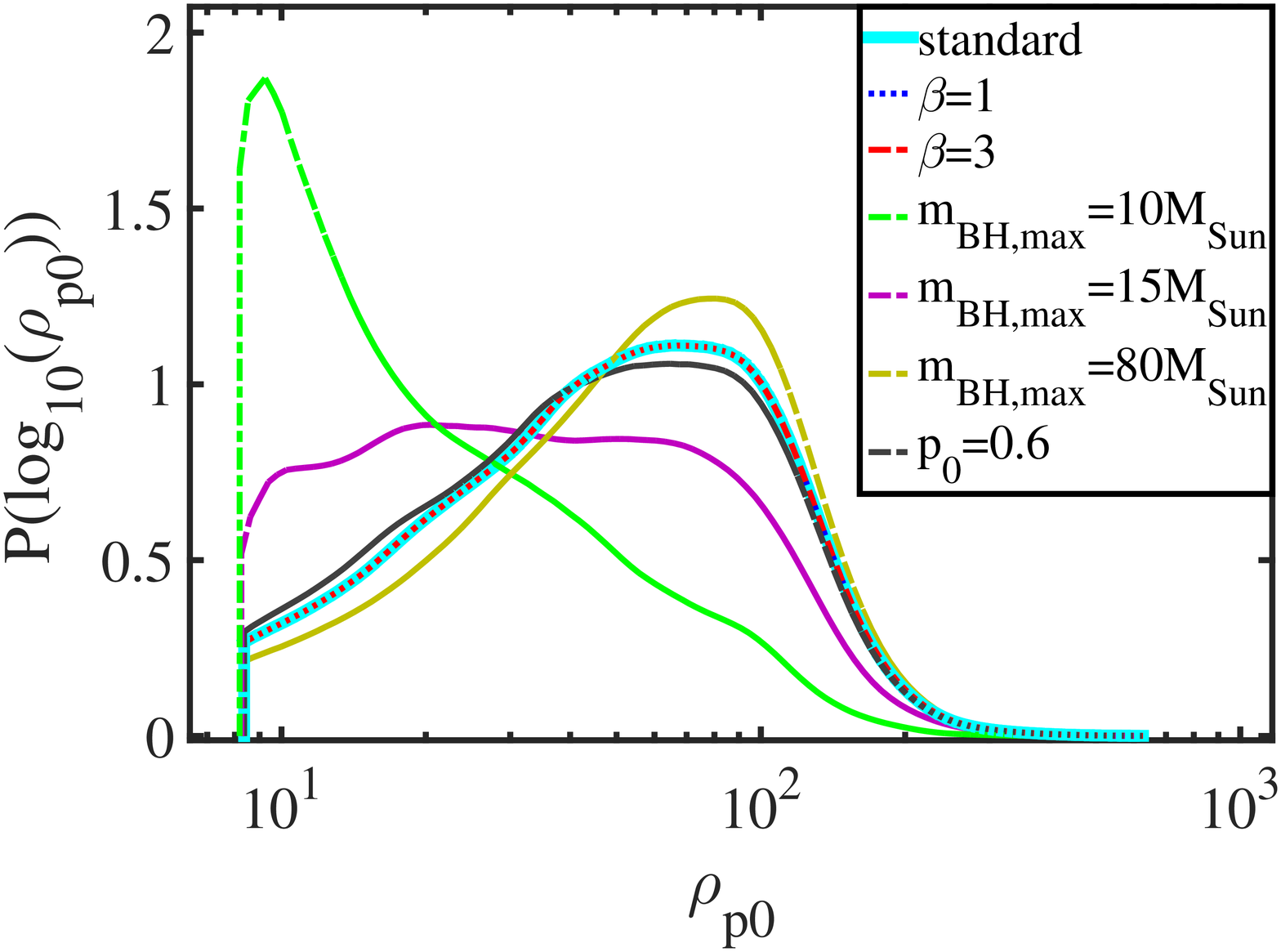}
    \includegraphics[width=85mm]{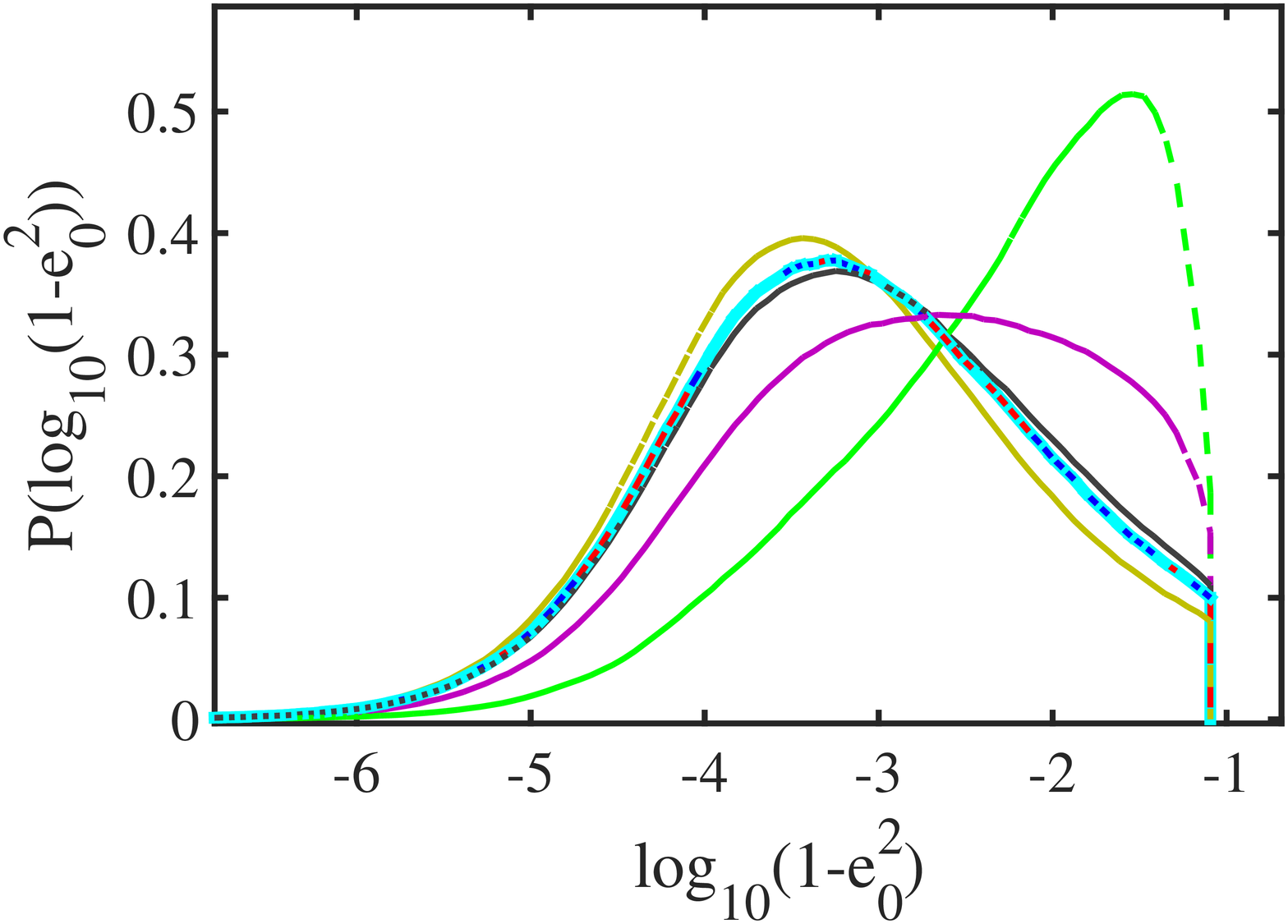}
\caption{ \label{fig:Distrho_Param} An example of how different multi-mass BH population parameters influence the PDF of $\rho_{\rm p0}$ (top panel) and $1-e_0^2$ (bottom panel). Results are shown for $10 \, \Msun - 10 \, \Msun$ EBBHs forming in a Milky Way-size nucleus. Solid thick lines correspond to a fiducial multi-mass BH population with a PMF $dN/dm_{\rm BH} \propto m_{\rm BH}^{-2.35}$, $m_\mathrm{BH,max} = 30 \, \Msun$, and $p_0 =0.5$, while dashed-dotted and dotted lines correspond to modified parameter values for a multi-mass BH population as shown in the legend of the \emph{top panel}. Consistently with analytic expectations, only $m_{\rm BH,max}$ significantly influences the PDF of $\rho_{\rm p0}$ and $1-e_0^2$.  }
\end{figure}

\item \label{i:D} In simulations with different $M_{\mathrm{SMBH}}$ (Figure \ref{fig:SMBHdeprhopaeps0}), we find that EBBHs form with lower $\rho_{\rm p0}$ and $e_0$ around more massive SMBHs, independently of the binary component masses and the BH population parameters. This is due to the fact that the average relative velocity with which BHs form EBBHs in the GN increases with $M_{\rm SMBH}$ (Figure \ref{fig:w_Avr_mSMBH}). Thus, by considering Equations (\ref{eq:rhoscaling}) and (\ref{eq:epsscaling}), we find that both $e_0$ and $\rho_{\mathrm{p}0}$ systematically decrease with $M_{\rm SMBH}$.

\item \label{i:E} In a single-mass BH population, $10 \, \Msun-10 \, \Msun$ EBBHs form with lower $\rho_{\rm p0}$ and $e_0$ than those in a multi-mass BH population. Moreover, the PDF of $\rho_{\rm p0}$ and $\epsilon_0$ are very similar for EBBHs forming in a single-mass BH population and EBBHs formed by the most massive BHs in a multi-mass BH population, which is due to the similarity between the corresponding EBBH merger rate distributions and $r_{\rm min}^{A,B}$ values (Figure \ref{fig:DisPAB}).
\end{enumerate}

 Eccentric binaries emit a GW signal with a broad spectrum of frequencies. Because the LIGO/VIRGO-type GW instruments are sensitive above $f_{\rm aLIGO} = 10 \, \mathrm{Hz}$, we assume that those EBBHs form within the aLIGO band in the $\epsilon_0$--$\rho_{\rm p0}$ plane for which $f_{\rm GW}>f_{\rm aLIGO}$, where $f_{\rm GW}$ is the peak GW frequency associated with the harmonic. This leads to the maximal emission of GW radiation being estimated as
\begin{equation}  \label{eq:fGW}
  f_{\rm GW} = \frac{2 (1+e)^{1.1954}}{(1-e^2)^{3/2}} \frac{M_{\rm tot}^{1/2}}{2 \pi a^{3/2}} \, ;
\end{equation}
 see \citet{Wen2003}. Here, $f_{\rm GW}$ defines the peak GW frequency. For any initial condition, the function $a\equiv a(e)$ is given by \citet{Peters1964}. Because EBBHs form with $e_0 \sim 1$, in this case, after combining Equation (\ref{eq:fGW}) with Equation (\ref{eq:rho0}), $f_{\rm GW}$ can be expressed as $f_{\rm GW,0} = (2^{0.3046} \pi M_{\rm tot} \rho_{\rm p0}^{3/2})^{-1}$. Thus, those EBBHs form within the aLIGO band for which $\rho_{\rm p0}$ does not exceed the limit
\begin{align}  \label{eq:rhoaLIGO}
  \rho_{\rm p0,aLIGO} &=  (2^{0.3046} \pi M_{\rm tot} f_{\rm aLIGO})^{-2/3} \nonumber\\&= 40.9
  \left(\frac{M_{\mathrm{tot}}}{20\,\Msun}\right)^{-2/3} \left( \frac{f_{\rm aLIGO}}{10\,\mathrm{Hz}}
  \right)^{-2/3}  \, .
\end{align}
 Note that this lies in the expected range of the $\rho_{\mathrm{p}0}$ distribution of GW capture sources given in Equation (\ref{eq:rhop0uni}).

 Because $\rho_{\mathrm{p}0}$ extends to relatively high values, a significant fraction of EBBHs may form with $f_{\rm GW}$ below the sensitive frequency band of aLIGO. For ground-based GW detectors, the lower bound of the sensitive frequency band is around $\sim 1 \, \mathrm{Hz}$, for the Einstein Telescope\footnote{\url{http://www.et-gw.eu/}} (ET, \citealt{Punturoetal2010}); for space-based GW observatories, the upper bound of the sensitive frequency band for the Laser Interferometer Space Antenna\footnote{\url{https://www.elisascience.org/}} (LISA) is around $\sim 1 \, \mathrm{Hz}$ \citep{AmaroSeoaneetal2013}. Figure \ref{fig:DistFormPlane10Multi} shows the regions where EBBHs form in the ET and LISA frequency bands, respectively.

\begin{table}
   \caption{ \label{tab:FracM_SMBH} The fraction of EBBHs having residual eccentricities larger than $0.1$ when their GW signals enter the aLIGO band ($F_{{\rm res} > 0.1}$), of EBBHs forming within the aLIGO band ($F_{\rm aLIGO}$), and of EBBHs having eccentricities larger than $0.1$ at the last stable orbit (LSO, $F_{ {\rm LSO} > 0.1}$). We use \emph{Single} to denote rows corresponding to a single-mass BH population, otherwise a fiducial multi-mass BH population is considered with a PMF $dN/dm_{\rm BH} \propto m_{\rm BH}^{-2.35}$, $m_\mathrm{BH,max} = 30 \, \Msun$, and $p_0 = 0.5$; and $M_{\SgrA}$ denotes the mass of the SMBH for a Milky Way-size nucleus. Systems were evolved from their initial orbit using the evolution equations of \citet{Peters1964}. $F_{\rm aLIGO}$ increases with $M_{\rm SMBH}$ at fixed component masses for both type of BH populations, which arises due to the decrease of $\rho_{\rm p0}$ against $M_{\rm SMBH}$. Consequently, a lower-and-lower fraction of EBBHs are needed to evolve from their initial orbit up until their GW frequency $f_{\rm GW}$ \citep{Wen2003} reaches the aLIGO band, which increases $F_{ {\rm res} > 0.1}$, and $F_{e_{\rm LSO} > 0.1}$ with $M_{\rm SMBH}$ as well. Moreover, we find that $F_{ {\rm res} > 0.1}$ ($F_{\rm aLIGO}$, and $F_{ {\rm LSO} > 0.1}$) is higher for EBBHs of lower $q$ at fixed $M_{\rm tot}$, and the discrepancy between $F_{{\rm res} > 0.1}$ ($F_{\rm aLIGO}$, and $F_{ {\rm LSO} > 0.1}$) of different $q$ increases with $M_{\rm tot}$. }
\centering
   \begin{tabular}{@{}ccccc}
   \hline
    $m_A - m_B \, \, (\Msun)$ & $M_{\rm SMBH} \, \, (\Msun)$ & $F_{ {\rm res} > 0.1}$ &
    $F_{\rm aLIGO}$  &  $F_{ {\rm LSO} > 0.1}$  \\
   \hline\hline
   $5 - 5$  &  $10^5$  &  $88 \%$  &  $35 \%$   &  $3 \%$ \\
   $10 - 10$  &  $10^5$  &  $69 \%$  &  $25 \%$  &  $4 \%$ \\
   $10 - 10$ (Single) &  $10^5$  &  $96 \%$  &  $78 \%$  &  $35 \%$ \\
   $5 - 15$  &  $10^5$  &  $73 \%$  &  $27 \%$  &  $4 \%$ \\
   $10 - 30$  &  $10^5$  &  $68 \%$  &  $32 \%$  &  $12 \%$ \\
   $20 - 20$  &  $10^5$  &  $69 \%$  &  $31 \%$  &  $12 \%$ \\
   $30 - 30$ &  $10^5$  &  $84 \%$  &  $51 \%$  &  $32 \%$ \\
   $5 - 5$  &  $10^6$  &  $96 \%$ &  $48 \%$  &  $5 \%$ \\
   $10 - 10$  &  $10^6$  &  $85 \%$  &  $35 \%$  &  $7 \%$ \\
   $10 - 10$ (Single)  &  $10^6$  &  $97 \%$  &  $79 \%$  &  $36 \%$ \\
   $5 - 15$  &  $10^6$  &  $87 \%$  &  $36 \%$  &  $6 \%$ \\
   $10 - 30$  &  $10^6$  &  $78 \%$  &  $38 \%$  &  $15 \%$ \\
   $20 - 20$  &  $10^6$  &  $80 \%$  &  $38 \%$  &  $16 \%$ \\
   $30 - 30$ &  $10^6$  &  $86 \%$  &  $54 \%$  &  $36 \%$ \\
   $5 - 5$  &  $M_{\SgrA}$  &  $98 \%$  &  $58 \%$  &  $6 \%$ \\
   $10 - 10$  &  $M_{\SgrA}$   &  $91 \%$  &  $41 \%$  &  $7 \%$ \\
   $10 - 10$ (Single)  &  $M_{\SgrA}$  &  $98 \%$  &  $80 \%$  &  $37 \%$ \\
   $5 - 15$  &  $M_{\SgrA}$  &  $93\%$  &  $44 \%$  &  $7 \%$ \\
   $10 - 30$  &  $M_{\SgrA}$  &  $81 \%$  &  $43 \%$  &  $17 \%$ \\
   $20 - 20$  &  $M_{\SgrA}$  &  $82 \%$  &  $41 \%$  &  $17 \%$ \\
   $30 - 30$  &  $M_{\SgrA}$  &  $88 \%$  &  $56 \%$  &  $37 \%$ \\
   $5 - 5$  &  $10^7$  &  $99 \%$  &  $64 \%$  &  $7 \%$ \\
   $10 - 10$  &  $10^7$  &  $88 \%$  &  $45 \%$  &  $8 \%$ \\
   $10 - 10$ (Single)  &  $10^7$  &  $99 \%$  &  $81 \%$  &  $38 \%$ \\
   $5 - 15$  &  $10^7$  &  $96 \%$  &  $49 \%$  &  $9 \%$ \\
   $10 - 30$  &  $10^7$  &  $88 \%$  &  $46 \%$  &  $18 \%$ \\
   $20 - 20$  &  $10^7$  &  $86 \%$  &  $43 \%$  &  $18 \%$ \\
   $30 - 30$  &  $10^7$  &  $90 \%$  &  $57 \%$  &  $38 \%$ \\
   \hline\hline
   \end{tabular}
\end{table}

 \begin{table}
   \caption{ \label{tab:FracM_SMBH_2} Same as Table \ref{tab:FracM_SMBH}, but for a source population with $m_\mathrm{BH, max} = 80 \, \Msun$. }
\centering
   \begin{tabular}{@{}ccccc}
   \hline
    $m_A - m_B \, \, (\Msun)$ & $M_{\rm SMBH} \, \, (\Msun)$ & $F_{ {\rm res} > 0.1}$ &
    $F_{\rm aLIGO}$  &  $F_{ {\rm LSO} > 0.1}$  \\
   \hline\hline
   $5 - 5$   &  $10^5$  &  $89 \%$  &  $37 \%$  &  $3 \%$ \\
   $10 - 10$ &  $10^5$  &  $68 \%$  &  $24 \%$  &  $4 \%$ \\
   $5 - 15$  &  $10^5$  &  $70 \%$  &  $24 \%$  &  $4 \%$ \\
   $10 - 30$ &  $10^5$  &  $51 \%$  &  $16 \%$  &  $5 \%$ \\
   $20 - 20$ &  $10^5$  &  $50 \%$  &  $9 \%$  &  $4 \%$ \\
   $30 - 30$ &  $10^5$  &  $43 \%$  &  $13 \%$  &  $6 \%$ \\
   $50 - 50$ &  $10^5$  &  $41 \%$  &  $11 \%$  &  $10 \%$ \\
   $20 - 80$ &  $10^5$  &  $56 \%$  &  $13 \%$  &  $12 \%$ \\
   $60 - 60$ &  $10^5$  &  $50 \%$  &  $13 \%$  &  $15 \%$ \\
   $50 - 70$ &  $10^5$  &  $49 \%$  &  $12 \%$  &  $14 \%$ \\
   $70 - 70$ &  $10^5$  &  $67 \%$  &  $11 \%$  &  $26 \%$ \\
   $80 - 80$ &  $10^5$  &  $63 \%$  &  $12 \%$  &  $29 \%$ \\
   $5 - 5$   &  $10^6$  &  $96 \%$  &  $47 \%$  &  $5 \%$ \\
   $10 - 10$ &  $10^6$  &  $81 \%$  &  $29 \%$  &  $5 \%$ \\
   $5 - 15$  &  $10^6$  &  $86 \%$  &  $30 \%$  &  $5 \%$ \\
   $10 - 30$ &  $10^6$  &  $66 \%$  &  $21 \%$  &  $7 \%$ \\
   $20 - 20$ &  $10^6$  &  $63 \%$  &  $13 \%$  &  $6 \%$ \\
   $30 - 30$ &  $10^6$  &  $55 \%$  &  $16 \%$  &  $7 \%$ \\
   $50 - 50$ &  $10^6$  &  $58 \%$  &  $14 \%$  &  $12 \%$ \\
   $20 - 80$ &  $10^6$  &  $51 \%$  &  $15 \%$  &  $13 \%$ \\
   $60 - 60$ &  $10^6$  &  $57 \%$  &  $15 \%$  &  $18 \%$ \\
   $50 - 70$ &  $10^6$  &  $55 \%$  &  $14 \%$  &  $17 \%$ \\
   $70 - 70$ &  $10^6$  &  $62 \%$  &  $15 \%$  &  $28 \%$ \\
   $80 - 80$ &  $10^6$  &  $68 \%$  &  $15 \%$  &  $34 \%$ \\
   $5 - 5$   &  $10^7$  &  $99 \%$  &  $67 \%$  &  $7 \%$ \\
   $10 - 10$ &  $10^7$  &  $94 \%$  &  $40 \%$  &  $7 \%$ \\
   $5 - 15$  &  $10^7$  &  $96 \%$  &  $43 \%$  &  $8 \%$ \\
   $10 - 30$ &  $10^7$  &  $79 \%$  &  $29 \%$  &  $10 \%$ \\
   $20 - 20$ &  $10^7$  &  $78 \%$  &  $18 \%$  &  $8 \%$ \\
   $30 - 30$ &  $10^7$  &  $69 \%$  &  $21 \%$  &  $10 \%$ \\
   $50 - 50$ &  $10^7$  &  $61 \%$  &  $16 \%$  &  $15 \%$ \\
   $20 - 80$ &  $10^7$  &  $67 \%$  &  $20 \%$  &  $18 \%$ \\
   $60 - 60$ &  $10^7$  &  $68 \%$  &  $18 \%$  &  $22 \%$ \\
   $50 - 70$ &  $10^7$  &  $63 \%$  &  $17 \%$  &  $21 \%$ \\
   $70 - 70$ &  $10^7$  &  $66 \%$  &  $18 \%$  &  $29 \%$ \\
   $80 - 80$ &  $10^7$  &  $72 \%$  &  $18 \%$  &  $36 \%$ \\
   \hline\hline
   \end{tabular}
\end{table}

\begin{table}
   \caption{\label{tab:Frac_BHPopParam} Same as Table \ref{tab:FracM_SMBH} for $10 \, \Msun -10 \, \Msun$ EBBHs, but for different values of the parameters shown in the first column. We find that only $m_\mathrm{BH,max}$ influences the results significantly: higher maximum BH masses in the cluster lead to much lower LSO eccentricities among mergers with fixed binary component masses. }
\centering
   \begin{tabular}{@{}cccccc}
   \hline
    Parameter  &  $F_{ {\rm res} > 0.1}$  &  $F_{\rm aLIGO}$  & $F_{ {\rm LSO} > 0.1}$  \\
   \hline\hline
     standard  &  $91 \%$  &  $41 \%$  &  $7 \%$ \\
     $\beta = 1$  &  $91 \%$  &  $41 \%$  & $7 \%$ \\
    $\beta = 3$  &  $91 \%$  &  $41 \%$  &  $7 \%$ \\
    $m_{\rm BH,max} = 10 \, \Msun$  &  $98 \%$  &  $80 \%$  & $37 \%$ \\
    $m_{\rm BH,max} = 15 \, \Msun$  &  $94 \%$ &  $59 \%$  & $16 \%$ \\
    $m_{\rm BH,max} = 80 \, \Msun$  &  $90 \%$  &  $35 \%$  & $6 \%$ \\
    $p_0 = 0.6$  &  $92 \%$  &  $43 \%$  & $8 \%$ \\
   \hline\hline
   \end{tabular}
\end{table}

\begin{figure}
    \centering
    \includegraphics[width=85mm]{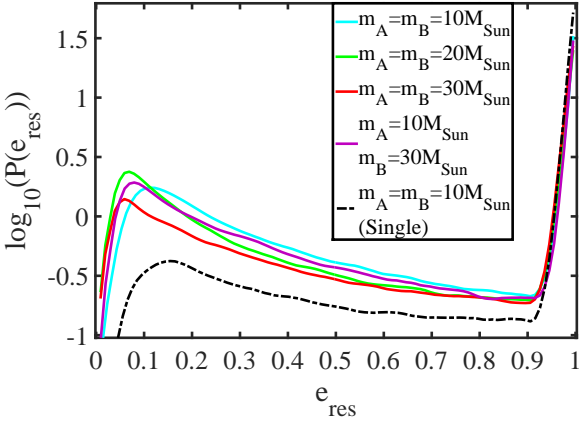}
\caption{ \label{fig:Dist_fp10} The PDF of residual eccentricity for different component-mass EBBHs when their emitted GW signals first peak at $10 \, \mathrm{Hz}$ or when they form at a higher frequency (cf. Figures \ref{fig:Distrhop0} and \ref{fig:DisteLSO} for the initial and final eccentricity distribution). We assume EBBHs forming through GW capture in a Milky Way-size nucleus. Different lines correspond to different component masses in a multi-mass population and in a single-mass population as in Figure \ref{fig:Distrhop0}. The sharp peaks beyond $e_{\rm res} \sim 0.9$ correspond to EBBHs forming within the aLIGO band. For masses $(10 \, \Msun - 10 \, \Msun, 20 \, \Msun - 20 \, \Msun, 30 \, \Msun - 30 \, \Msun, 10 \, \Msun - 30 \, \Msun)$, respectively, in a multi-mass BH population, we find that $(41,41,56,43)\%$ of EBBHs form in the aLIGO band. This number is $80\%$ for a single-mass BH population. We find that residual eccentricities at $10 \, \mathrm{Hz}$ are larger than $0.1$ for $(91,82,88,81)\%$ of $(10 \, \Msun - 10 \, \Msun, 20 \, \Msun - 20 \, \Msun, 30 \, \Msun - 30 \, \Msun, 10 \, \Msun - 30 \, \Msun)$ EBBHs and $98 \%$ in a single-mass BH population (see Table \ref{tab:FracM_SMBH}). }
\end{figure}

 An important question for GW detections is the fraction of objects that are eccentric when they emit GWs in the aLIGO band. In Tables \ref{tab:FracM_SMBH}, \ref{tab:FracM_SMBH_2}, and \ref{tab:FracM_SMBH_Belcz}, we show the fraction of highly eccentric GW capture sources that form with a GW spectrum that peaks above $10 \, \mathrm{Hz}$ ($F_{\rm aLIGO}$), the fraction of EBBHs that have $e> 0.1$ when the GW spectrum first peaks above $10 \, \mathrm{Hz}$ ($F_{ {\rm res} > 0.1}$), and the fraction of EBBHs that have $e > 0.1$ when the binary reaches the LSO ($F_{ {\rm LSO} > 0.1}$) for various binary and SMBH masses. Depending on the component masses and the SMBH mass, we find in source populations with $m_\mathrm{BH, max}=30 \, \Msun$, $80 \, \Msun$, or the \citet{Belczynskietal2014_2} PMF, that the fraction $F_{\mathrm{res} >0.1} = 68 - 99 \%$, $41 - 99 \%$, or $68 - 99 \%$, respectively, if the EBBHs are at least moderately eccentric, with $e > 0.1$ when the spectrum first peaks above $10 \, \mathrm{Hz}$ (see discussion in Section \ref{subsec:DistFGW10Hz}), and $F_{\rm aLIGO} = 25 - 64 \%$, $9 - 67 \%$, or $27 - 86 \%$,  respectively, of highly eccentric sources form in the aLIGO band (these sources have $e_0 > 0.9$, see Figure \ref{fig:Dist_fp10}). However, most of these sources circularize to below $e<0.1$ by the time the binary approaches the LSO. We find that the fractions of sources in this formation channel with $e > 0.1$ at the LSO are between $F_{\mathrm{LSO} > 0.1} = 3 - 38 \%$, $3 - 36 \%$, and $3 - 36 \%$, respectively, for EBBHs forming in a source population with $m_{\mathrm{BH,max}} = 30 \, \Msun$, $80 \, \Msun$, and the \citet{Belczynskietal2014_2} PMF as shown in Tables \ref{tab:FracM_SMBH}, \ref{tab:FracM_SMBH_2}, and \ref{tab:FracM_SMBH_Belcz}. In particular, sources that remain eccentric until the LSO have masses comparable to the maximum BH mass in the cluster, while lower-mass binaries are systematically more circular near the LSO. We discuss $F_{\rm aLIGO}$ and $F_{\mathrm{LSO} >0.1}$ separately in Sections \ref{subsec:DistFGW10Hz} and \ref{subsec:DistLSO}.

 To check the robustness of these predictions, we show the fractional rates of eccentric sources for different BH mass function exponents $\beta$ and mass segregation parameters $p_0$ in Table \ref{tab:Frac_BHPopParam}. We find that (i) $F_{\rm aLIGO}$ is significantly influenced by $m_\mathrm{BH,max}$ and slightly by $p_0$ and $\beta$; (ii) $F_{\rm aLIGO}$ is higher for EBBHs of lower $q$ at fixed $M_{\rm tot}$; (iii) $F_{\rm aLIGO}$ increases with $M_{\rm SMBH}$; and (iv) $F_{\rm aLIGO}$ is lower for EBBHs forming in a single-mass BH population than similar EBBHs forming in a multi-mass BH population. These results may be explained by arguments listed in \ref{i:B},  \ref{i:C}, \ref{i:D}, and \ref{i:E}, above. However, $F_{\rm aLIGO}$ is not similar for EBBHs forming in a single-mass BH population and EBBHs formed by the most massive BHs in a multi-mass BH population--nor does it systematically increase with the component masses, because $\rho_{\rm p0,aLIGO}$ depends on the component masses as well.

 Note, further, that the fraction of EBBHs forming in the LISA band increases with the total mass of the binary up to $\sim 23 \%$ ($\sim 19 \%$, $\sim 15 \%$) for $80 \, M_{\odot} - 80 \, M_{\odot}$ for $10^5 \, M_{\odot}$ ($10^6 \, M_{\odot}$, $10^7 \, M_{\odot}$) GNs. However, we find that these EBBHs emit a negligible fraction of their total signal power in the LISA band. All EBBHs merge within the aLIGO band for the considered ranges of $M_{\rm SMBH}$, $m_{\rm BH}$, and BH populations.

\begin{table}
   \caption{ \label{tab:FracM_SMBH_Belcz} Same as Table \ref{tab:FracM_SMBH}, but for a multi-mass BH population with the \citet{Belczynskietal2014_2} PMF and $p_0 = 0.5$.  }
\centering
   \begin{tabular}{@{}ccccc}
   \hline
    $m_A - m_B \, \, (\Msun)$ & $M_{\rm SMBH} \, \, (\Msun)$ & $F_{ {\rm res} > 0.1}$ &
    $F_{\rm aLIGO}$  &  $F_{ {\rm LSO} > 0.1}$  \\
   \hline\hline
   $3 - 3$  &  $10^5$  &  $97 \%$  &  $50 \%$   &  $3 \%$ \\
   $5 - 5$  &  $10^5$  &  $89 \%$  &  $41 \%$   &  $4 \%$ \\
   $10 - 10$  &  $10^5$  &  $70 \%$  &  $27 \%$  &  $5 \%$ \\
   $5 - 15$  &  $10^5$  &  $76 \%$  &  $33 \%$  &  $6 \%$ \\
   $20 - 20$  &  $10^5$  &  $75 \%$  &  $37 \%$  &  $14 \%$ \\
   $26 - 26$  &  $10^5$  &  $84 \%$  &  $53 \%$  &  $31 \%$ \\
   $3 - 3$  &  $10^6$  &  $99.2 \%$  &  $67 \%$   &  $5 \%$ \\
   $5 - 5$  &  $10^6$  &  $96 \%$ &  $51 \%$  &  $6 \%$ \\
   $10 - 10$  &  $10^6$  &  $87 \%$  &  $41 \%$  &  $7 \%$ \\
   $5 - 15$  &  $10^6$  &  $89 \%$  &  $40 \%$  &  $9 \%$ \\
   $20 - 20$  &  $10^6$  &  $83 \%$  &  $47 \%$  &  $23 \%$ \\
   $26 - 26$  &  $10^6$  &  $90 \%$  &  $55 \%$  &  $32 \%$ \\
   $3 - 3$  &  $M_{\SgrA}$  &  $99.5 \%$  &  $76 \%$   &  $6 \%$ \\
   $5 - 5$  &  $M_{\SgrA}$  &  $98 \%$  &  $64 \%$  &  $7 \%$ \\
   $10 - 10$  &  $M_{\SgrA}$   &  $93 \%$  &  $46 \%$  &  $9 \%$ \\
   $5 - 15$  &  $M_{\SgrA}$  &  $93 \%$  &  $45 \%$  &  $10 \%$ \\
   $20 - 20$  &  $M_{\SgrA}$  &  $87 \%$  &  $52 \%$  &  $22 \%$ \\
   $26 - 26$  &  $M_{\SgrA}$  &  $91 \%$  &  $57 \%$  &  $34 \%$ \\
   $3 - 3$  &  $10^7$  &  $99.8 \%$  &  $86 \%$   &  $7 \%$ \\
   $5 - 5$  &  $10^7$  &  $99 \%$  &  $69 \%$  &  $9 \%$ \\
   $10 - 10$  &  $10^7$  &  $95 \%$  &  $49 \%$  &  $10 \%$ \\
   $5 - 15$  &  $10^7$  &  $99 \%$  &  $52 \%$  &  $12 \%$ \\
   $20 - 20$  &  $10^7$  &  $93 \%$  &  $58 \%$  &  $27 \%$ \\
   $26 - 26$  &  $10^7$  &  $93 \%$  &  $61 \%$  &  $36 \%$ \\
   \hline\hline
   \end{tabular}
\end{table}

 The findings of this section for the characteristics of distributions also stand when using the \citet{Belczynskietal2014_2} PMF, which is due to the fact that these characteristics do not depend significantly on $r_{\rm min}^{A,B}$ (Section \ref{subsec:FormRate}). Furthermore, the results presented in Table \ref{tab:FracM_SMBH} are very similar to those presented in Table \ref{tab:FracM_SMBH_Belcz} for similar EBBHs, which arises due the fact that the radial distribution $P_{AB}(r)$ and the BH mass range are very similar for both types of multi-mass BH populations. This is consistent with analytic expectation presented at the end of Section  \ref{subsec:FormRate}.

 Additionally, our calculations show that only a negligible fraction ($\la 1 \%$) of the EBBHs interact with a third object during the eccentric inspiral. This implies that analytic expressions can be derived to determine the two-dimensional PDFs in the $\epsilon_0$--$\rho_{\rm p0}$ plane (Appendix \ref{TWORPEPS}) and similarly for the marginalized 1D $\rho_{\rm p0}$ and $\epsilon_0$ distributions, (Appendix \ref{PERICDIST} and \ref{ED}).

\subsection{Eccentricity Distribution When the GW Signal First Enters the aLIGO Band}
\label{subsec:DistFGW10Hz}

 We evolve systems from their initial orbital parameters using the evolution equations of \citet{Peters1964}. For many systems, the binary has a GW frequency in the aLIGO band at formation. However, systems that form outside of the aLIGO band may also enter the aLIGO band before merger. Figure \ref{fig:Dist_fp10} displays examples for the PDF of residual eccentricity $e_{\rm res}$ when the peak GW signal frequency $f_\mathrm{GW}$ (Equation \ref{eq:fGW}) first reaches $10\,\mathrm{Hz}$. In all cases, we find that (i) a sharp peak occurs in the PDF at $e_{\rm res} \sim 1$, which corresponds to EBBHs forming within the aLIGO band at $f_\mathrm{GW} > 10 \, \mathrm{Hz}$; and (ii) a peak forms at moderately small ($0.05-0.15$) eccentricities. The latter (ii) can be explained as follows: $e_{\rm res}$ can be obtained by solving the equation $f_{\rm aLIGO} = f_{\rm GW}$, which can be rewritten as
\begin{equation}  \label{eq:erho_fGW10Hz}
  f_{\rm aLIGO} = \left[(1 + e_{\rm res})^{0.3046} \pi M_{\rm tot} \rho_{\rm p}^{3/2}(e_{\rm res},e_0,
  \rho_{\rm p0})\right]^{-1} \, ,
\end{equation}
 where $\rho_{\rm p}$ is given by Equation (\ref{eq:rhoe}). By definition, $e_{\rm res} \equiv e_0$ for EBBHs with $\rho_{{\rm p}0} \leqslant \rho_\mathrm{p0,aLIGO}$ (Equation \ref{eq:rhoaLIGO}). Solving Equation (\ref{eq:erho_fGW10Hz}) for $e_{\rm res}$ by setting $\rho_{{\rm p}0} \geqslant \rho_\mathrm{p0,aLIGO}$, we find that $e_{\rm res}$ drops off quickly at $\rho_{{\rm p}0} \sim \rho_\mathrm{p0,aLIGO}$ and its tail starts at $e_{\rm res} \sim 0.5$, which implies that a significant fraction of EBBHs with $\rho_{{\rm p}0} \geqslant \rho_\mathrm{p0,aLIGO}$ have $e_{\rm res} \leqslant 0.5$ when they enter the aLIGO band. Furthermore, EBBHs that formed with $\rho_\mathrm{p0} \ga \rho_\mathrm{p0, aLIGO} $ exhibit a peak in the PDF of $e_{\rm res}$ at $e_{\rm res} \sim 0.05-0.15$.

 Tables \ref{tab:FracM_SMBH}--\ref{tab:FracM_SMBH_Belcz} present the total fraction of EBBHs with residual eccentricities larger than $0.1$ when their emitted GW signals first exceed $10 \, \mathrm{Hz}$ ($F_{{\rm res} > 0.1}$), for various values of fixed component masses. We find that (i) $F_{{\rm res} > 0.1}$ is significantly influenced by $m_{\rm BH,\max}$, but it is not sensitive to the mass function exponent $\beta$ or the mass segregation parameter $p_0$; (ii) $F_{ {\rm res} > 0.1}$ is higher for EBBHs of lower $q$ at fixed $M_{\rm tot}$, and the difference  between $F_{ {\rm res} > 0.1}$ of different $q$ becomes more substantial with increasing $M_{\rm tot}$; (iii) $F_{{\rm res} > 0.1}$ increases with $M_{\rm SMBH}$; and (iv) $F_{ {\rm res} > 0.1}$ is lower for EBBHs forming in a single-mass BH population than for similar EBBHs forming in a multi-mass BH population. These findings arise due to \ref{i:C}, \ref{i:D}, \ref{i:E}, and \ref{i:B} because $\rho_{ \rm p0}$ is the only free parameter that determines $e_{\rm res}$ when the component masses are fixed. However, $F_{ {\rm res} > 0.1}$ is not similar for EBBHs forming in a single-mass BH population versus EBBHs formed by the most massive BHs in a multi-mass BH population; nor does it increase systematically with the component masses, because $e_{\rm res}$ depends on the component masses as well.

 The detectability of a compact binary is related to how much energy is radiated in the frequency range where the detector is sensitive. An aLIGO type detector is most sensitive at $\sim 100 \, \mathrm{Hz}$, and the sensitivity decreases rapidly over $\sim 100 \, \mathrm{Hz}$. If a compact binary forms with a peak GW frequency above this limit, then most of the GW energy would be radiated at higher frequencies than $\sim 100 \, \mathrm{Hz}$, which may make such sources difficult to detect. Utilizing the fact that EBBHs are expected to form with $e_0 \sim 1$ and $\rho_{\mathrm{p}0} \geqslant 8$ (Section \ref{subsec:DistIniParam}), Equation (\ref{eq:rhoaLIGO}) can be used to estimate the upper limit for $M_{\rm tot}$ at which EBBHs still form with $f_{\rm GW} \gtrsim 100 \, \mathrm{Hz}$. We find this limit to be $M_{\rm tot} \sim 23 \, M_{\odot}$, and the maximum $\rho_{\mathrm{p}0}$ with which EBBHs with $M_{\rm tot} \lesssim 23 \, M_{\odot}$ form with $f_{\rm GW} \gtrsim 100 \, \mathrm{Hz}$ can be given as
 \begin{equation}  \label{eq:rho100Hz}
  \rho_{\rm p0,100Hz} = 8.816 \left(\frac{M_{\mathrm{tot}}}{20\,\Msun}\right)^{-2/3} \, .
\end{equation}
 Because EBBHs with $M_{\rm tot} \sim 23 \, M_{\odot}$ form with generally high $\rho_{\mathrm{p}0}$, a negligible fraction of low-mass EBBHs form with $f_{\rm GW} \gtrsim 100 \, \mathrm{Hz}$ over the considered ranges of $p_0$, $M_{\rm SMBH}$, $m_{\rm BH}$, and the BH PMFs; see Table \ref{tab:FracM_over100Hz} for examples. EBBHs, which form below the aLIGO band, merge within the aLIGO band for $m_{\rm BH}<10^3\Msun$. Therefore, EBBH GW capture sources in GNs are typically within the frequency range of aLIGO-type GW detectors.

 Similarly to Section \ref{subsec:DistIniParam}, we find that the PDF of $e_{\rm res}$ is qualitatively very similar for the power-law and \citet{Belczynskietal2014_2} PMF, and the results shown in Table \ref{tab:FracM_SMBH} are very similar to those presented in Table \ref{tab:FracM_SMBH_Belcz} for similar EBBHs.

 Overall, we find that, in a source population with $m_\mathrm{BH,max}=30 \, \Msun$, $80 \, \Msun$, or the \citet{Belczynskietal2014_2} PMF, respectively, at least $F_{ {\rm res} > 0.1 } = 65 - 86 \%$, $38 - 58 \%$, and $66 - 89 \%$ of EBBHs have residual eccentricities larger than $0.1$ when their GW signals first reach $10 \, \mathrm{Hz}$ or form with higher frequencies. The actual value within this range is generally set by the binary component masses, the mass segregation parameter, the exponent of the BH mass function, and $M_{\rm SMBH} \in [10^5 \, \Msun, 10^7 \, \Msun]$. Larger values correspond to higher $M_{\rm SMBH}$. Similarly, for a single-mass BH population, we find that the fraction of sources with a residual eccentricity is higher: $F_{ {\rm res} > 0.1 } \geqslant 96 \%$, as shown in Table \ref{tab:FracM_SMBH}.

\begin{table}
   \caption{ \label{tab:FracM_over100Hz} The fraction of EBBHs forming with $f_{\rm GW} \gtrsim 100 \, \mathrm{Hz}$ for various BH PMFs and component masses ($M_{\rm tot} \lesssim 23 \, M_{\odot}$, see Section \ref{subsec:DistFGW10Hz} for details). Monte Carlo results shown here correspond to those in Tables \ref{tab:FracM_SMBH}, \ref{tab:FracM_SMBH_2}, and \ref{tab:FracM_SMBH_Belcz}. }
\centering
   \begin{tabular}{@{}ccccc}
   \hline
    $m_A - m_B \, \, [\Msun]$  &  $M_{\rm SMBH} \, \, [\Msun]$  &  Table \ref{tab:FracM_SMBH}  &
    Table \ref{tab:FracM_SMBH_2}  &  Table \ref{tab:FracM_SMBH_Belcz}  \\
   \hline\hline
   $3 - 3$  &  $10^5$  &  - &  -   &  $7 \%$ \\
   $5 - 5$  &  $10^5$  &  $4 \%$  &  $4 \%$   &  $4 \%$ \\
   $10 - 10$  &  $10^5$  &  $0.5 \%$  &  $0.4 \%$  &  $0.5 \%$ \\
   $5 - 15$  &  $10^5$  &  $0.6 \%$  &  $0.5 \%$  &  $0.9 \%$ \\
   $3 - 3$  &  $10^6$  &  -  &  -   &  $10 \%$ \\
   $5 - 5$  &  $10^6$  &  $5 \%$ &  $6 \%$  &  $6 \%$ \\
   $10 - 10$  &  $10^6$  &  $0.9 \%$  &  $1 \%$  &  $0.9 \%$ \\
   $5 - 15$  &  $10^6$  &  $0.9 \%$  &  $0.7 \%$  &  $1.2 \%$ \\
   $3 - 3$  &  $M_{\SgrA}$  &  -  &  -   &  $12 \%$ \\
   $5 - 5$  &  $M_{\SgrA}$  &  $6 \%$  &  -  &  $8 \%$ \\
   $10 - 10$  &  $M_{\SgrA}$  &  $1 \%$  &  -  &  $1 \%$ \\
   $5 - 15$  &  $M_{\SgrA}$  &  $1 \%$  &  -  &  $1.4 \%$ \\
   $3 - 3$  &  $10^7$  &  -  &  -   &  $15 \%$ \\
   $5 - 5$  &  $10^7$  &  $7 \%$  &  $8 \%$  &  $9 \%$ \\
   $10 - 10$  &  $10^7$  &  $1.2 \%$  &  $1.2 \%$  &  $1.2 \%$ \\
   $5 - 15$  &  $10^7$  &  $1.1 \%$  &  $1.3 \%$  &  $1.8 \%$ \\
   \hline\hline
   \end{tabular}
\end{table}

\subsection{Eccentricity Distribution at the Last Stable Orbit}
\label{subsec:DistLSO}

\begin{figure}
    \centering
    \includegraphics[width=85mm]{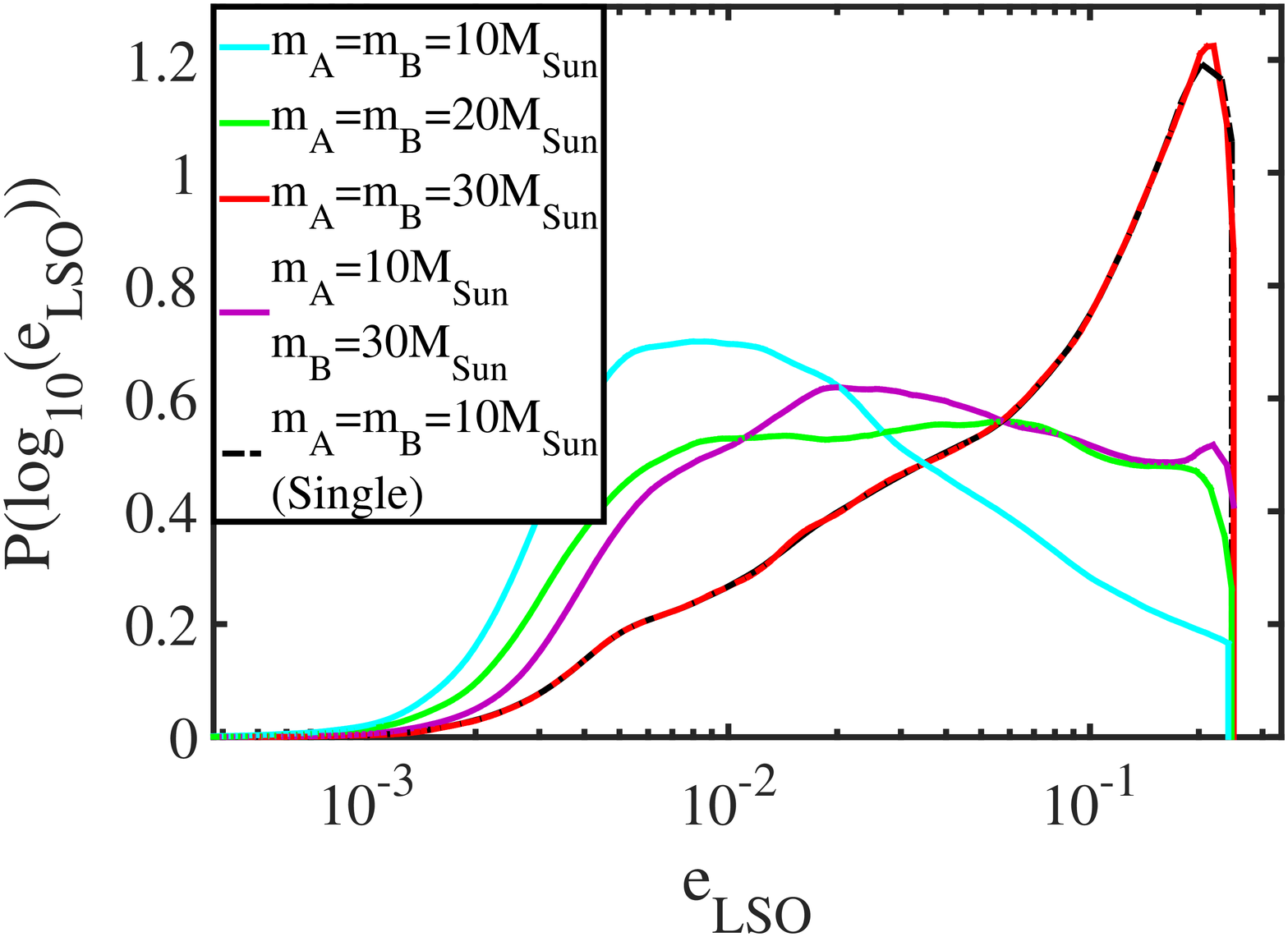}
    \includegraphics[width=85mm]{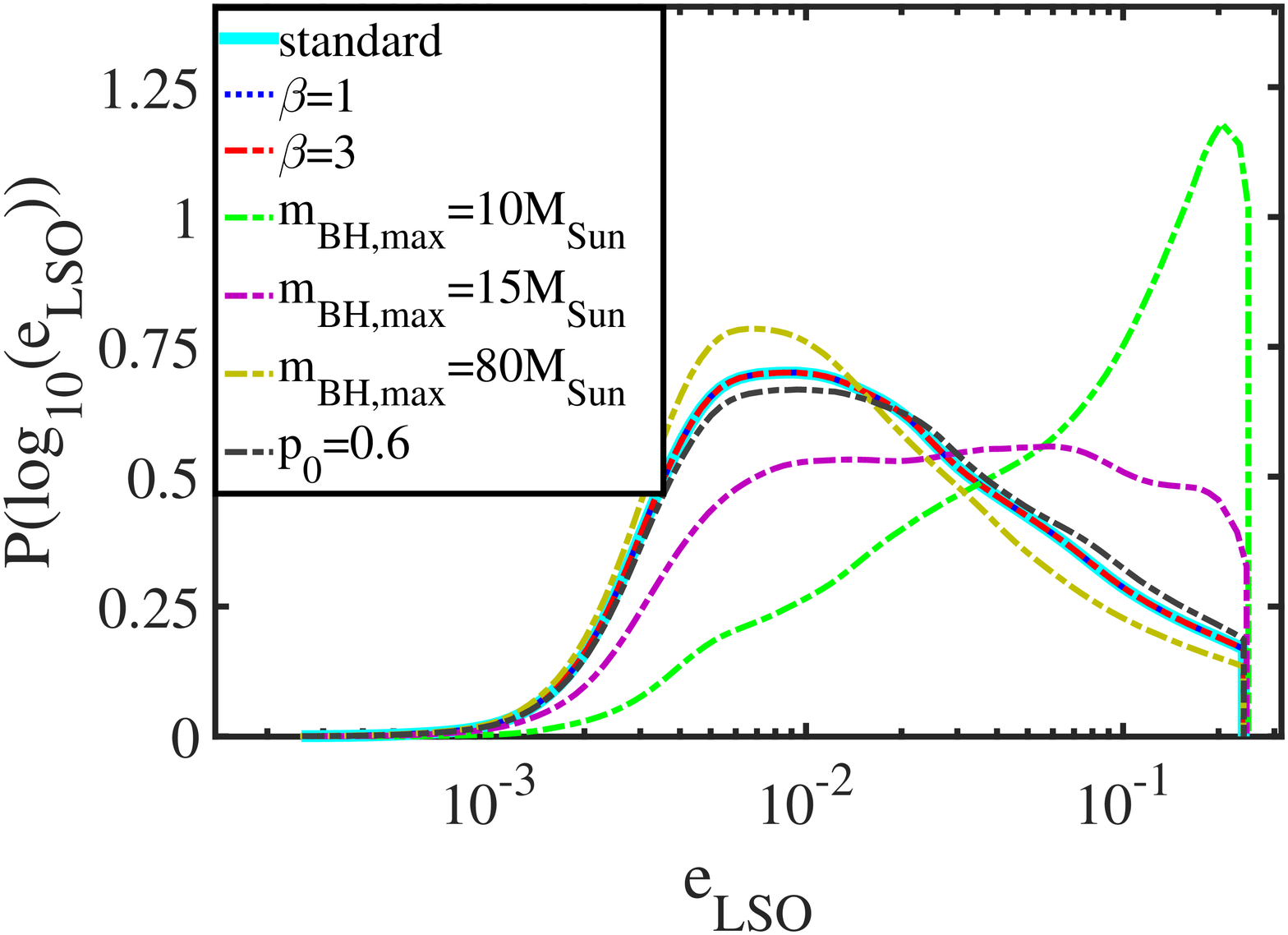}
    \includegraphics[width=85mm]{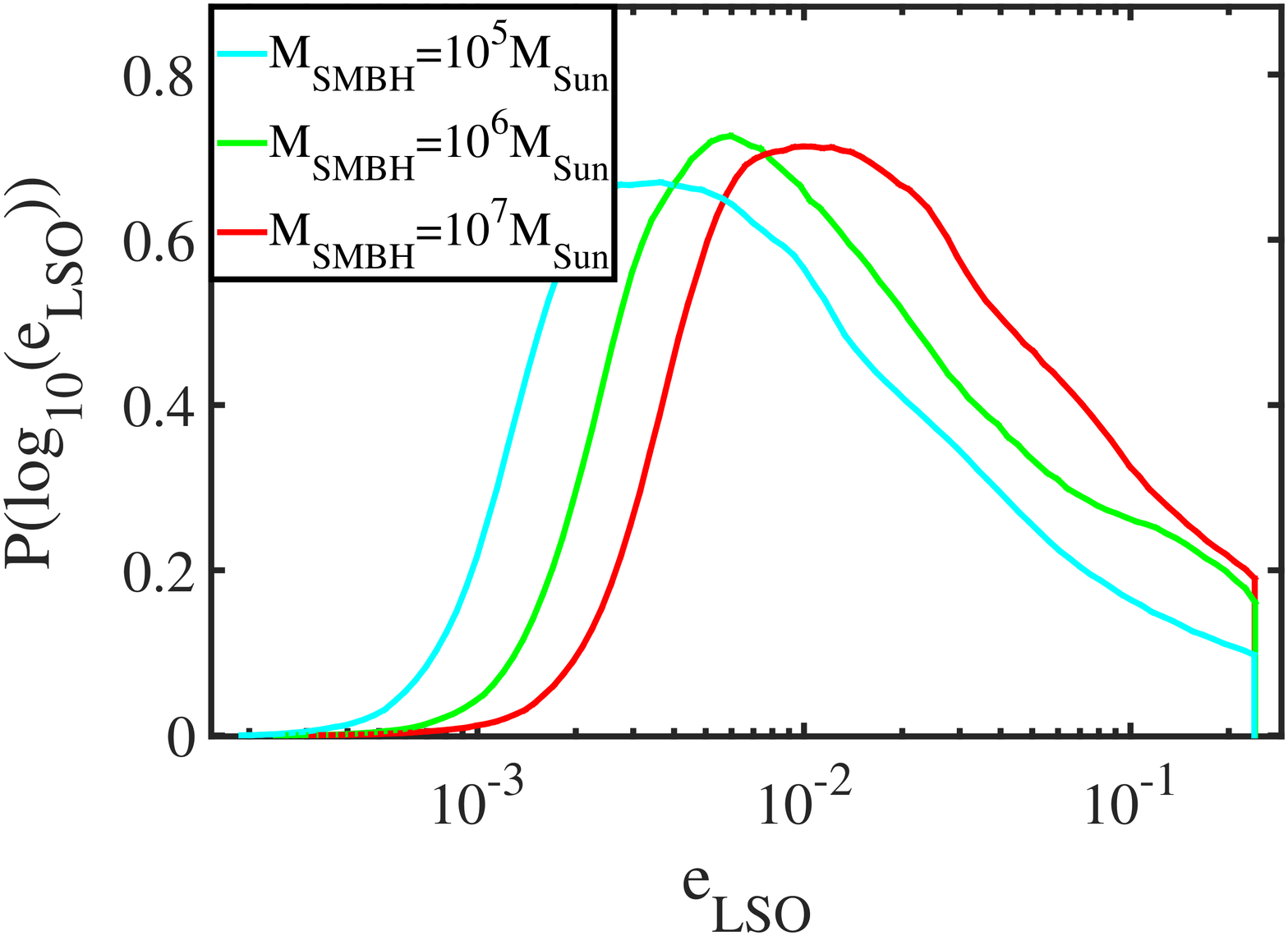}
\caption{ \label{fig:DisteLSO} Top panel: the distribution of eccentricity at the LSO ($e_{\rm LSO}$, cf. Figures \ref{fig:Distrhop0} and \ref{fig:Dist_fp10} for the initial eccentricity distribution and that at $10 \, \mathrm{Hz}$). Different lines correspond to different component masses in a multi-mass BH population and in a single-mass BH population as in Figure \ref{fig:Distrhop0}. In a multi-mass BH population, more massive EBBHs have systematically higher $e_{\rm LSO}$, and EBBHs forming in a single-mass BH populations have higher $e_{\rm LSO}$ than similar EBBHs forming in a multi-mass BH population. Middle panel: the influence of different assumptions on the results. We vary the mass function exponent $\beta$, the maximum mass of the multi-mass BH population $m_\mathrm{BH,max}$, and the mass segregation parameter $p_0$. We find that only $m_{\rm BH,max}$ influences the distribution of $e_{\rm LSO}$ significantly. Bottom Panel: the $M_{\rm SMBH}$ dependence of $e_{\rm LSO}$ distribution (cf. Figure \ref{fig:SMBHdeprhopaeps0} for $e_0$). }
\end{figure}

 Let us now examine the distribution of the final eccentricity at LSO shown in Figure \ref{fig:DisteLSO}. These numerical results show that (i) higher $e_{\rm LSO}$ values correspond to higher-mass BHs in the distribution; (ii) the PDF of $e_{\rm LSO}$ is affected significantly by $m_\mathrm{BH,max}$, but it is not influenced significantly by the mass distribution exponent or mass-segregation parameters $\beta$ and $p_0$; (iii) the $M_{\rm tot}$ influences the result more significantly than the mass ratio, a lower $q$ shifts the $e_{\rm LSO}$ distribution to slightly  lower values, as shown; (iv) EBBHs form with higher $e_{\rm LSO}$ around more massive SMBHs (Tables \ref{tab:FracM_SMBH}, \ref{tab:FracM_SMBH_2}, and \ref{tab:FracM_SMBH_Belcz}); and (v) EBBHs forming in a single-mass BH populations have an $e_{\rm LSO}$ distribution similar to that of the most massive EBBHs in a multi-mass BH population. Because $e_{\rm LSO}$ is a decreasing function of $\rho_{\mathrm{p}0}$ (Equation \ref{eq:eLSOrhop0}), these results may be explained respectively by arguments for the distribution of $\rho_{\mathrm{p}0}$ listed in \ref{i:A}, \ref{i:B}, \ref{i:C}, \ref{i:D}, and \ref{i:E} above.

 The characteristic scale of $e_{\mathrm{LSO}}$ may be understood using the characteristic value for $\rho_{\mathrm{p}0}$ given by Equation (\ref{eq:rhop0uni}) and the relation between $\rho_{\mathrm{p}0}$ and $e_{\mathrm{LSO}}$ given by Equation (\ref{eq:eLSOrhop0}). For mergers at a fixed radius $r$, the distribution of $\rho_{\mathrm{p}0}$ is flat for $\rho_{\mathrm{p}0} \leqslant \rho_{\mathrm{p}0,\mathrm{uni}}$, implying that $P(e_{\rm LSO}) \propto e_{\rm LSO}^{-8/5}$ for $e_{\mathrm{LSO}} \gtrsim e_{\mathrm{LSO,peak}}$, where
\begin{align}  \label{eq:eLSOuni}
  e_\mathrm{LSO,peak}  \approx &  0.2682 \rho_{{\rm p}0,\rm uni}^{8/5}
\nonumber\\ =&
0.0252 \left( 4 \eta \right)^{-16/35}
    \left( \frac{v_{\max}}{1,000\,\rm{km/s}}\right)^{32/35}
\nonumber\\ =&
  0.2703 \left( 4\eta \right)^{-16/35}
    \left( \frac{r}{1000\,M_{\rm SMBH}}  \right)^{-16/35} \, .
\end{align}
 The distribution on a logarithmic scale follows $P(\ln e_{\mathrm{LSO}}) = e_{\mathrm{LSO}} P(e_{\mathrm{LSO}}) \propto e_{\mathrm{LSO}}^{-3/5}$ for values $e_{\mathrm{LSO}} \gtrsim e_{\mathrm{LSO,peak}}$. The characteristic values of $e_{\mathrm{LSO}}$ may be determined for GW capture events with different masses by combining Equations (\ref{eq:eLSOrhop0}) and (\ref{eq:rhop0uni}). For the most massive EBBHs, we find that roughly $50 \%$ of them are within $r \sim 4 \times 10^{-3} \, \mathrm{pc}$ in a Milky Way-size nucleus, which corresponds to $2000 \, M_{\rm SgrA*}$, where $e_{\mathrm{LSO, peak}}$ is $\sim 0.19$. However, for the least massive EBBHs with $m_A = m_B = 5 \, \Msun$, $50 \%$ of them are within $1.2 \, \mathrm{pc} \sim 5.8 \times 10^6 \, M_{\rm SgrA*}$ for which $e_{\mathrm{LSO,peak}}\sim 0.005$. Thus, the peak of the $e_{\rm LSO}$ distribution is high for heavy members of a given population because they are in the close vicinity of the SMBH, while the low-mass members have a much lower $e_{\rm LSO}$ because they are typically much farther out. Thus, the measurement of the $e_{\rm LSO}$ distribution for different masses may be used to detect the mass segregation of the sources within their host environment.

 Conversely, solving Equation (\ref{eq:eLSOuni}) for the velocity dispersion $\sigma \sim v_{\rm max}/\sqrt{2}$ gives
 \begin{equation}\label{eq:sigma-eLSO}
    \sigma \sim 258\,\frac{\rm km}{\rm s}\,(4\eta)^{1/2} \left(\frac{e_{\rm LSO,peak}}{0.01}\right)^{35/32} \, .
 \end{equation}
 Thus, the measurement of the peak eccentricity of a GW capture binary at LSO gives an estimate of the typical velocity dispersion of the source environment.

 In all of our simulations, we have ignored those encounters in which BHs undergo a direct head-on collision without forming an EBBH (i.e. $\rho_{{\rm p}0} \geqslant 8$ in all cases, see Figure \ref{fig:Distrhop0}). This causes the PDFs of $e_{\rm LSO}$ to cut off at $e_{\rm LSO} \sim 0.23$; see the top panel of Figure \ref{fig:DisteLSO}.

 By examining the $e_{\rm LSO}$ distribution for various SMBH mass, BH mass, and BH population parameters, we find that approximately $\lesssim 71 \%$, $\lesssim 14 \%$, and $\lesssim 0.2\%$ of the most massive EBBHs have eccentricities lower than $0.1$, $10^{-2}$, and $10^{-3}$ at LSO, respectively, for the considered ranges of SMBH mass and BH population parameters. The corresponding respective numbers for the lowest-mass EBBHs are $\lesssim 97 \%$, $\lesssim 66 \%$, and $\lesssim 4 \%$. For the \citet{Belczynskietal2014_2} PMF, the corresponding numbers for the highest-mass EBBHS are $\lesssim 69 \%$, $\lesssim 13 \%$, and $\lesssim 0.2 \%$, and for the lowest-mass EBBHs are $\lesssim 97 \%$, $\lesssim 65 \%$, and $\lesssim 5 \%$. In the case of a single-mass BH population, approximately $\lesssim 60 \%$, $\lesssim 16 \%$, and $\lesssim 0.24 \%$ of EBBHs have eccentricities lower than  $0.1$, $10^{-2}$, and $10^{-3}$ at LSO, respectively, for SMBH masses between $10^5 \, \Msun$ and $10^7 \, \Msun$.\footnote{Note that these results correspond to $10^5 \, \Msun$ SMBHs because EBBHs with lowest $e_{\rm LSO}$ values form in these SMBHs; see Figure \ref{fig:DisteLSO}. The results are insensitive to the upper bound of the SMBH masses.} Overall, we find that a negligible fraction of EBBHs have eccentricities lower than $10^{-3}$ at LSO over the considered ranges of BH mass, SMBH mass, and BH population parameter ranges. Note that, for initially highly eccentric precessing BH binaries, the expected relative measurement accuracy of $e_{\rm LSO}$ is $\lesssim 5 \%$ for $30 \, \Msun - 30 \, \Msun$ ($10 \, \Msun - 10 \, \Msun$) precessing EBBHs with $\rho_{\mathrm{p}0} \leqslant 50$ ($\rho_{\mathrm{p}0} \leqslant 100$) for the aLIGO-AdV-KAGRA detector network \citep{Gondanetal2017}. This implies that the predicted range of $e_{\rm LSO}$ may typically be distinguished from zero in future GW detections.

 Similarly to Section \ref{subsec:DistIniParam}, we find that the PDF of $e_{\rm LSO}$ is qualitatively similar for the \citet{Belczynskietal2014_2} PMF, and results presented in Table \ref{tab:FracM_SMBH} are very similar to those presented in Table \ref{tab:FracM_SMBH_Belcz} for similar EBBHs.

 Note that the distribution of $e_\mathrm{LSO}$ has been previously calculated in \citet{OLearyetal2009} for a single-mass population. We find two main differences with respect to those results. First, the PDF peaks at a slightly higher value ($\sim 0.2$ for a single-mass distribution, instead of $\sim 0.1$ as in \citealt{OLearyetal2009}), and it drops off quickly beyond the peak instead of taking values at higher eccentricities. The somewhat higher $e_{\rm LSO}$ arises due to the fact that the minimum radius at which the formation of EBBHs is still considered is about $4.5$ times lower than that in \citet{OLearyetal2009}; this implies higher $w$, and lower $\rho_{\rm p0}$ and consequently larger $e_{\rm LSO}$. Note that we assumed a slightly greater massive SMBH mass than \citet{OLearyetal2009}, which also increases the values of $e_{\rm LSO}$. In our case, the PDF of $e_{\rm LSO}$ drops off quickly beyond the peak of the distribution at $e_{\rm LSO} \sim 0.23$ because we have ignored those encounters for which BHs coalesce before EBBHs form (i.e. $\rho_{\mathrm{p}0} \leqslant 8$ in all cases; see Sections \ref{sec:PIOP} and \ref{subsec:DistIniParam} for details), while those binaries are included in the plotted $e_{\rm LSO}$ distributions in \citet{OLearyetal2009}. However, in a multi-mass distribution, the peak $e_{\mathrm{LSO}}$ values are even lower than in \citet{OLearyetal2009}, as shown in Figure \ref{fig:DisteLSO}.

\begin{figure*}
    \centering
    \includegraphics[width=85mm]{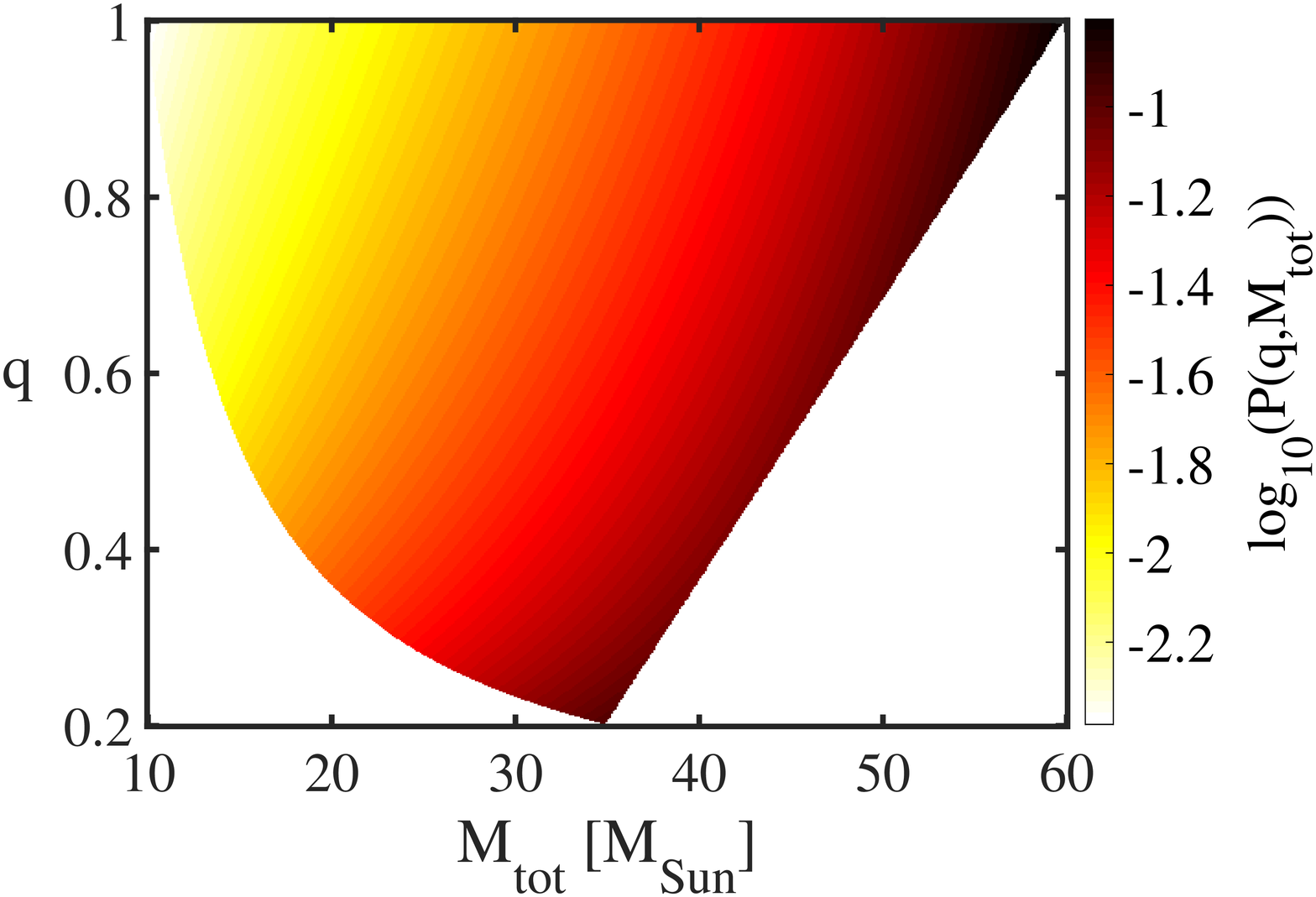}
    \includegraphics[width=85mm]{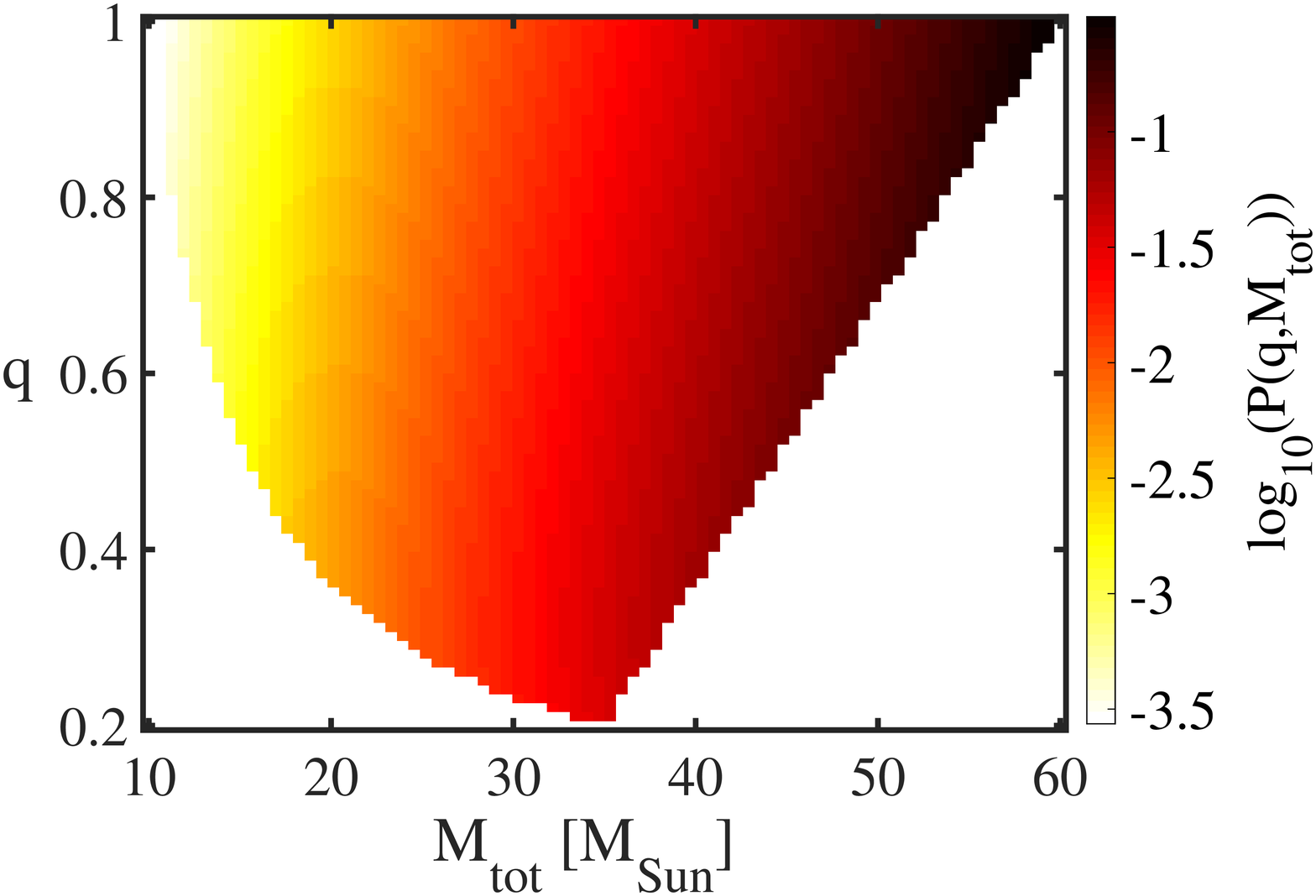}
\\
    \includegraphics[width=85mm]{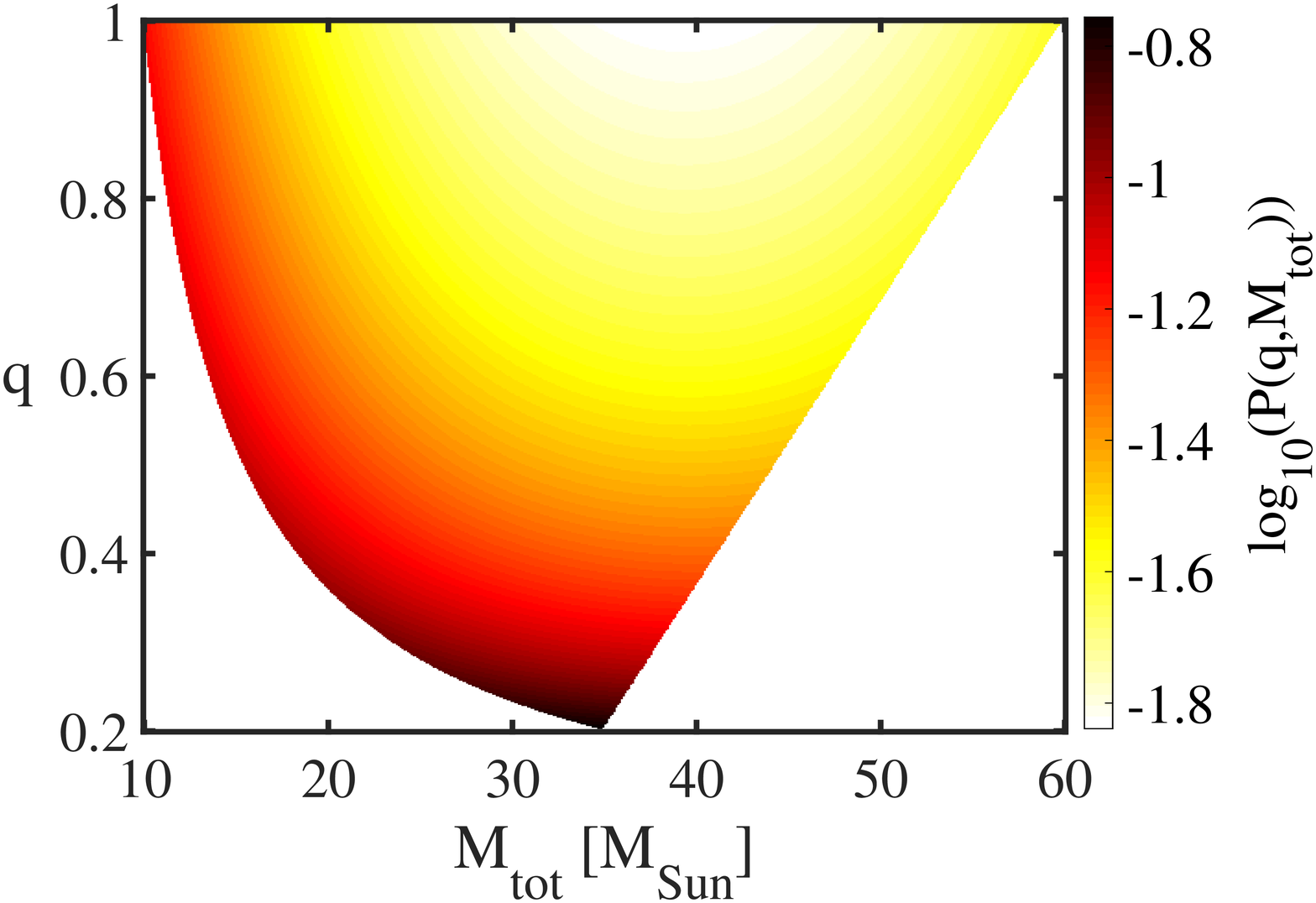}
    \includegraphics[width=85mm]{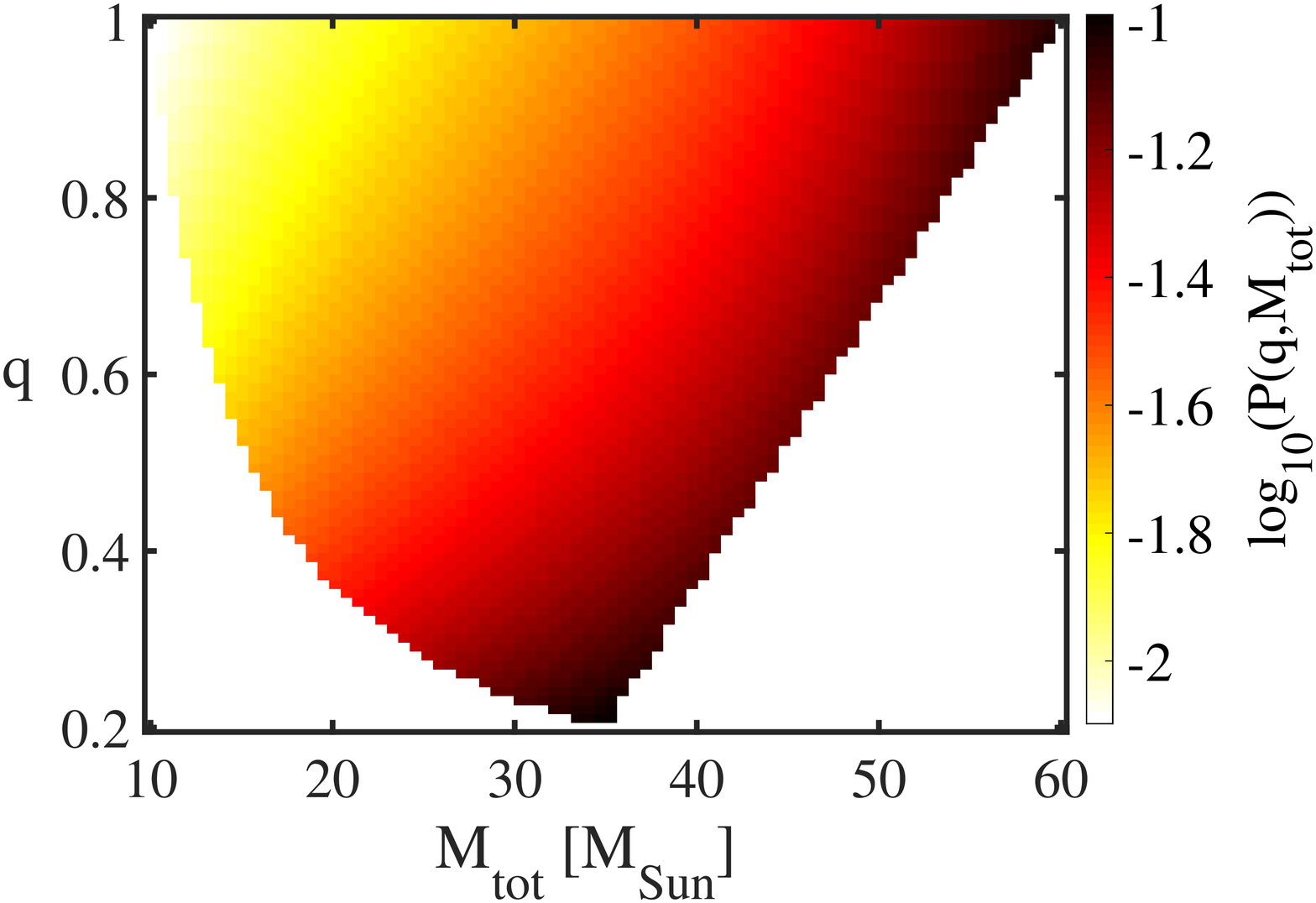}
\\
    \includegraphics[width=85mm]{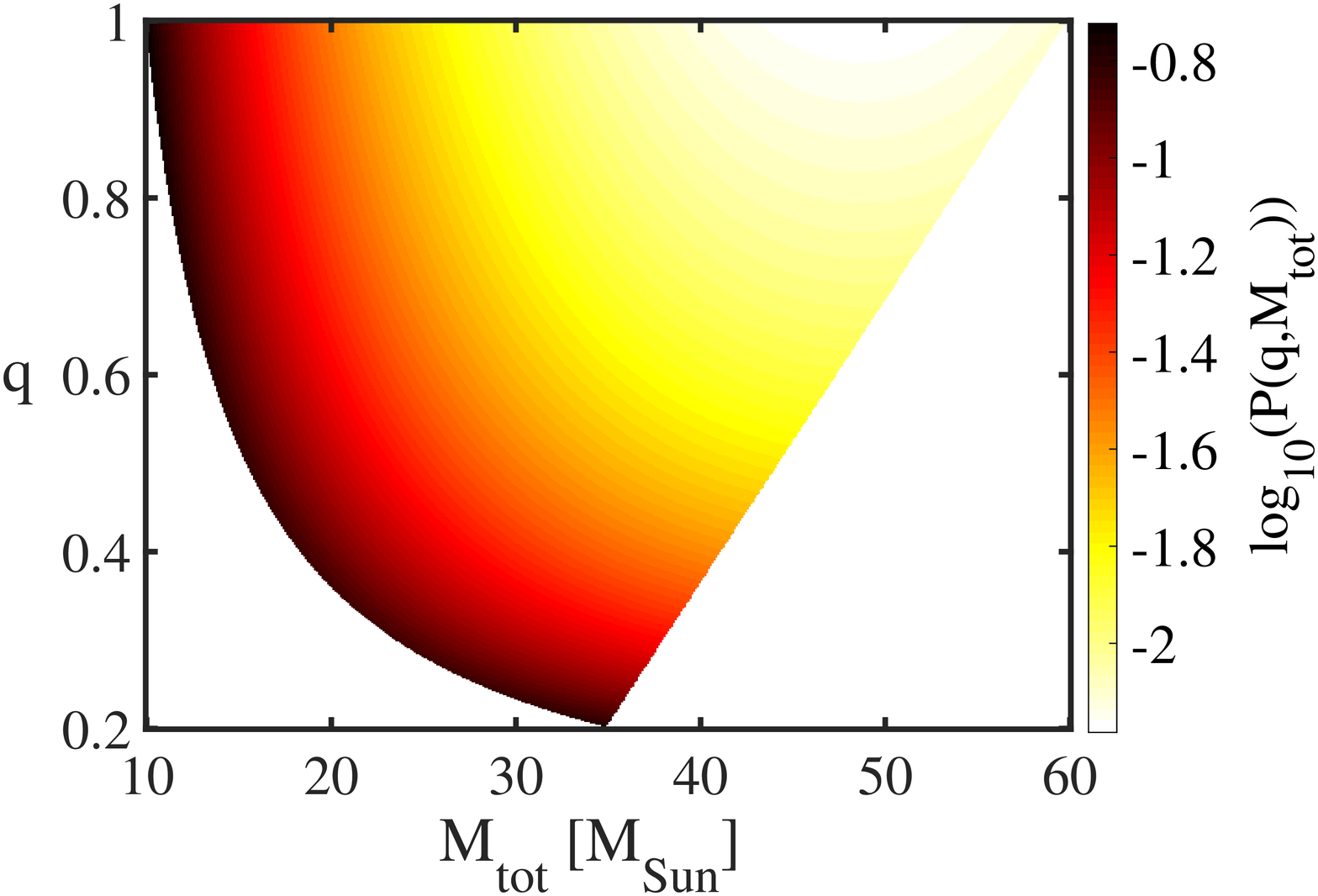}
    \includegraphics[width=85mm]{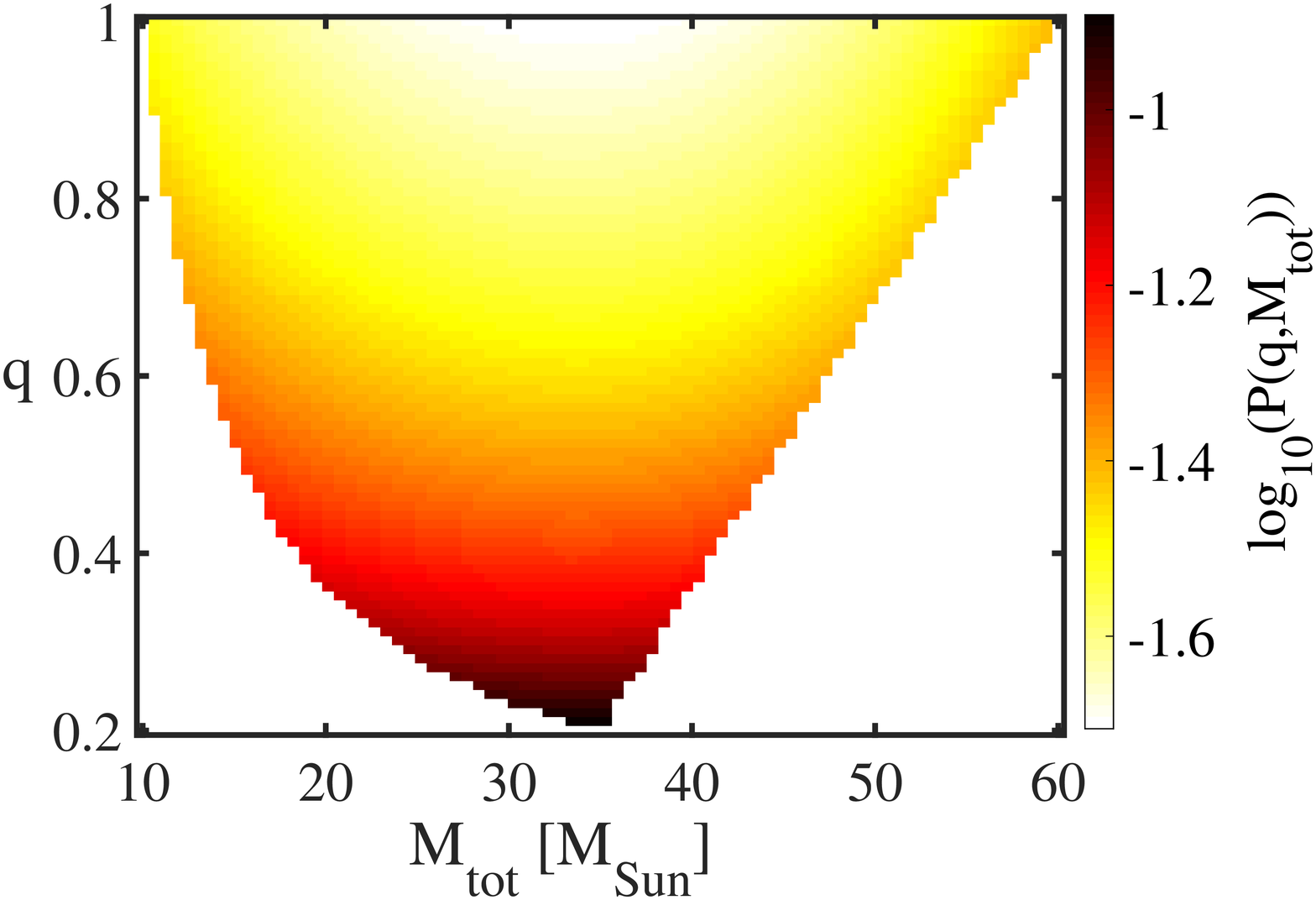}
\caption{ \label{fig:NoPMFDist} Left panels show the two-dimensional PDF of merger rates of BH binaries that form due to GW emission in a single nuclear star cluster, as a function of binary mass ratio and total mass. Right panels show the PDF of detection rates for Advanced LIGO at design sensitivity, which includes the mergers from all detectable galaxies in the universe. Different rows of panels correspond to different BH PMFs $dN/dm_{\rm BH}=m_{\rm BH}^{-\beta}$, where $\beta = 1$ (top), 2.35 (middle), and 3 (bottom), respectively. The sharp boundaries are due to the assumption $5 \, \Msun \leqslant m_{\rm BH} \leqslant 30 \, \Msun$ for both BHs forming the binary. The segregation parameter is  $p_0 = 0.5$. For different maximum BH mass or $p_0$ assumptions (not shown) the merger rate distribution in a single cluster (left panel) is changed only through a rescaling of the $M_{\rm tot}$ axis (see Equation \ref{eq:P(M,q)}). The merger rates in a single nuclear star cluster do not depend significantly on the SMBH mass.} 
\end{figure*}

\begin{figure*}
    \centering
    \includegraphics[width=85mm]{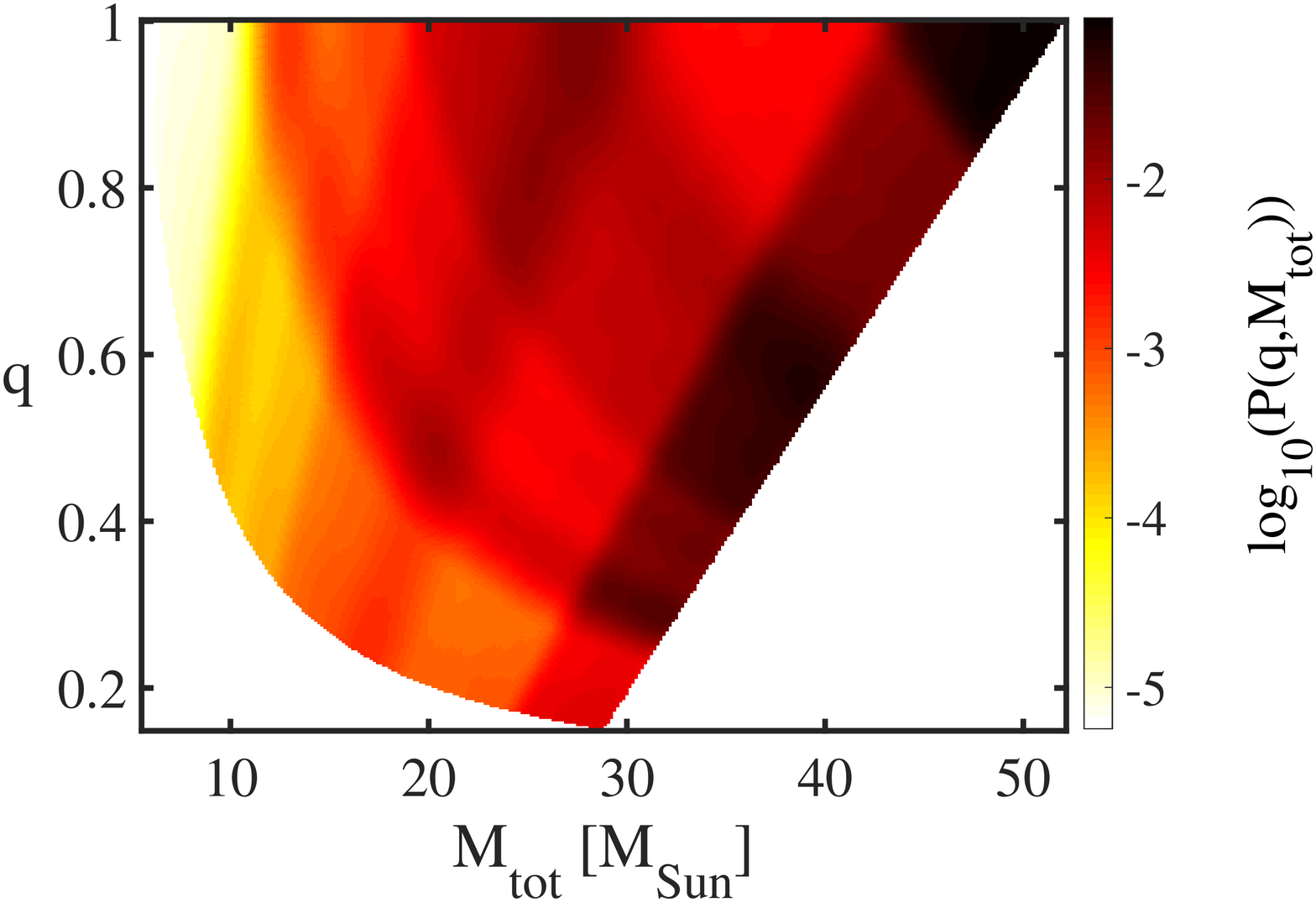}
    \includegraphics[width=85mm]{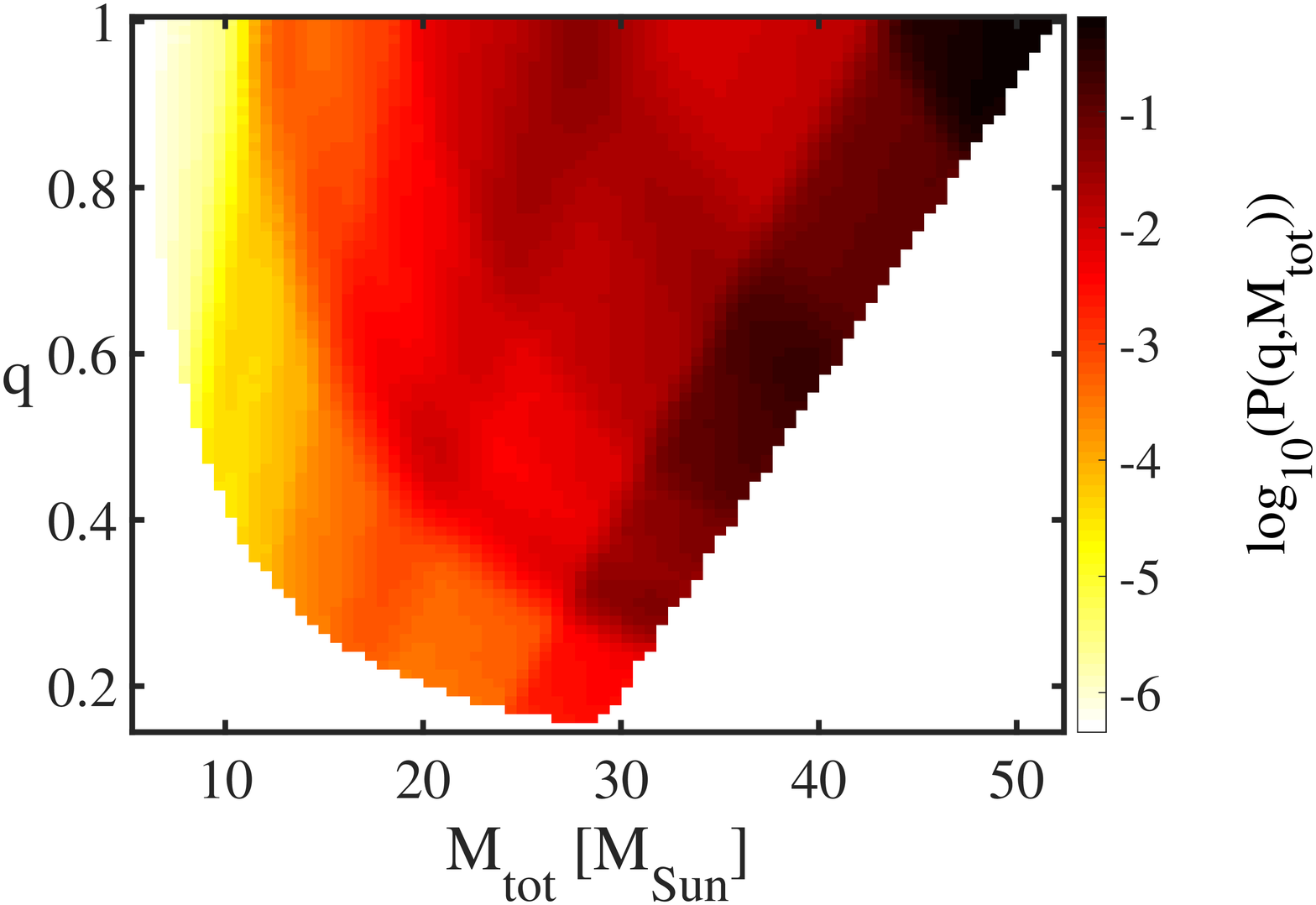}
\caption{ \label{fig:MergRate_BelczModel} Same as Figure \ref{fig:NoPMFDist}, but for the \citet{Belczynskietal2014_2} PMF. The left panel does not depend significantly on the SMBH mass.} 
\end{figure*}

\subsection{Total Mass and Mass Ratio Distributions}
\label{subsec:CompMassDepFormRate}

 Let us now examine the two-dimensional distribution of the merger rate in the $q$--$M_{\rm tot}$ plane, $\left\langle \partial^2 \Gamma/ \partial  M_{\rm tot} \partial q \right\rangle$, which affects the mass-dependent detection rates. This is derived by integrating $ \partial^3 \Gamma / \partial r \partial m_A\partial m_B$ (Equation \ref{eq:radiusPDF}) over the allowed range of $r$, from $r_{\rm min} ^{A,B}$ to $r_{\rm max}$, for each fixed $m_A$ and $m_B$ component mass. In Appendix \ref{sec:MMDGN}, we show that this leads to an analytical expression\footnote{This expression includes GW captures and direct head-on collisions. See Equation (\ref{eq:Ratemulti-mass}) for the case where head-on collision are ignored.}.  Up to a constant normalization factor,
\begin{align}
  P(m_A,m_B) \approx & \eta^{\frac{2}{7}-\beta} M_{\rm tot}^{2-2\beta}
  \left(1 - \frac{2}{3} \overline{M}_{\rm tot} + \frac{4 \eta}{9} \overline{M}_{\rm tot}^2  \right)
  \frac{ 1 - \Lambda^{\overline{M}_{\rm tot}-\frac{11}{14}} }{\frac{11}{14}
  -\overline{M}_{\rm tot}}
\end{align}
 if $m_{\rm BH,min} \leqslant m_{A,B} \leqslant m_{\rm BH,max}$, and zero otherwise, where we introduced the Coulomb logarithm\footnote{In the following expressions, we ignore the possible mass dependence of $\Lambda$.}
\begin{equation} \label{eq:CoulombLog}
 \ln \Lambda=\ln\frac{r_{\max}}{r_{\min}^{A,B}} \, ,
 \end{equation}
 and $\overline{M}_{\rm tot}$ for the dimensionless total mass
\begin{equation}
   \overline{M}_{\rm tot}=p_0 \frac{M_{\rm tot}}{m_{\rm BH,max}} \, ,
\end{equation}
 where $p_0\sim 0.5$, for which $m_{\rm BH, min}/m_{\rm BH, max} \leqslant \overline{M}_{\rm tot} \leqslant 1$.

 The likelihood\footnote{i.e., conditional probability} of merger for any two given BHs in the cluster, with mass $m_A$ and $m_B$ is proportional to $P(m_A,m_B)$ divided by the prior probability proportional to $\mathcal{F}(m_A) \mathcal{F}(m_B) 
 \propto m_A^{-\beta} m_B^{-\beta} = \eta^{-\beta}M_{\rm tot}^{-2\beta}$, which gives
\begin{equation}
  \mathcal{L}(m_A,m_B) \approx \eta^{\frac{2}{7}} M_{\rm tot}^{2} \left(1 - \frac{2}{3}
  \overline{M}_{\rm tot} + \frac{4 \eta}{9} \overline{M}_{\rm tot}^2  \right)
  \frac{ 1 - \Lambda^{\overline{M}_{\rm tot}-\frac{11}{14}} }{\frac{11}{14}
  -\overline{M}_{\rm tot}}
\end{equation}
 if $m_{\rm BH,min} \leqslant m_{A,B} \leqslant m_{\rm BH,max}$, and zero otherwise.

 Recently \citet{Kocsisetal2017} have introduced a useful indicator\footnote{not to be confused with the number density exponent $\alpha$ in Equation (\ref{eq:alpham})} to distinguish different source populations: 
\begin{align} \label{eq:alpha}
   \alpha & = -M_{\rm tot}^2 \frac{\partial^2 \ln P(m_A,m_B)}{\partial m_A \partial m_B}
  \nonumber \\
  & = \frac{10}{7} - \frac{\overline{M}_{\rm tot}^2}{\left(
  \overline{M}_{\rm tot}-\frac{11}{14}\right)^2 } + \frac{ \overline{M}_{\rm tot}^2
  \Lambda^{\overline{M}_{\rm tot}-\frac{11}{14}} \ln^2 \Lambda}{
 \left[ 1 - \Lambda^{\overline{M}_{\rm tot}-\frac{11}{14}}\right]^2} 
\end{align}
 if $m_{\rm BH,min} \leqslant m_{A,B} \leqslant m_{\rm BH,max}$, and zero otherwise. The $\alpha$ parameter is universal, in that it is independent of the underlying BH mass function. This is a monotonically decreasing function of $\overline{M}_{\rm tot}$ and ranges between $1.43 \geqslant \alpha \geqslant -20.35 + (\Lambda^{-3/28} - \Lambda^{3/28})^{-2} \ln^2 \Lambda$ for $0 \leqslant \overline{M}_{\rm tot} \leqslant 1$. Assuming a Milky Way-size nucleus and a fiducial multi-mass BH population with $dN/dm_{\rm BH} \propto m_{\rm BH}^{-2.35}$, $m_\mathrm{BH,max} = 30 \, \Msun$, and $p_0=0.5$, $\ln \Lambda$ is equal to $10.71$ and $10.17$ for the lightest and heaviest binaries, ergo $\alpha$ ranges between $-5.39 \la \alpha \la 1.36$ from the heaviest to lightest components of the EBBH population. The value of $\alpha$ is different for other source populations: $\alpha \sim 4$ for BH mergers in GCs \citep{OLearyetal2016}, $\alpha = 1$ for primordial BH binaries formed in the early universe \citep{Kocsisetal2017}, and $\alpha = 10/7 = 1.43$ for primordial BHs formed in dark matter halos through GW capture\footnote{Dark matter halos are collisionless systems, where mass segregation does not take place, so $\overline{M}_{\rm tot}=0$ in Equation (\ref{eq:alpha}).} \citep{Birdetal2016}.

 Next, we change variables from $m_A$ and $m_B$ to $q$ and $M_{\rm tot}$ using the Jacobi determinant $ \partial^3 \Gamma / \partial r\partial M_{\rm tot}\partial q = J \partial^3 \Gamma / \partial r \partial m_A\partial m_B$, where $J = M_{\rm tot}(1+q)^{-2}$ (see Equation (\ref{eq:mergrate_q_Mtot}) in Appendix \ref{sec:MMDGN}) 
\begin{align}  \label{eq:P(M,q)}
 P(M_{\rm tot},q) & \approx q^{\frac{2}{7}-\beta} M_{\rm tot} ^{3-2\beta}
 \left(1 - \frac{2}{3} \overline{M}_{\rm tot} + \frac{4 \, q }{9 \, (1+q^2)} \overline{M}_{\rm tot}^2 \right)
 \nonumber \\
 & \times \frac{ (1+q)^{2\beta - \frac{18}{7}} }{ \left(\frac{11}{14}-\overline{M}_{\rm tot}\right) }
  \left[ 1 - \left(\frac{r_{\min}^{A,B}}{r_{\max}}\right)^{\frac{11}{14}-\overline{M}_{\rm tot}} \right]
\end{align} 
 if $m_{\rm BH,min} \leqslant M_{\rm tot}/(1+q) \leqslant m_{\rm BH,max}$ and $m_{\rm BH,min} \leqslant M_{\rm tot} q/(1+q) \leqslant m_{\rm BH,max}$, and zero otherwise. These analytical results correspond to the distribution of all GW capture events, including direct collisions. The latter represents a small \mbox{($ \la 10 \%$)} fraction of the mergers. In the following, we plot the probability distribution functions for EBBHs that avoid a direct collision.

 The distribution of total detection rates from all galaxies is generally different from the merger rate density or the merger rate for a single nuclear star cluster, because the detection horizon $d_{\rm aLIGO}$ is different for different binary parameters. We calculate $d_{\rm aLIGO}$ for an eccentric inspiral waveform by calculating the signal-to-noise ratio following \citet{OLearyetal2009} \citep[see also][]{Gondanetal2017}. The detection rate distribution is then
 \begin{align}
 \frac{\partial^2 \mathcal{R}_{\rm aLIGO}}{\partial m_A \partial m_B}
 & =
 \int_{\rho_\mathrm{p0,min}}^{\rho_\mathrm{p0,max}} d\rho_\mathrm{p0} \int _{M_\mathrm{SMBH,min}}
 ^{M_\mathrm{SMBH,max}} dM_{\rm SMBH} V_{\rm aLIGO} \frac{d n_{\rm gal}}{dM_{\rm SMBH}}
  \nonumber \\
  & \times \left\langle \frac{\partial^3 \Gamma}{ \partial \rho_\mathrm{p0} \partial m_A
  \partial m_B} \right \rangle \, ,
 \end{align}
 where $d n_{\rm gal}/dM_{\rm SMBH}$ is the number density of galaxies in the universe with nuclear star clusters that host an SMBH of mass $M_{\rm SMBH}$, and $M_\mathrm{SMBH,min}$ and $M_\mathrm{SMBH,max}$ are the lower and upper bounds of the SMBH mass range of interest (Section \ref{subsec:GNModels}). The detectable volume is $V_{\rm aLIGO} = \frac43 \pi d_{\rm aLIGO}^3$, if $d_{\rm aLIGO}$ is much less than the Hubble scale. We substitute Equation (\ref{eq:radiusPDF}) for the partial event rates for different $\rho_\mathrm{p0}$, and change variables from $m_A$ and $m_B$ to $q$ and $M_{\rm tot}$ by using the Jacobian as in Equation (\ref{eq:P(M,q)}).

\begin{figure*}
    \centering
    \includegraphics[width=85mm]{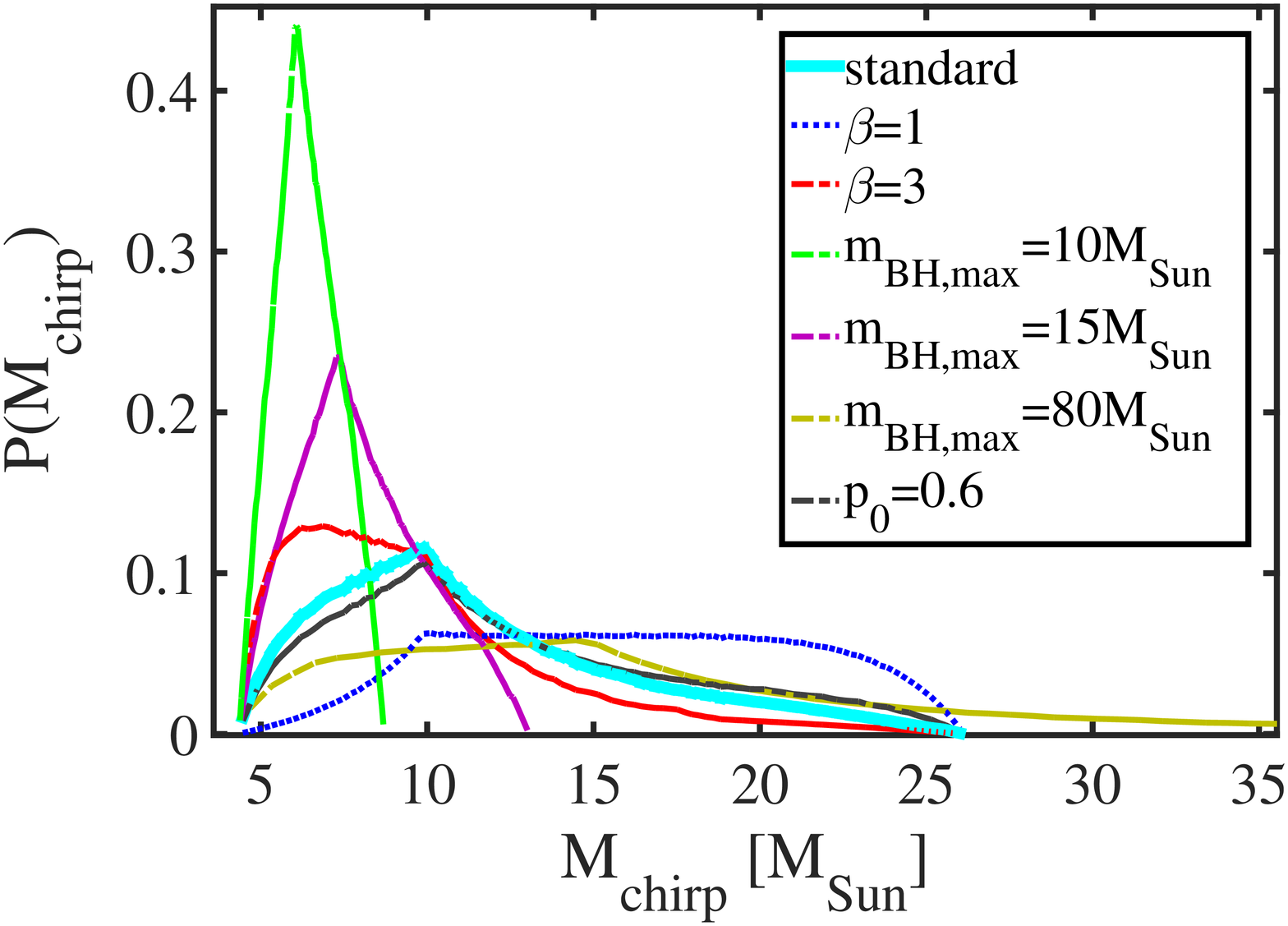}
    \includegraphics[width=85mm]{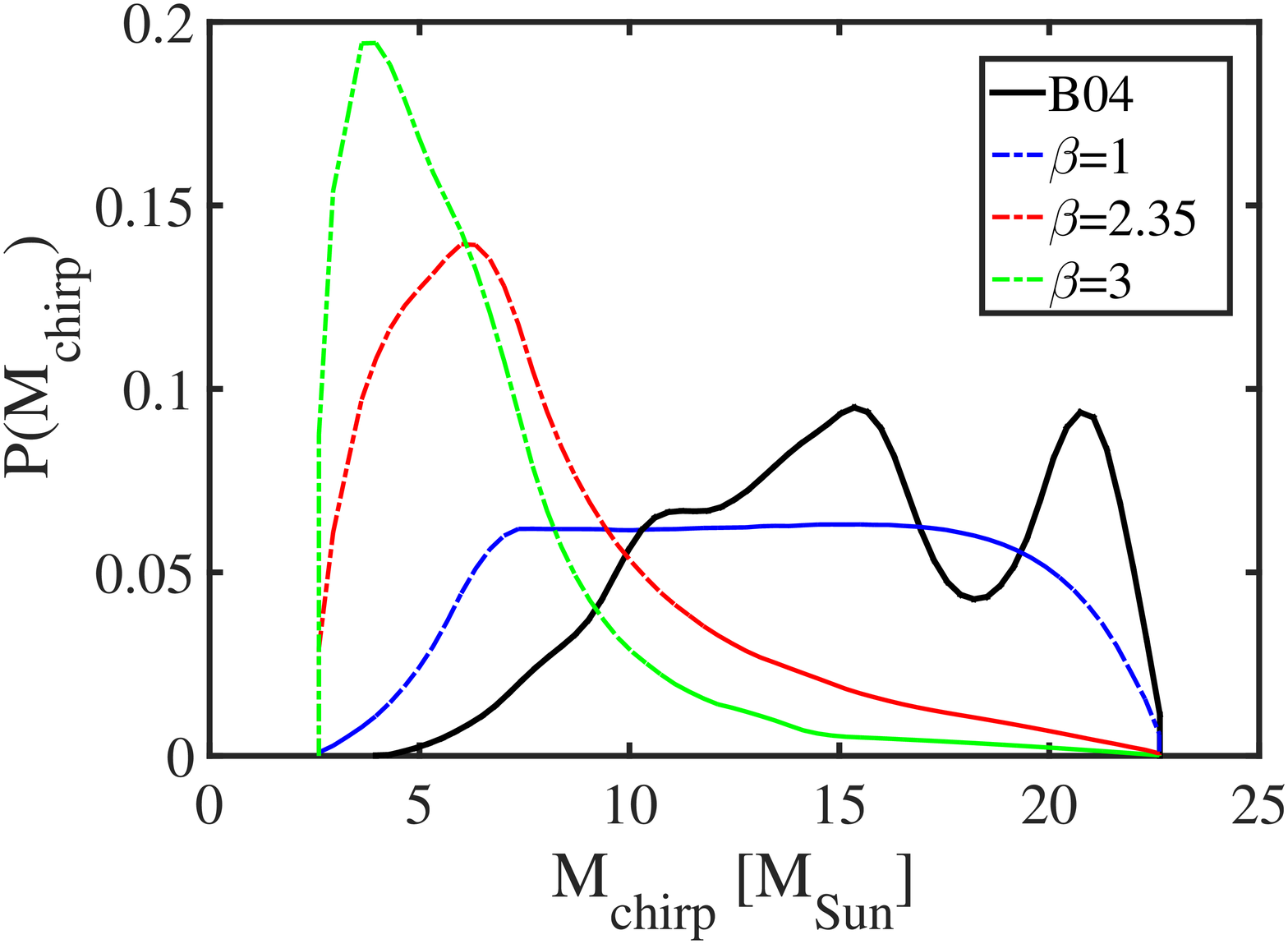}
 \\
    \includegraphics[width=85mm]{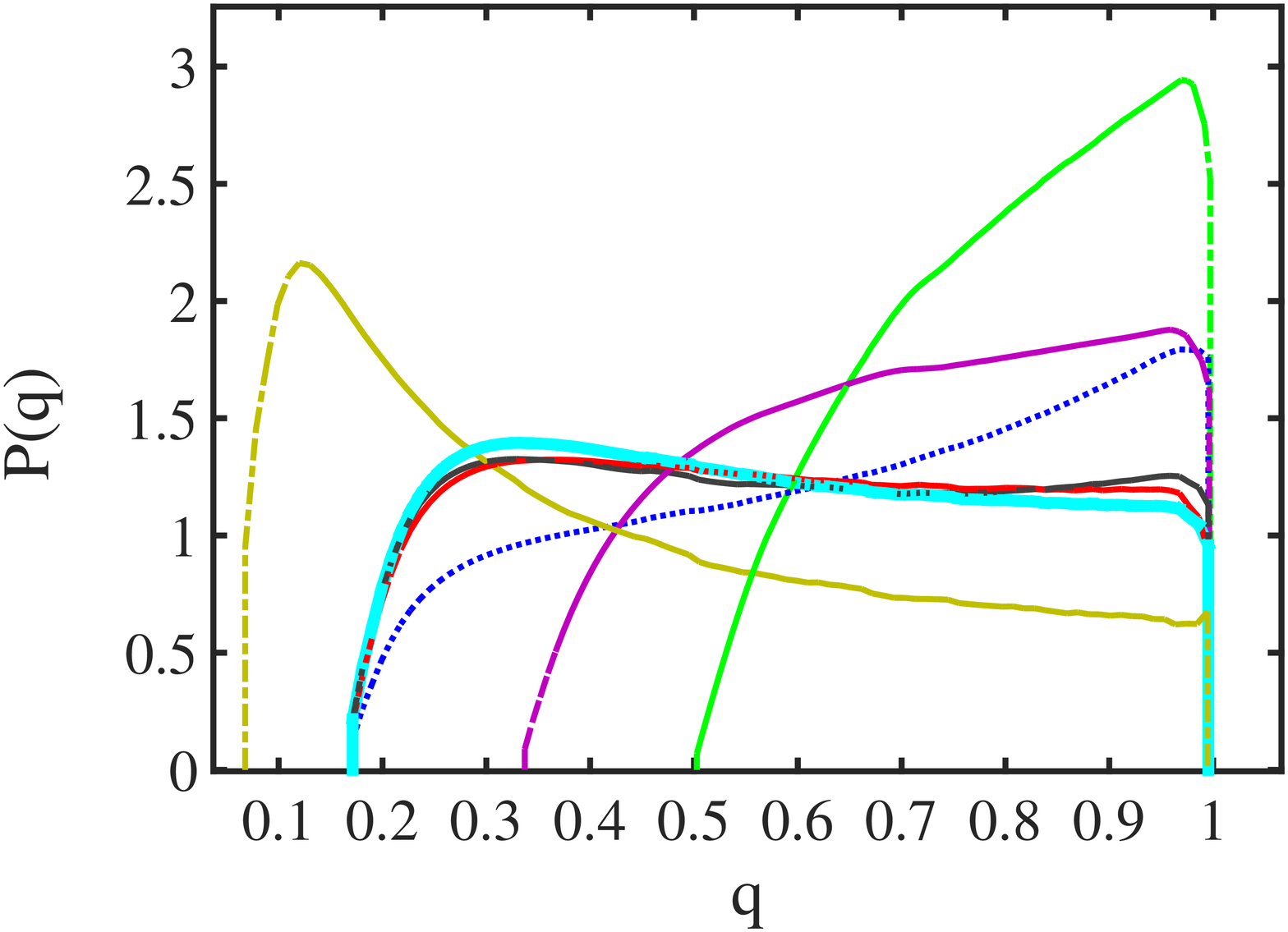}
    \includegraphics[width=85mm]{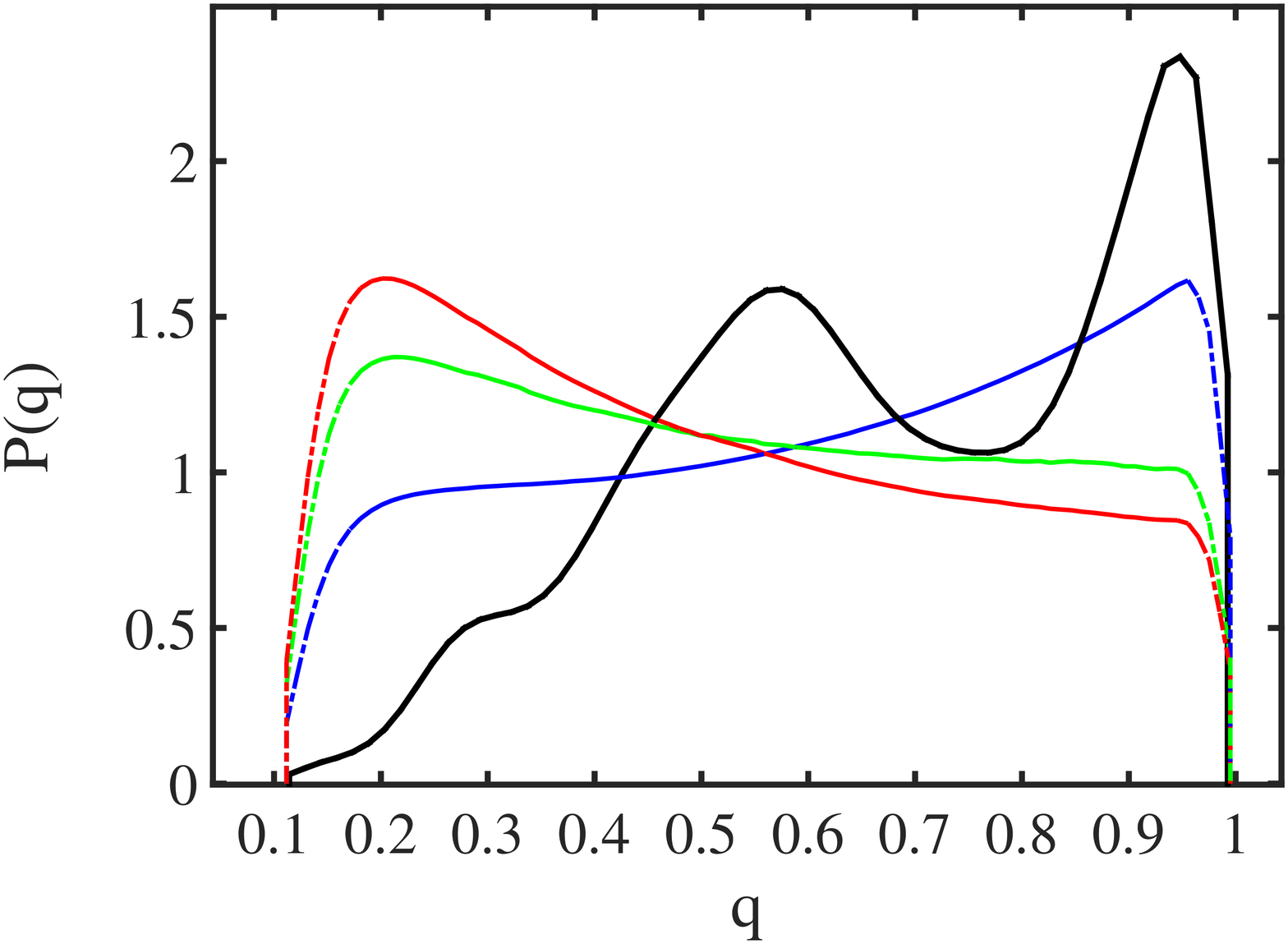}
 \\
    \includegraphics[width=85mm]{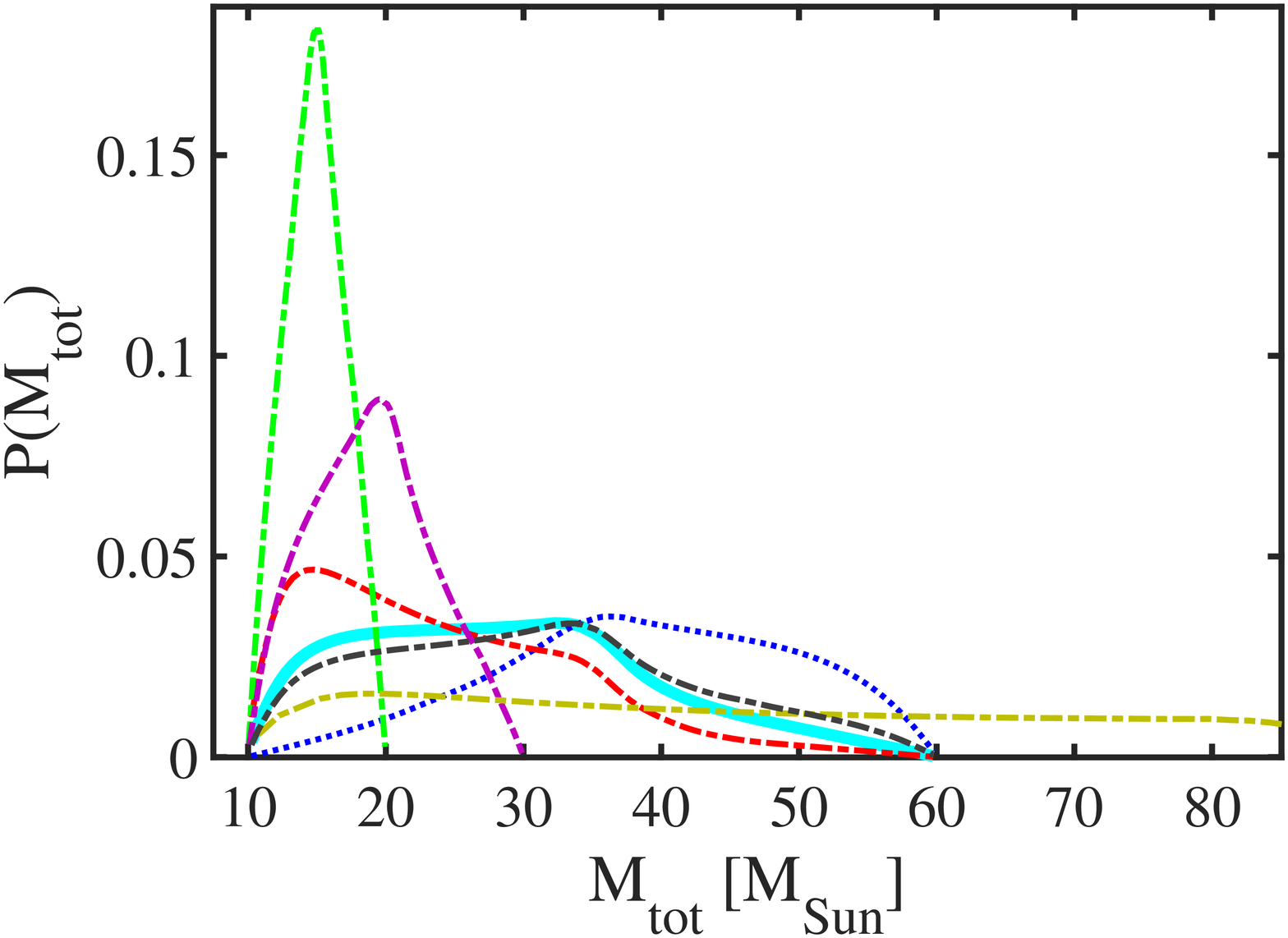}
    \includegraphics[width=85mm]{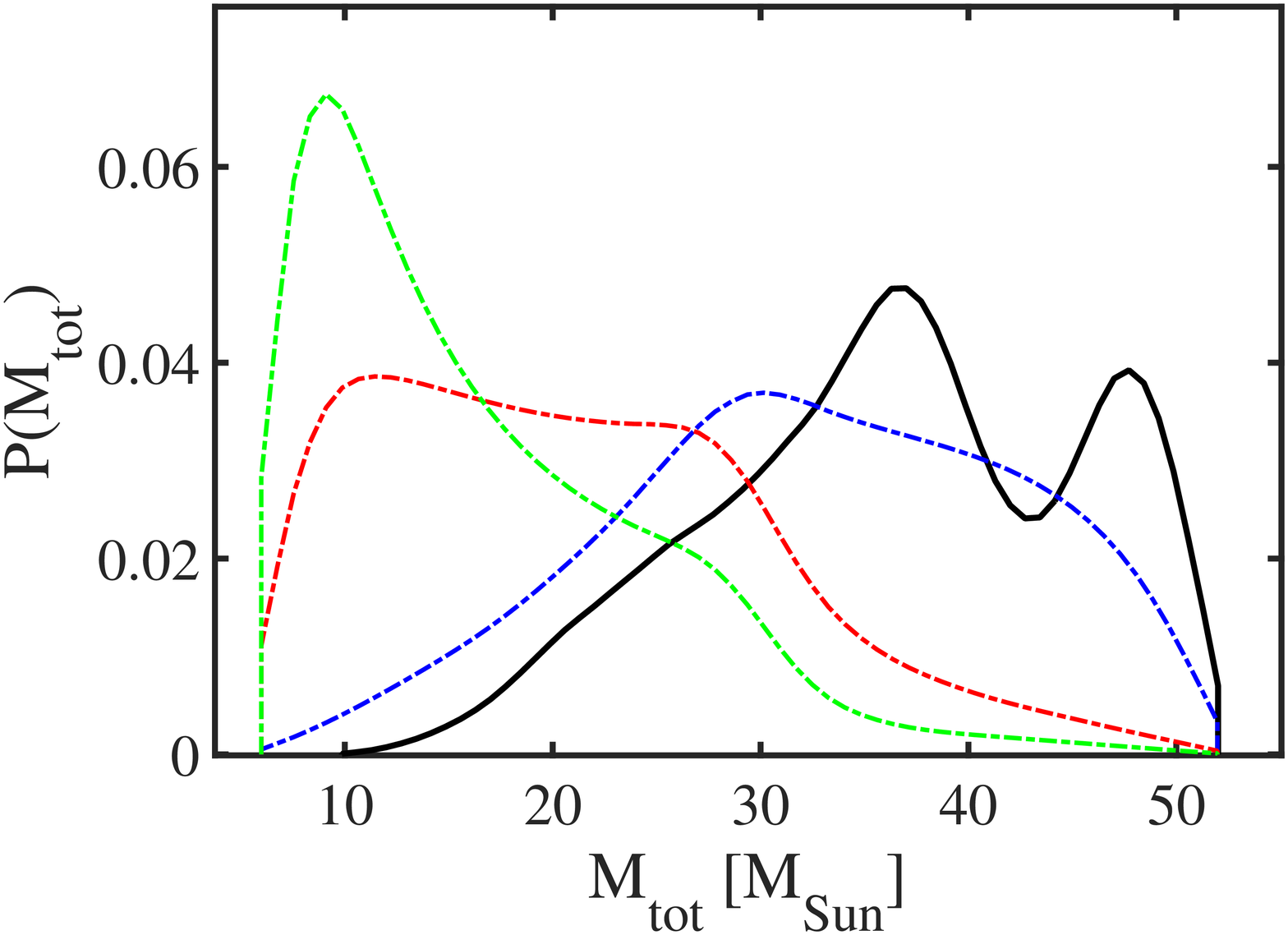}
\caption{ \label{fig:MassFuncs_PowerLaw_Belcz} Left panels: the PDF of chirp mass (top panel), mass ratio (middle panel), and total mass (bottom panel) for merging EBBHs in a single Milky Way-size nucleus. We show how different parameters of a multi-mass BH population influence the PDFs. Solid thick lines correspond to our fiducial multi-mass BH population with a PMF $dN/dm_{\rm BH} \propto m_{\rm BH}^{-\beta}$ with $\beta=2.35$, $p_0=0.5$, and $m_\mathrm{BH,max} = 30 \, \Msun$. Other line styles show the distributions for other parameters as labeled in the legend of the top panel. Right panels: the PDF of chirp mass (top panel), mass ratio (middle panel), and total mass (bottom panel) for merging EBBHs in a single Milky Way-sized nucleus. Solid lines correspond to a multi-mass BH population with the \citet{Belczynskietal2014_2} PMF and $p_0 = 0.5$ (labeled as B04 in the legend). Dashed-dotted lines correspond to a multi-mass BH population with a PMF $dN/dm_{\rm BH} \propto m_{\rm BH}^{-\beta}$ with $\beta=1, \, 2.35, \, 3$, $p_0=0.5$, and the BH mass range corresponds to the \citet{Belczynskietal2014_2} PMF. For all panels the results do not depend significantly on the assumed SMBH mass. } 
\end{figure*}

 Figures \ref{fig:NoPMFDist} and \ref{fig:MergRate_BelczModel} show the two-dimensional PDF of merging EBBH event rates as a function of $q$ and $M_{\rm tot}$ for power-law BH PMFs and for the \citet{Belczynskietal2014_2} PMF, respectively. The left panels show the distribution in a single Milky Way-size nucleus, and the right panels show the distribution for all galaxies in the universe that are detectable by aLIGO at design sensitivity. The top, middle, and bottom panels correspond to different BH PMF exponents $\beta=1$, $2.35$, and $3$, respectively, where the range of BH masses is assumed to be between $5 \, \Msun \leqslant m_{\rm BH} \leqslant 30 \, \Msun$. Because the two-dimensional PDF of merger rate in a single GN does not depend significantly on the assumed SMBH mass (Appendix \ref{sec:MMDGN}), the left panels display the distribution of the merger rate density for single GNs in the universe. Low-mass binaries dominate the merger rate density if $\beta \gtrsim 3/2$ (see Equation \ref{eq:P(M,q)}) and high-mass binaries close to $M_{\rm tot} \sim 2m_{\rm BH, max}$ dominate the merger rate density for a top-heavy mass function with $\beta=1$. However, because aLIGO is more sensitive to higher-mass mergers with mass ratios closer to 1\footnote{That is, the maximum luminosity distance of detection increases with $M_{\rm tot}$ and $q$ over the considered ranges of $M_{\rm tot}$ and $\rho_{\rm p0}$, see Figure 11 in \citet{OLearyetal2009} or Appendix \ref{sec:aLIGOeventrate}}, the detection rate distributions shown in the right panels are skewed to higher total masses and equal mass ratios. This observational bias causes the detection rates to be highest at $M_{\rm tot} \gtrsim m_{\rm BH, max}$ up to $2 m_{\rm BH, max}$ for $\beta =1$ and $2.35$. For $\beta= 3$, the distribution is peaked near $M_{\rm tot} \sim m_{\rm BH, max}$ with unequal mass ratios. In all cases, equal mass mergers at the low end of the mass distribution $M_{\rm tot} \sim 2 m_{\rm BH, min}$ are highly disfavored for $\beta \leqslant 3$. Note that the mass ratio dependence of the detection rate for fixed $M_{\rm tot}$ is known analytically because $d_{\rm aLIGO} \propto \eta^{1/2} = q^{1/2}(1+q)^{-1}$; see Equation (\ref{eq:dLmax}) or \citet{OLearyetal2009}. Thus, $\partial^2 \mathcal{R}_{\rm aLIGO}/ \partial q \partial M_{\rm tot} \propto q^{(25/14) - \beta} (1+q)^{2\beta - (39/7) }$.

\begin{figure}
    \centering
    \includegraphics[width=85mm]{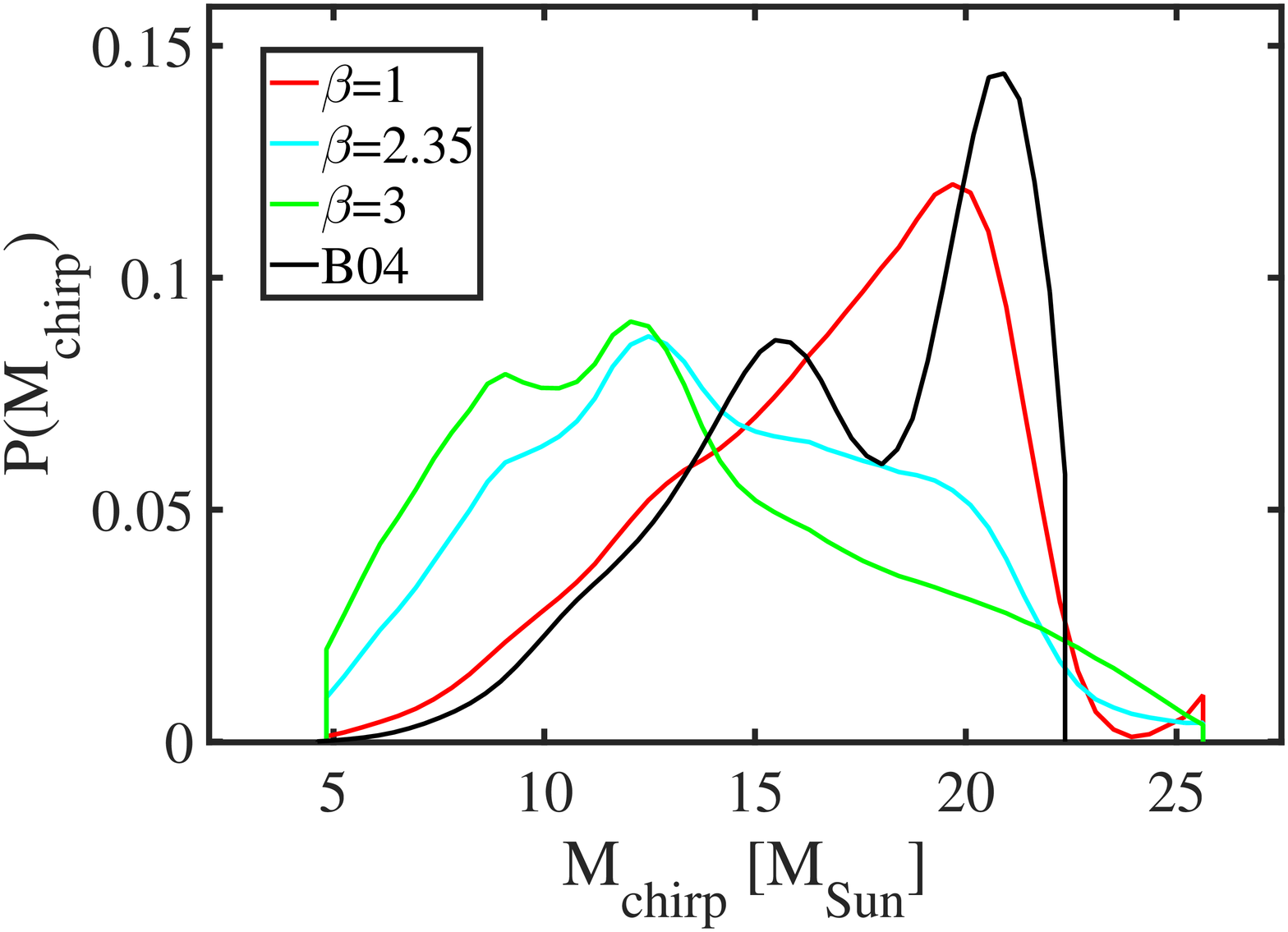}
    \includegraphics[width=85mm]{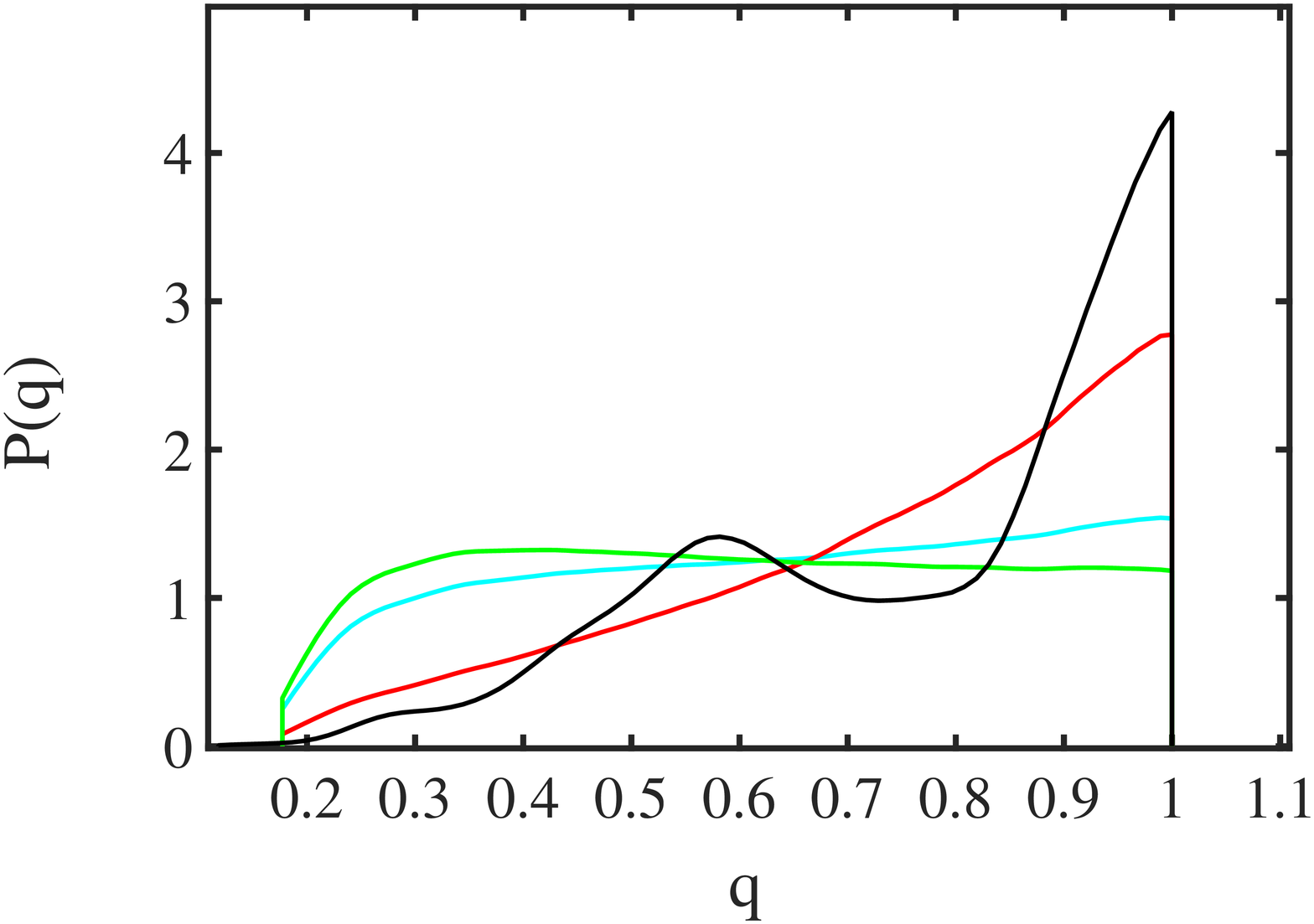}
    \includegraphics[width=85mm]{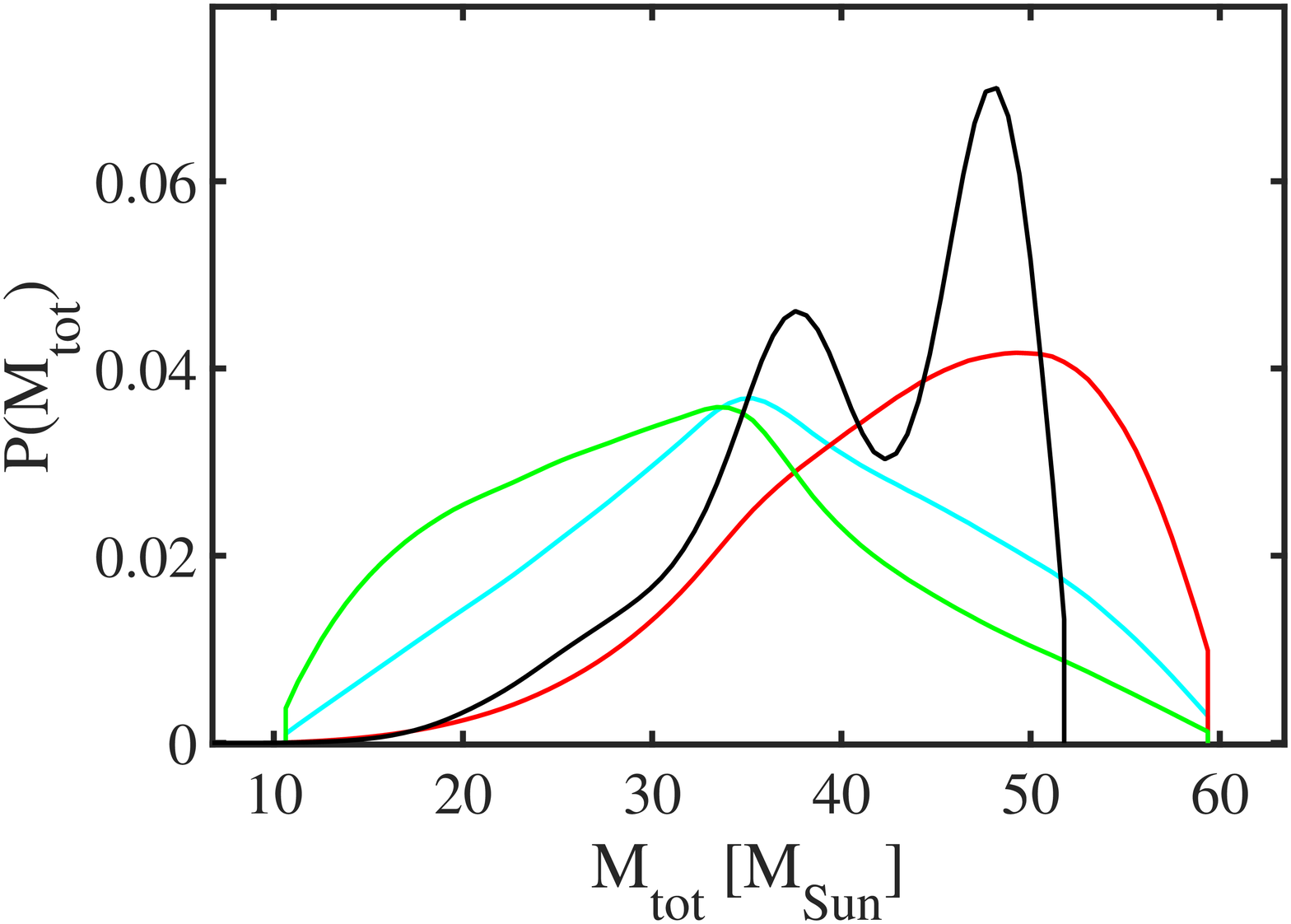}
\caption{ \label{fig:MassDists_DetRate} The 1D marginalized distribution of the merger rate for all galaxies in the universe detectable by aLIGO at design sensitivity, as a function of chirp mass (top panel), mass ratio (middle panel), and binary total mass, respectively. Here, we assumed a fiducial multi-mass BH population with PMFs $dN/dm_{\rm BH} \propto m_{\rm BH}^{-\beta}$, where $\beta = 1,\, 2.35,\, 3$ as labeled in the legend, and $m_\mathrm{BH,max} = 30 \, \Msun$, and $p_0 =0.5$. Furthermore, we also considered the \citet{Belczynskietal2014_2} PMF with $p_0 =0.5$ (black curve). } 
\end{figure}

 The different rows of panels in Figure \ref{fig:MassFuncs_PowerLaw_Belcz} show the 1D marginalized distribution of the merger rate in a single galaxy in a Milky Way-size nucleus as a function of chirp mass, binary total mass, and mass ratio, respectively. The left and right panels show results for multi-mass BH populations with power-law PMFs and \citet{Belczynskietal2014_2} PMF. The mass ratio distribution of the merger rate density is widely distributed, with comparable component-mass mergers being a factor of $\sim 2$ more likely than mergers with uneven component masses when $\beta \sim 1$, while uneven mass mergers are $20\%$ more likely than comparable mass mergers for $\beta \sim 2.35$. These panels also show how the marginalized merger rate density depends on other model parameters as follows. Varying $m_{\rm BH, \max}$ changes the boundaries of the $P(q,M_{\rm tot})$ distributions. However, Equation (\ref{eq:P(M,q)}) shows that $P(q, M_{\rm tot})$ is independent of $m_{\rm BH, \max}$ and the value of the mass segregation parameter $p_0$, other than for a rescaling of the $M_{\rm tot}$ mass units. The mass distributions are also insensitive to $M_{\rm SMBH}$ (Appendix \ref{sec:MMDGN}).

 Finally, Figure \ref{fig:MassDists_DetRate} shows examples for the 1D marginalized distribution of the merger rate for all galaxies in the universe that are detectable by aLIGO at design sensitivity, as a function of chirp mass, binary total mass, and mass ratio, respectively. We used a fiducial multi-mass BH population with PMFs $dN/dm_{\rm BH} \propto m_{\rm BH}^{-\beta}$, where $\beta = 1,\, 2.35,\, 3$; $m_\mathrm{BH,max} = 30 \, \Msun$; and $p_0 =0.5$. The black line shows results for the \citet{Belczynskietal2014_2} PMF with $p_0 =0.5$. The PDFs presented in these panels are skewed to higher total masses and equal mass ratios, compared to those presented in Figure \ref{fig:MassFuncs_PowerLaw_Belcz} because aLIGO is more sensitive to higher-mass mergers with higher mass ratios.

\section{Summary and Conclusion}
\label{sec:Summary}

 In this paper, we determined the expected distribution of intrinsic parameters of merging BH binaries that form in GN from initially unbound BHs due to GW emission during close encounters. The main intrinsic parameters describing these binaries are their component masses, the initial impact parameter, and initial relative velocity before the encounter. Due to the large velocity dispersion in GN compared to objects in GCs, the galactic disk, or the halo, the objects in GNs must approach one another at much smaller distances to form a binary. The implication is that the initial pericenter distance of the binary is relatively small when the initial eccentricity is beyond $e_0>0.9$. We extended the study of  \citet{OLearyetal2009} by calculating the eccentricity distribution for mergers among different component masses in a multi-mass distribution, and examined the eccentricity distribution at three particular places during the evolution: at formation, at the instant when the peak GW frequency reaches $10 \, \mathrm{Hz}$, and at the LSO. We have also examined how different parameters describing a multi-mass BH distribution affect the distribution of merger rates as a function of masses, eccentricity, and initial pericenter distance. Many of our numerical results and the identified trends are validated by analytical expressions (Appendices \ref{TWORPEPS}--\ref{ED}).

The main findings of this work can be summarized as follows:
\begin{enumerate}[label=(\roman*),ref=(\roman*)]

\item We identified the region in the initial eccentricity--initial dimensionless pericenter distance plane where EBBHs may form through GW capture, and determined the two-dimensional PDF therein. The PDF of initial orbital parameters (Figure \ref{fig:Distrhop0}) shows that EBBHS form with high eccentricities ($0.9 \lesssim e_0 \lesssim 0.9999$), and the distribution of initial pericenter distance drops off quickly beyond $20-80 \, M_{\rm tot}$. The reason is that BHs form EBBHs with high relative velocities in nuclear star clusters around SMBHs (Figure \ref{fig:w_Avr_mSMBH}). The eccentricity makes GW capture sources in these environments qualitatively different from other types of inspiraling GW  sources.

\item We determined the radial distribution from the central SMBH for GW capture sources with different component masses. We found that, on a logarithmic radial scale, the merger rate distribution is approximately independent of radius for the most massive components, $P(\ln r) \propto r^{-3/14}$ (Equation \ref{eq:formrateBHBH_multi}), between the radius of influence and the inner boundary set by GW-driven infall into the SMBH (i.e. $\sim 3 \, \mathrm{pc}$ and of order $10^{-4}\, \mathrm{pc}$, respectively, in Milky-Way type galaxies). This implies that roughly $50 \%$ of the high-mass events are formed in the innermost $4 \times 10^{-3} \, {\rm pc}$, and approximately $97 \%$ within $1.2 \, {\rm pc}$. The rate of GW capture sources among the least massive objects within the nuclear star cluster is skewed to higher radii on a logarithmic scale, $P(\ln r) \propto r^{11/14}$, for $m_{\rm BH} \ll m_{\mathrm{BH,max}}$ (Equation \ref{eq:formrateBHBH_multi}). These events are preferentially closer to the radius of influence; only $2 \%$ and roughly $50 \%$ of them are within $4 \times 10^{-3} \, {\rm pc}$ and $1.2 \, {\rm pc}$, respectively.

\item We found that a negligible fraction ($1 \% \la$) of EBBHs interact with a third object during the eccentric inspiral.

\item We determined how the distributions of orbital parameters depend on the mass of the central SMBH, on the binary masses, and on the parameters of the BH population (i.e. the highest possible BH mass, the exponent of the BH mass function, and the mass segregation parameter). We found that the initial pericenter distance systematically decreases with $M_{\rm SMBH}$ and the component masses, while the eccentricity at any stage of the time evolution systematically increases with $M_{\rm SMBH}$ and the component masses. As a consequence, more massive EBBHs around more massive SMBHs reach a $10 \, {\rm Hz}$ frequency limit with higher eccentricity, and they have higher eccentricity at the LSO (see Tables \ref{tab:FracM_SMBH}, \ref{tab:FracM_SMBH_2}, and \ref{tab:FracM_SMBH_Belcz}). The highest possible BH mass of the BH population affects the binary parameters of GW capture sources. Other parameters of the BH population do not influence the eccentricity distribution at formation, at $10\, \mathrm{Hz}$, or at the LSO.

\item We found that almost all EBBHs form below the frequency range where aLIGO type detectors are the most sensitive, and so GW capture sources typically enter the aLIGO type detectors' sensitive frequency band for stellar mass BHs (Table \ref{tab:FracM_over100Hz}).

 \item We found that most of the sources become circularized with eccentricities $e_{\mathrm{LSO}} \sim 1 - 10 \%$ when they approach the LSO. The fraction of sources with $e_{\rm LSO} > 0.1$ is $3 - 30 \%$ if the maximum possible BH mass in the population is $30 \, \Msun$: this fraction of eccentric mergers among BHs with masses less than $30 \, \Msun$ is less than $10\%$ if the highest mass in the population reaches $80 \, \Msun$ (Table \ref{tab:FracM_SMBH_2}). Because the current GW detectors are only sensitive to the final part of the signal, the currently known six detections cannot constrain the fraction of sources formed in this initially highly eccentric merger channel. Only 1 in 30 detections with $e_{\rm LSO} > 0.1$ may be expected if the heaviest stellar BH mass in GNs is $80 \, \Msun$. The heaviest BHs in such a population are expected to have higher eccentricities; $29 - 36 \%$ of them have $e_{\rm LSO}>0.1$, but their detection may require algorithms sensitive to non-circular orbit waveform features.

\item The prospects for detecting eccentric mergers are expected to improve significantly with the ongoing detector upgrades. We found that the fraction of sources with an eccentricity larger than 0.1 when the peak GW frequency reaches $10\, \mathrm{Hz}$ is $84 - 90\%$ ($43 - 69\%$) for $30\, \Msun - 30 \, \Msun$ binaries in a BH population with maximum mass of $30\Msun$ ($80\Msun$). For lower-mass binaries, these numbers are higher, e.g. around $68 - 94\%$ of $10 \, \Msun - 10 \, \Msun$ binaries (Tables \ref{tab:FracM_SMBH}, \ref{tab:FracM_SMBH_2}, and \ref{tab:FracM_SMBH_Belcz}). Thus, when Advanced LIGO and VIRGO reach their design sensitivity at $10 \, \mathrm{Hz}$, more than half of GW capture sources in GN are initially eccentric in the detector band. The measurement of the eccentricity and mass distribution will make way to identify the fraction of sources formed in this channel.

\item A negligible fraction of EBBHs have eccentricities at LSO lower than $10^{-3}$ (Figure \ref{fig:DisteLSO}). In a companion paper, \citet{Gondanetal2017}, we have shown that the expected relative measurement accuracy of $e_{\rm LSO}$ for the aLIGO--AdV--KAGRA GW detector network is $\la 5 \%$ when EBBHs form with relatively small $\rho_{{\rm p}0}$ values at a distance of $100 \, \mathrm{Mpc}$\footnote{Eccentric BH binaries form through GW capture in GN with initial dimensionless pericenter distance $\rho_{{\rm p}0}=r_{\rm p0}/M_{\rm tot} \la 50$ ($\rho_{{\rm p}0} \la 100$) for masses $30 \, \Msun -30 \, \Msun$ ($10 \, \Msun -10 \, \Msun$); see Figure \ref{fig:Distrhop0}.}. Other binary formation and evolution channels typically produce binaries with eccentricities at or below $10^{-3}$ when their GW signals enter the aLIGO band \citep{Kowalskaetal2011,Cholisetal2016,Rodriguezetal2016a,Rodriguezetal2016b,SamsingRamirezRuiz2017,SilsbeeTremaine2017,RandallXianyu2018}. The measurement of eccentricity for a BH merger will be a smoking gun signature of sources in high velocity dispersion environments such as in the inner regions of GNs (Equation \ref{eq:sigma-eLSO}).

\item We conclude that Advanced LIGO/VIRGO detections may confirm or rule out this binary formation channel. However, algorithms using circular binary templates will be ineffective, for the majority of sources, in searching for GW signals of EBBHs forming through GW capture in GNs. This especially affects the heaviest BHs because they are the most eccentric among this source population. The current GW detection limits of the heaviest black holes \citep{Abbottetal2017_4,FishbachHolz2017} may be impacted by the inefficiency of the applied search algorithms to eccentric sources.

 \item We determined the PDFs of merger rates in single nuclear star clusters in terms of component masses (i.e. total binary mass and mass ratio) and similarly for the detection rates for sources with the Advanced LIGO horizon (Figure \ref{fig:NoPMFDist}). The distributions are sensitive to the BH mass function exponent. The highest possible BH mass in GNs affects the total mass and mass ratio distribution just by a rescaling of the mass unit (Equation \ref{eq:P(M,q)}), but the mass distribution of merger rates are otherwise insensitive to the maximum BH mass in GNs. In particular, mass functions steeper/shallower than $dN/dm_{\rm BH} \propto m_{\rm BH}^{-1.5}$ lead the merger rate density to be dominated by the lowest/highest mass components in the cluster. However, the detection rate distribution is skewed toward higher masses due to Advanced LIGO/VIRGO's higher sensitivity to heavier BHs for mass function exponents shallower than $dN/dm_{\rm BH} \propto m_{\rm BH}^{-3}$.

 \item We calculated the value of the universal dimensionless parameter $\alpha = -M_{\rm tot}^2 \partial^2 \ln \mathcal{R} /\partial m_A \partial m_B$, which characterizes the physical origin of a source population independently of the underlying BH mass function \citep{Kocsisetal2017}. We found that, for GW capture sources in a mass-segregated cusp, it follows a monotonically decreasing function of total binary mass (Equation \ref{eq:alpha}) ranging between $1.43 \geqslant \alpha \geqslant -20.35 + (\Lambda^{-3/28} - \Lambda^{3/28})^{-2} \ln^2 \Lambda$, where $\ln \Lambda$ is the Coulomb logarithm (Equation \ref{eq:CoulombLog}). For $\ln \Lambda\sim 10.5$, this parameter is approximately $\alpha \sim -5.39$ and $1.36$ for the heaviest and lightest components, respectively. This is very different from the $\alpha$ of other GW source populations because $\alpha \sim 4$ for BH mergers in GCs \citep{OLearyetal2016}, $\alpha=1$ for primordial BH binaries formed in the early universe  \citep{Kocsisetal2017}, and $\alpha=1.43$ for BH binaries that form due to GW capture in otherwise dynamically collisionless systems, such as black holes in dark matter halos \citep{Birdetal2016}.

 \item The mass ratio distribution is wide, for mass functions $dN/dm_{\rm BH} \propto m_{\rm BH}^{-1}$ to $dN/dm_{\rm BH} \propto m_{\rm BH}^{-3}$ and for the \citet{Belczynskietal2014_2} mass function (Figures \ref{fig:NoPMFDist} and \ref{fig:MergRate_BelczModel}). For these mass functions, the detection of equal-mass mergers are highly disfavored for low-mass BHs, but equal mass mergers are possible among high-mass BH sources. Unequal mass mergers are possible for a wider range of total masses (Figure \ref{fig:MassFuncs_PowerLaw_Belcz}).

\item Because GW capture sources are formed roughly independent of BH spins and spin directions at pericenter distances exceeding a few tens of $M_{\rm tot}$, the spin direction distribution for these sources is expected to be isotropic.

\end{enumerate}

 We worked under the assumption of a relaxed cusp and assumed spherical symmetry. In the Milky Way's center, the distribution of massive young stars does not represent such a relaxed profile, but forms a clockwise disk structure with a number density distribution $n(r) \sim r^{-2.9}$, which is steeper than the model assumed in \citep{Bartkoetal2009,Yeldaetal2014}. The stellar distribution around SMBHs with masses $M_{\rm SMBH}>10^7\Msun$ does not have time to relax within a Hubble time. If BHs in these regions form steeper radial profiles, EBBHs form closer to the  center of the GN in higher-velocity dispersion regions. This implies that the eccentricity distribution of mergers is skewed toward higher eccentricities. However, these massive GN are expected to be subdominant, in terms of EBBH rates, because the rates per single GN approximately do not depend on the SMBH mass and low-mass SMBHs are more common (see \citet{Sijackietal2015} and references therein). However, the mass-segregated cusp may have a different radial dependence, even in equilibrium, than that assumed in this paper. \citet{Keshetetal2009} found a steeper number density dependence for the heavy objects if they are subdominant in mass. \citet{Fragioneetal2018} also found a steeper profile due to disrupted binaries. Recently \citet{Szolgyenetal2018} have shown that the heavy objects form a flattened distribution. The mass-segregated radial profile may be different in such configurations \citep{Fouvryetal2018}, leading to a modified EBBH merger rate distribution.

 In conclusion, the measurement of the parameter distribution of merging compact objects may be useful to distinguish this formation mechanism from other mechanisms. Search methods targeting EBBH sources forming through GW capture in GN hosts may be optimized for the predicted distribution of binary parameters.

\section*{Acknowledgment}

 We thank the anonymous referee for constructive comments, which helped improve the quality of the paper. This project has received funding from the European Research Council (ERC) under the European Union's Horizon 2020 Programme for Research and Innovation under grant agreement ERC-2014-STG No. 638435 (GalNUC), and from the Hungarian National Research, Development, and Innovation Office under grant NKFIH KH-125675. P\'eter Raffai was supported by the \'UNKP-17-4 New National Excellence Program of the Ministry of Human Capacities of Hungary. This work was performed in part at the Aspen Center for Physics, which is supported by National Science Foundation grant PHY-1607761.

\appendix

\section{The Region of EBBH Formation Inside the Galactic Nucleus}
\label{sec:CalRmin}

 The PDFs of $\epsilon_0$ and $\rho_\mathrm{p0}$ may be sensitive to the value of the innermost radius $r_{\rm min}^{A,B}$, where BHs may reside. Here, we determine $r_{\rm min}^{A,B}$ under the following assumptions.
\begin{enumerate}[label=(\Alph*),ref=(\Alph*)]
   \item \label{i:critA} If $t_{\rm rlx} \leqslant t_{\rm H}$, the system reaches collisional equilibrium, i.e. the number density profiles of stellar populations (e.g. WDs, MSs, NS, and BHs) relax into a mass-segregated, steady-state, power-law density cusp within a Hubble time $t_{\rm H}=10^{10}$ years.

   \item \label{i:critB} If $t_{\rm rlx} \leqslant t_{\rm GW}$, the relaxation time of BHs $t_{\rm rlx}$ is equal to or less than the infall of a BH into the SMBH $t_{\rm GW}$.
 \end{enumerate}

 We next discuss the implications of these criteria in separate subsections.

\subsection{Criterion A, $t_{\rm rlx} \leqslant t_{\rm H}$}
\label{subsec:CritI}

 We assume the GN to be relaxed if all of its stellar components (e.g. WDs, MSs, NSs, and BHs) are relaxed within the age of the galaxy, $t_{\rm H}$. The evolution of the number density distribution of stellar populations is governed by the Fokker-Plank equation, which describes the energy and angular momentum diffusion and dynamical friction due to two-body interactions among all objects in the system \citep{BahcallWolf1976,BahcallWolf1977,Freitagetal2006,HopmanAlexander2006b,AlexanderHopman2009,OLearyetal2009,BarOr2013,BarOrAlexander2016,Alexander2017}. For each stellar component of mass $m$, we use the local Chandrasekhar two-body energy relaxation time to reach steady state \citep{AlexanderLivio2004,OLearyetal2009,BarOr2013,AlexanderPfuhl2014,Alexander2017}
\begin{equation}  \label{eq:trlx}
  t_{\rm rlx} (r,m, M_{\rm SMBH}) = 0.34 \frac{\sigma ^3_* (r) }{G^2 n_{\rm tot}(r)
  \langle M^2 \rangle (r) \mathrm{ln} \Lambda}
\end{equation}
 \citep{Spitzer1987}. Here, $\sigma _* (r)$ denotes the 1D velocity dispersion, which is of the form
\begin{equation}  \label{eq:sigma}
 \sigma _* ^2 (r) = \frac{ v_{\rm circ}^2 }{ 1 + \alpha (m) }
\end{equation}
 \citep{Alexander1999,AlexanderPfuhl2014}, where $v_{\rm circ}$ denotes the circular velocity at radius $r$\footnote{$v_{\rm circ} = \sqrt{G M_{\rm SMBH}/r}$}, and $\alpha(m)$ is the exponent of the number density distribution of mass $m$ objects surrounding the SMBH\footnote{Section \ref{subsec:PropStellPop} gives $\alpha$ for the considered stellar populations.}. The Coulomb logarithm $\ln \Lambda$ in Equation (\ref{eq:trlx}) can be approximated as $\Lambda \sim M_{\rm SMBH} / m$ in the region where the gravitational force is dominated by the central SMBH (i.e. when $r_{\rm G} \leqslant r \leqslant r_{\rm max}$) \citep{BarOr2013}. In Equation (\ref{eq:trlx}), $\langle M^2 \rangle$ is the second moment of the mass function for the mixture of stellar populations at each radius:
 \begin{equation}  \label{eq:M2avr}
 \langle M^2 \rangle(r) = n(m,r)^{-1} \int dm \, n(m,r) \, m^2 \, ,
 \end{equation}
 where $n(m,r)$ denotes the number density distribution of mass $m$ objects. Thus, $\langle M^2 \rangle$ can be given for a mixture of the WD, MS, NS, and BH populations. In this paper, we carry out our calculations separately for an idealized single-mass BH population and for a multi-mass population. For the single-mass case, we have
\begin{align}  \label{eq:M2avrsingle}
  \nonumber n_{\rm tot}(r)\langle M^2 \rangle (r) & =  n_{\rm MS}(r) \, m_{\rm MS}^2 +  n_{\rm WD} (r)
   \, m_{\rm WD}^2 +   n_{\rm NS} (r) \, m_{\rm NS}^2
   \\
  & +  n_{\rm BH,s} (r)  \, m_{\rm BH}^2 \, ,
\end{align}
 and for the general multi-mass BH population
\begin{align}  \label{eq:M2avrmulti}
   \nonumber n_{\rm tot}(r)\, \langle M^2 \rangle (r) & = n_{\rm MS}(r) \, m_{\rm MS}^2 + n_{\rm WD} (r)
   \, m_{\rm WD}^2 +  n_{\rm NS} (r) \, m_{\rm NS}^2
  \\
  & + \int_{m_{\rm BH,min}}^{ m_{\rm BH,max} } dm_{\rm BH} \, n_{\rm BH,m} (r,m_{\rm BH}) \, m_{\rm BH}^2 \, ,
\end{align}
 where $n_{\rm MS}$, $n_{\rm WD}$, $n_{\rm NS}$, and $n_{\rm BH, s, m}$ are given in Section \ref{subsec:PropStellPop}. We look for the solution to Criterion \ref{i:critA} in the range of radii between the gravitational radius of the central SMBH \mbox{($r_{\rm GR} = 2 GM_{\rm SMBH}/c^2$)} and the radius of influence $r_{\rm max}$.

 We find numerically that $t_{\rm rlx}$ increases with $r$, $m$, and $M_{\rm SMBH}$. Because $r<r_{\max}$ and \mbox{$m_{\rm  BH} \leqslant m_\mathrm{BH,max}$}, we find that Criterion \ref{i:critA} may only be satisfied if $M_{\rm SMBH} \lesssim 10^7 \, \Msun$ for all multi-mass BH parameters considered. A more detailed analysis by \citet{BarOr2013} reached a similar conclusion. Therefore, we restrict the upper bound of the SMBH mass range of interest to be $10^7 \, \Msun$.

\subsection{Criterion B, $t_{\rm rlx} \leqslant t_{\rm GW}$}
\label{subsec:CritII}

 We use \citet{Peters1964} to calculate the evolution of the semi-major axis for an eccentric orbit
 \begin{align}  \label{eq:dadt}
 \nonumber t_{\rm GW} (a, m_{\rm BH}, M_{\rm SMBH},e) & = \left\arrowvert a \left(\frac{da}{dt}\right)^{-1} \right\arrowvert
 \\
  & = \frac{5 \, c^5 \, a^4 }{64\, G^3 \, \mu \, M_{\rm tot}^2} \frac{(1-e^2)^{7/2}}{\left(1 +
  \frac{73}{24} e^2 + \frac{37}{96} e^4 \right)} \, ,
 \end{align}
 where now, when discussing the infall of an object of mass $m$ into the SMBH, $M_{\rm tot} = M_{\rm SMBH}$ and $\mu=m_{\rm BH}$. At the boundary of Criterion \ref{i:critB}, we set this equal to the relaxation time, where for concreteness\footnote{The results are not sensitive to this assumption because the relaxation time is very weakly sensitive to radius, $t_{\rm rlx} \propto r^{\alpha-1.5}$.} we substitute the semi-major axis in Equation (\ref{eq:trlx}). Note that this refers to two-body relaxation, which governs the evolution of the semi-major axis. The eccentricity distribution relaxes faster due to coherent gravitational torques arising between elliptical orbits, on the so-called resonant relaxation timescale \citep{RauchTremaine1996,KocsisTremaine2011}.

 The GW inspiral timescale $t_{\rm GW}$ is smaller for more massive objects or for higher orbital eccentricity around the SMBH for a fixed semi-major axis. This leads to the preferential removal of the more massive BHs and the more eccentric orbits for a fixed semi-major axis, creating a loss-cone around the SMBH. However, resonant relaxation tends to mix the eccentricity distribution of orbits around the SMBH toward an isotropic thermal distribution where $dN/de \propto e$. For such a distribution, $81\%$ of the orbits have $e < 0.9$, and $90\%$ have $e < 0.95$.

\begin{figure}
    \centering
    \includegraphics[width=85mm]{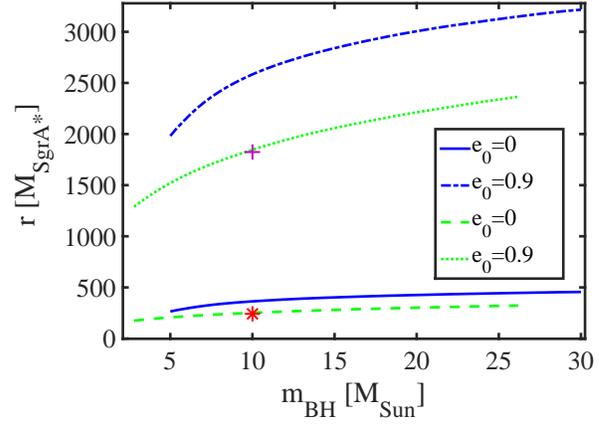}
\caption{ \label{fig:trlx_tGW}
 The radius $r^{\rm rlx,GW}$ at which the GW inspiral timescale equals the two-body relaxation time as a function of the mass of the object orbiting around the SMBH. We considered a Milky Way-sized nucleus here, and curves correspond to two eccentricities: $e=0$ and $0.9$. The solid ($e =0$) and dashed-dotted ($e=0.9$) curves correspond a fiducial multi-mass BH population with a PMF $dN/dm_{\rm BH} \propto m_{\rm BH}^{-2.35}$, $m_\mathrm{BH,max} = 30 \, \Msun$, and  $p_0=0.5$. The two symbols correspond to $r^{\rm rlx,GW}$ for a single mass BH population with $10 \, \Msun$ (star for $e_0=0$, cross for $e_0 = 0.9$). The dashed ($e=0$) and dotted ($e=0.9$) curves correspond a multi-mass BH population with the  \citet{Belczynskietal2014_2} PMF and $p_0 = 0.5$.} 
\end{figure}

 To account simply for these effects, we calculate the $r_{\rm rlx,GW}$ radius for each BH mass $m_{\rm BH}$ at which Criterion \ref{i:critB} is satisfied in two ways. We set the eccentricity in Equation (\ref{eq:dadt}) either to zero or to $e=0.9$. These results set upper and lower bounds on $r_{\rm rlx,GW}$ for at least $81\%$ of the population. However, because resonant relaxation is ignored in this calculation, the result for $r_{\rm rlx,GW}$ using this estimate is larger than its true value for eccentric orbits. We refer the reader to \citet{BarOrAlexander2016} for a detailed analysis. For a Milky Way-size nucleus (Figure \ref{fig:trlx_tGW}), we find that, for $e=0$ ($e=0.9$), $t_{\rm rlx} \leqslant t_{\rm GW}$ is satisfied at $r \gtrsim 260 \, M_{\rm SgrA*}$ ($r \gtrsim 1980 \, M_{\rm SgrA*}$) for $m_{\rm BH} = 5 \, \Msun$ and $r \gtrsim 460 \, M_{\rm SgrA*}$ ($r \gtrsim 3220 \, M_{\rm SgrA*}$) for $m_{\rm BH}=30 \, \Msun$, respectively. For EBBHs with fixed component masses, we find that $r_{\rm min}^{A,B}$ is approximately $2 \times$ lower for the \citet{Belczynskietal2014_2} PMF than for the same BH mass range  with a power-law PMF over the considered ranges of $M_{\rm SMBH}$, $p_0$, $m_\mathrm{BH}$, and $\beta$. The boundary varies monotonically with the mass of the objects and $e$. In the following, we adopt the cutoff radii
\begin{equation}  \label{eq:rmin}
  r_{\rm min} (m_{\rm BH}) = r^{\rm rlx,GW}_{\rm circ} (m_{\rm BH}) \, ,
\end{equation}
 which corresponds to circular orbits. We assume that, at \mbox{$r>r_{\rm min}$}, the objects that are removed by the SMBH due to GW emission are also continuously replenished from the outside, due to relaxation to maintain a steady-state density. The innermost radius of the GN at which BHs with masses $m_A$ and $m_B$ can form EBBHs is defined as
\begin{equation}  \label{eq_rminAB}
  r_{\rm min}^{A,B} = \mathrm{max}(r_{\rm min}(m_A),r_{\rm min}(m_B)) \, .
\end{equation}
 We note that this minimum radius is weakly sensitive to the parameters of both multi-mass BH populations.

 The results for $r_{\min}$ are robust and only weakly sensitive to the assumptions on the mass distribution of stellar populations. For instance, by considering a multi-mass MS population with a power-law PMF $m_{\rm MS}^{-2.35}$, a highest possible stellar mass of $10 \, \Msun$\footnote{This assumption is consistent with most of the supersolar-mass stars in the Galactic center with lifetimes higher than $10^7$ years \citep{Bartkoetal2010}, and we ignore the dynamical contributions of shorter-lived massive stars. The results are not sensitive to this assumption.} and radial density exponent $\alpha_{\rm MS} = 1.4$, we find that $r_{\min}$ and thus $r_{\rm min}^{A,B}$ slightly ($\leqslant 15 \%$) differs from that calculated for a single-mass MS population. This can be explained as follows: $r_{\min}$ can be given by changing the variable $a$ to $r_{\min}$ in Equation (\ref{eq:dadt}), setting this equation equal to Equation (\ref{eq:trlx}), substituting $e$ by zero, and solving the obtained equation for $r_{\min}$.  Thus, we find that $r_{\min} \propto  [ n_{\rm tot}(r) \langle M^2 \rangle (r) ]^{-2/11}$. Here, $n_{\rm tot}(r)$ is dominated by the MS population in the GN \citep{HopmanAlexander2006b,AlexanderHopman2009}, therefore parameters of the WD, NS, and BH populations could affect $r_{\min}$ only through $\langle M^2 \rangle (r)$. Because BHs are more  massive than MSs (or WDs and NSs), $\langle M^2 \rangle (r)$ is dominated by the BH population. Because $r_{\min}  \propto \left( \langle M^2 \rangle (r) \right)^{-2/11}$, the parametrization of the BH population has a minor effect on $r_{\min}$ and $r_{\rm min}^{A,B}$. As a consequence, the region inside the GN, where BHs are expected to form EBBHs, is weakly sensitive to the mass distribution of stellar populations. Note, however, that results for the BH population significantly depend on $m_{\rm BH,max}$, due to Equation (\ref{eq:formrateBHBH_multi}).

 From numerical investigations, we find that $r_{\rm min}$ is best-fit by a power-law relationship $r_{\rm min} \propto M_{\rm SMBH}^{B_{\rm SMBH}}$, where $B_{\rm SMBH} \approx 0.816 - 0.832$ for a multi-mass BH population with $dN/dm_{\rm BH} \propto m_{\rm BH}^{-\beta}$, depending on $m_{\rm BH}$, $\beta$, $p_0$, and $m_{\rm BH, max}$. Moreover, $B_{\rm SMBH} \approx 0.83$ for a single--mass BH population, and $B_{\rm SMBH} \approx 0.826$ for a multi-mass BH population with the \citet{Belczynskietal2014_2} mass function.

\section{ Analytic estimates of merger rate distributions}
\label{sec:AppA}

 In this appendix, we analytically derive the EBBH merger rate distributions in a single GN without resorting to Monte Carlo techniques. Generally, the scattering rates of objects of two masses $m_A$ and $m_B$ are proportional to $n_A n_B \sigma w$, where $n_A$ and $n_B$ are the corresponding number densities, and $\sigma$ is the cross section to form an EBBH in a close encounter. The latter depends on the magnitude of relative velocity $w = |\mathbf{v}_A - \mathbf{v}_B|$ and impact parameter $b$. We organize the derivation as follows. First, we define the phase-space distribution function, assuming an isotropic distribution, in Appendix \ref{App:AppGN}. In Appendix \ref{App:HandRelvel}, we derive the relative velocity distribution. In Appendix \ref{PDGWCO}, we calculate the radial distribution of GW capture and collision rates. In Appendices \ref{TWORPEPS}--\ref{ED}, we derive the EBBH merger rate distribution as a function of initial binary parameters, including the two-dimensional merger rate distribution in the $\epsilon_0$--$r_\mathrm{p0}$ plane, and the marginalized 1D merger rate distribution as a function of $r_\mathrm{p0}$ and $\epsilon_0$, respectively. In Appendix \ref{sec:MMDGN}, we determine the total-mass- and mass-ratio-dependent merger rate distributions, and the merger rate distribution of other component-mass-dependent variables (e.g. chirp mass, mass ratio, total mass, reduced mass, and symmetric mass ratio).

\subsection{Relative Velocities in Galactic Nuclei}
\label{App:AppGN}

 In this paper, we consider an approximately spherical (isotropic) distribution of compact objects around an SMBH. The equilibrium one-body phase space distribution of objects can be given by the Bahcall-Wolf model \citep{BahcallWolf1976,BahcallWolf1977,Keshetetal2009} as
\begin{equation}  \label{e:app:f(r,v)}
  f( \mathbf{r}, \mathbf{v}, m) = C(m) E(r,v)^{p(m)} \, ,
\end{equation}
 where the $C(m)$ normalization constant and $p(m)$ exponent depend on the mass of the component as shown in Equation (\ref{eq:pm}), and $E$ is the binding energy per unit mass:
\begin{equation}  \label{e:app:E(r,v)}
     E(r,v) = \frac{\Msmbh}{r} - \frac{v^2}{2} \, .
\end{equation}
 Equation (\ref{e:app:f(r,v)}) is valid outside the ``loss cone'' \citep{ShapiroLightman1976,SyerUlmer1999}, where objects are not removed by the SMBH, and interior to the radius of influence of the SMBH. We approximate this region with \mbox{$r_{\rm min} \leqslant r \leqslant r_\mathrm{max}$} and \mbox{$0 \leqslant v \leqslant \vmax(r)$}, where $r_{\rm min}$ is defined to be the inner radius of the GN at which the density cusp of BHs exhibit a cutoff (Appendix \ref{sec:CalRmin}), $r_\mathrm{max}$ is the radius of influence (Section \ref{subsec:GNModels}), and $\vmax(r)$ is the local escape velocity at radius $r$. The $C(m)$ normalization constant in Equation (\ref{e:app:f(r,v)}) can be given by the total number of BHs at fixed mass $m$:
\begin{align}  \label{APc3}
  \frac{dN}{dm} =  \int_{r_{\rm min}(m)}^{r_\mathrm{max}} dr \, 4\pi r^2 \int_{0}^{v_\mathrm{
  max}(r)} dv \, 4\pi v^2 f( \mathbf{r}, \mathbf{v},m) \, .
\end{align}

 Let us now change the integration variable from $v$ to \mbox{$\vbar=v/v_{\max}(r)$}, where $v_{ \max}(r)$ is given by Equation (\ref{eq:vmax}). From Equations (\ref{e:app:f(r,v)})--(\ref{APc3}), we find that the corresponding phase space distribution function is
\begin{equation}  \label{e:f(r,vbar)}
    f( r, \vbar, m) = C(m)\, n(r,m)\, \varphi(\vbar,m) \, ,
\end{equation}
 where
\begin{equation} \label{e:nphi}
 n(r,m) \equiv r^{-\alpha(m)}  \quad{\rm and}\quad \varphi(\vbar,m) \equiv (1-\vbar^2)^{p(m)} \, ,
\end{equation}
 if $r_{\rm min} \leqslant r \leqslant r_\mathrm{max}$, and if $0 \leqslant \vbar \leqslant 1$. The $\alpha_m$ exponent in Equation (\ref{e:nphi}) satisfies
\begin{equation} \label{eq:alpha_m}
 \alpha(m) = \frac{3}{2} + p(m) = \frac32 + p_0 \frac{ m }{ m_\mathrm{BH,max} } \, ,
\end{equation}
 where we have used Equation (\ref{eq:pm}) and we set $p_0=0.5$. Thus, $C(m)$ can be expressed in Equation (\ref{e:f(r,vbar)}) as
\begin{equation}  \label{eq:Cfact}
 C(m)  = N_{\rm BH} \mathcal{F}(m) C_{r}(m) C_{\vbar}(m) \, ,
\end{equation}
 where $N_{\rm BH}$ is the total number of BHs in the GN (Equation \ref{eq:encBH}), $\mathcal{F}(m)$ is the BH mass function with unit integral (Equation \ref{e:f(m)}), $C_{r}(m)$ and $C_{\vbar}(m)$ are normalization constants defined such that $C_{r}(m) n(r,m)$, and $C_{\vbar}(m) \phi(\vbar, m)$ have unit integrals over $r$ and $\vbar$, respectively, for fixed $m$ (Equation \ref{e:f(r,vbar)}),
\begin{align}
  \label{eq:Cr}  C_{r}(m) &= \frac{3-\alpha(m)}{4\pi}\frac{ 1 }{r_\mathrm{max}^{3-\alpha(m)} -
  r_\mathrm{min}^{3-\alpha(m)}(m) }\,,
  \\
  \label{eq:Cp}  C_{\vbar}(m) &= \frac{4^{\alpha(m)-1}\alpha(m) [2\alpha(m)-1] \,
  \Gamma^2[\alpha(m)]}{ \pi^{2} \, \Gamma[2 \, \alpha(m)]} \, .
\end{align}
 Here, $\Gamma$ is the Euler Gamma function. We next utilize the distribution function in Equation (\ref{e:f(r,vbar)}).

\subsection{Relative Velocity Distribution }
\label{App:HandRelvel}

 Here, we derive the distribution function of the magnitude of relative velocity $w$ of two types of objects $A$ and $B$ that have with fixed mass $m_A$ and $m_B$, respectively. Let us denote the \mbox{three-dimensional} individual velocity distributions by $f_A(\mathbf{v}_A)$ and $f_B (\mathbf{v}_B)$, meaning that the number of objects in a volume element $d^3 v_{i}$ around $\mathbf{v}_i$ is $dN_i = f(\mathbf{v}_i) \, d^3 v_i$ for $i=A$ or $B$. The two-body velocity space is six-dimensional. We assume that both velocity vectors have isotropic distributions. The distribution of the magnitude of relative velocity is defined as
\begin{align}  \label{APb3}
    F_{AB}(w) = \int d^3 v_A \, f_A(\mathbf{v}_A) \int d^3 v_B \, f_B(\mathbf{v}_B)
    \delta (w-|\mathbf{v_A}-\mathbf{v_B}|) \, ,
\end{align}
 where $\delta(\cdot)$ is the Dirac-$\delta$ function. We simplify this equation in the following.

 Let us adopt an arbitrarily oriented spherical coordinate system for the $\mathbf{v}_A$ integral; for the  $\mathbf{v}_B$ integral, let us measure angles relative to $\mathbf{v}_A$. That is, we adopt $(\theta_B, \phi_B)$ spherical coordinates in which the $\theta_B = 0$ axis is oriented along $\mathbf{v}_A$, while $(\theta_B, \phi_B)$ defines $\mathbf{v}_B$. The $\theta_A$, $\phi_A$, and $\phi_B$ integrals in Equation (\ref{APb3}) can be independently evaluated because the integrand does not depend on these variables. We carry out the $\theta_B$ integral by changing to the variable $z = \cos  \left( \theta_B \right)$ as
\begin{align}
  F_{AB}(w) &= 16\pi^2 \int_0^{\infty} d v_A \, v_A^2 f_B(\mathbf{v}_A)
  \int_0^{\infty} d v_B \, v_B^2 f_A(\mathbf{v}_B)
\nonumber\\
  &\times \frac{1}{2}\int _{-1} ^{1} dz \, \delta ( w-\sqrt{ v^2_A +v^2_B -2
  v_A  v_B  z  } ) \, .  \label{e:FAB0}
\end{align}
 We denote the last line with $\Phi$, and evaluate it using the identities of the Dirac-$\delta$ function
\begin{align}   \label{APb5}
   \int_I f(x) \delta(x) \, dx & =
  \left\{ \begin {array}{ll}
   f(0)   &  \textrm{ if }\, 0 \in I   \\
   0   &  \textrm{ otherwise }   \\
   \end {array} \right. \nonumber\\
&= \int _J  x'(z) f[x(z)] \delta [x(z)] \, dz    \nonumber\\
    &= \int_J g(z) \delta[x(z)]  \, dz \nonumber\\ &=
  \left\{ \begin {array}{ll}
   g(z_0)/x'(z_0)    &  \textrm{ if }\, z_0 \in J  \\
   0   &  \textrm{ otherwise }   \\
   \end {array} \right.
\end{align}
 where $x(z)$ is an arbitrary monotonic differentiable function, $g(z) = x'(z) f[x(z)] $, where $x'(z)=dx/dz$, $z_0$ and $J$ are defined by $x(z_0)=0$ and $x[J] = I$. For $g(z)=1$ and \mbox{$x(z)=w-\sqrt{v_A^2+v_B^2-2v_A v_B z}$}, we get
\begin{align}
     \label{APb6}
\Phi &= \frac{ w }{ v_A v_B } \times
    \left\{
  \begin {array}{ll}
   1    &  \textrm{ if $|\mathbf{v}_A-\mathbf{v}_B| \leq w \leq |\mathbf{v}_A + \mathbf{v}_B| $} \, .  \\
   0   &  \textrm{ otherwise } \, .  \\
   \end {array} \right.
\end{align}
 Substituting in Equation (\ref{e:FAB0}), the support of the integral is the region for which $|\mathbf{v}_A - \mathbf{v}_B| \leqslant w \leqslant |\mathbf{v}_A + \mathbf{v}_B|$, $0 \leqslant v_A \leqslant v_A^\mathrm{max}$ and $0 \leqslant v_B \leqslant v_B^\mathrm{max}$; thus we find
\begin{align}
  F_{AB}(w) &= 8 \pi^2 w \int_0 ^{ \min( v_A^\mathrm{max} , v_B ^\mathrm{max} +w)}
  dv_A \, v_A f_A(v_A)  \nonumber\\
  &\quad\times \int_{ \min(1, |v_A-w|) }^{\min( v_B^\mathrm{max} ,
  v_A^\mathrm{max} +w)} dv_B \, v_B f_B(v_B) \, .
   \label{APb7}
\end{align}
 Now we consider the case where $f_A(\cdot)=f_B(\cdot)$. Let us switch to the dimensionless velocity variables $\vbar_i = v_i/v_\mathrm{max}(r)$ and the dimensionless relative velocity $\wbar = w/v_\mathrm{max}(r)$ ($0 \leqslant \wbar \leqslant 2$), where $v_\mathrm{max}(r)$ is the local escape velocity at radius $r$ given by Equation~(\ref{eq:vmax}). The distribution of $\wbar$, $F_{AB}(\wbar)$, can be obtained by the equation $ F_{AB}(w) d w = F_{AB}(\wbar) d \wbar$, which gives
\begin{align}\label{e:FAB}
    F_{AB}(\wbar) = 8 \pi^2 \wbar \int\limits_{0}^{1} d\vbar_A \, \vbar_A f(\vbar_A)
\int\limits_{ \min(1, |\vbar_A-\wbar|) }^{ \min(1, \vbar_A+\wbar)} d\vbar_B \, \vbar_B f(\vbar_B) \, .
\end{align}

 For the phase space distribution function of relaxed GNs, the integral over $d\vbar_B$ can be evaluated  analytically, which yields
\begin{align}  \label{e:FAB2}
  F_{AB}(\wbar) &= \frac{4\pi^2 \wbar}{p_B+1}
 \nonumber \\
 & \times \int_{\wbar-1}^1 d \vbar_A \, \vbar_A \left(1-\vbar_A^2\right)^{p_A}
 \left[1-(\vbar_A-\wbar)^2\right]^{p_B+1} \, .
 \end{align}
 The remaining integral over $\vbar_A$ can also be evaluated analytically for any integer or half-integer $p_A$ and $p_B$. In particular, for $p_A=p_B=0$, $\alpha_A = \alpha_B = 3/2$, $F_{AB}(\wbar) = \frac13\pi^2\wbar^2 (2-\wbar)^{2}(\wbar+4)$. Note that $\alpha=7/4$ for the equilibrium distribution for single-mass clusters \citep{BahcallWolf1976} and $1.5 < \alpha < 3$ for multi-mass mass-segregated models \citep{Keshetetal2009}.

\subsection{GW Capture and Collision Rates for Fixed Component Masses}
\label{PDGWCO}

 We consider the merger rates between two types of objects $A$ and $B$ with masses $m_A$ and $m_B$, drawn from phase space distribution given by Equation (\ref{e:app:f(r,v)}). The differential rate of merger among two types of objects $A$ with phase space distributions $f_A(\mathbf{r}_A, \mathbf{v}_A)$ at location $(\bm{r}_A,\bm{v}_A)$ in a phase space volume $d^3r_Ad^3v_A$, which we find similarly for objects $B$, is
\begin{equation}  \label{APd1}
 d^{12} \Gamma _{A B} = \sigma w f_A(\mathbf{r}_A, \mathbf{v}_A) f_B(\mathbf{r}_B,
 \mathbf{v}_B) \, d^3 r_A \, d^3 r_B \, d^3 v_A \, d^3 v_B \, ,
\end{equation}
 where $\sigma\equiv \sigma(w)$ is the GW capture cross section. We consider the formation of EBBHs in the radius range of $[r_{\rm min}^{A,B}, r_{\rm max}]$.

 Next, we examine the case of GW capture and direct collision events. Here, ``GW capture'' refers to events where initially unbound black holes (i.e. having asymptotically nonzero kinetic energy at infinite separation) are captured into a bounded system due to GW emission or if they suffer a direct collision. ``Direct collision'' refers to the case in which the impact parameter is sufficiently small such that the BHs collide directly in less than one orbit. In order to be captured into an eccentric binary with multiple orbits before merger, the impact parameter must be in the range $b_\mathrm{min}(w) < b < b_{\max}(w)$ given by Equation (\ref{eq:impactlim}). The cross section to form such a binary may be written as $\sigma(w) = \pi b_\mathrm{max}^2(w)-\pi b_\mathrm{min}^2(w)$. From Equation (\ref{eq:impactlim}), we find that the encounter rate for this event can be expressed as
 \begin{equation}  \label{e:sigmaw}
 \sigma w = \left[ \zeta_{\mathrm{capt}}(r) \,
 \xi_{\mathrm{capt}}(\wbar) - \zeta_{\mathrm{coll}}(r) \, \xi_{\mathrm{coll}} (\wbar) \right]
 \, \delta(\bm{r}_A-\bm{r}) \, \delta(\bm{r}_B-\bm{r}) \, ,
 \end{equation}
 where the Dirac $\delta$ functions ensure that bounded systems can form only when $\mathbf{r}_A \approx \mathbf{r}_B=\bm{r}$ for any $\bm{r}$ (``short-range encounters''). We also introduced
\begin{align}
 \zeta_{\mathrm{capt}}(r) &= \pi c_\eta^{2/7} M_\mathrm{tot}^2 v_\mathrm{max}^{-11/7}(r)\,,
 \quad \xi_{\mathrm{capt}}(\wbar) = \wbar^{-11/7}\,,  \label{eq:captzeta} \\
 \zeta_{\mathrm{coll}}(r) &=16 \pi  M_\mathrm{tot}^2 v_\mathrm{max}^{-1}(r)\,,
 \quad \xi_{\mathrm{coll}}(\wbar) = \wbar^{-1}\,, \label{eq:collzeta}
\end{align}
 and where
\begin{equation}\label{e:C0}
 c_\eta = \frac{340 \pi}{3} \eta \, .
\end{equation}

 The rate of eccentric inspirals is the difference between the capture and direct collision rates. In the following, we separately evaluate the merger rates of capture sources and direct collisions using the corresponding $\zeta$ and $\xi$ functions in Equations (\ref{eq:captzeta}) and (\ref{eq:collzeta}). Substituting into Equations (\ref{APd1}), the differential rate of mergers in a spherical shell of volume $d^3 r_{A,B} = 4 \pi r^2 dr$ with $r = |\mathbf{r}_A| \approx|\mathbf{r}_B|$ with velocities is $\bm{v}_A$ and $\bm{v}_B$ is
\begin{equation}  \label{eq:app:infformrdep}
  d^7 \Gamma_{AB} = 4 \pi r^2 \zeta(r) \xi(\wbar) f_A(r,\mathbf{v}_A) f_B(r,\mathbf{v}_B) \,
  d r \, d^3 v_A\, d^3 v_B \, .
\end{equation}

 The GW capture process depends on $\bm{v}_A$ and $\bm{v}_B$ only through the combination $w = |\bm{v}_A-\bm{v}_B|$, allowing us to make further simplifications. The merger rate distribution as a function of $w$ may be derived by multiplying $d^7 \Gamma_{AB}$ by $\delta (w-|\mathbf{v_A}-\mathbf{v_B}|)$ and integrating over $d^3v_A$ and $d^3 v_B$. Substituting into Equation (\ref{e:f(r,vbar)}), we get
\begin{align}   \label{APc8}
 \left\langle \frac{\partial^4 \Gamma }{ \partial r \partial w \partial m_A \partial m_B }
 \right\rangle_{\mathbf{v}_A,
 \mathbf{v}_B} & = 4 \pi r^2 \, \zeta(r) \, \xi(\wbar) \, n_A(r) \, n_B(r) \, C_A \, C_B
 \nonumber \\
 & \times \int d^3 v_A \, \varphi_A(\vbar_A) \int d^3 v_B \, \varphi_B(\vbar_B)
 \nonumber \\
 & \times \delta(w - | \mathbf{v}_A-\mathbf{v}_B|) \, .
\end{align}
 Using the results of Appendices \ref{App:HandRelvel}, this is
\begin{align}  \label{APc10}
 \left\langle \frac{\partial^4 \Gamma }{\partial r \partial \wbar \partial m_A \partial m_B}
 \right\rangle_{\mathbf{v}_A, \mathbf{v}_B} & = 4 \pi r^2 \, \zeta(r) \, \xi(\wbar) \,
 \nonumber \\
 & \times C_A \, C_B \, n_A(r) \, n_B(r) \, F_{AB}(\wbar) \, ,
\end{align}
 where $F_{AB}(\wbar)$ is the dimensionless relative velocity distribution given by Equation (\ref{e:FAB2}).

 The total merger rate within a spherical shell of radius $r$ for any relative velocity can be obtained for both capture and collision events as
\begin{align}  \label{e:Gamma_ABrfixed}
 \left\langle \frac{ \partial^3 \Gamma }{\partial r \partial m_A \partial m_B } \right\rangle
 _{\mathbf{v}_A, \mathbf{v}_B} & = 4 \pi r^2 \zeta(r) \,  C_A C_B n_A(r) \, n_B(r)
 \nonumber \\
 & \times  \int_0^{2} d\wbar \, \xi(\wbar) F_{AB}(\wbar) \, ,
\end{align}
 and the total merger rate for any encounter position is
\begin{align}  \label{e:Gamma_ABresult}
 \left\langle \frac{\partial^2 \Gamma }{ \partial m_A \partial m_B } \right\rangle & =
 4 \pi C_A C_B\int_{r_\mathrm{min}^{A,B}}^{r_\mathrm{max}} dr \, r^2 n_A(r) n_B(r) \zeta(r)
 \nonumber \\
 & \times \int_0^{2} d \wbar\, \xi(\wbar) F_{AB}(\wbar) \, .
\end{align}

 Although this result is already simple to evaluate numerically for the relative velocity distribution function in Equation (\ref{e:FAB2}) for capture and direct collision sources $\zeta$ and $\xi$ given by Equations (\ref{eq:captzeta}) and (\ref{eq:collzeta}), we find that the result is not sensitive to the relative velocity distribution, in the following sense. Following \citet{OLearyetal2009}, we may approximate the integral over the relative velocity distribution with the value for circular orbits, $\wbar = v_{\rm circ}(r)/ v_{\rm max}(r) = 2^{-1/2}$, to get
\begin{equation} \label{APd5}
 C_{\vbar A} \, C_{\vbar B} \int_0^{2} d\wbar \, \xi(\wbar) F_{AB}(\wbar) \approx
 \xi(2^{-1/2})  \, .
\end{equation}
 Here, $C_{\vbar A,B}$ are given by Equation (\ref{eq:Cp}). We find that this approximation is valid up to an error less than $18 \%$ for a wide range of BH masses between $5 \, \Msun$ and $80\,\Msun$.

 Thus, Equations (\ref{e:Gamma_ABrfixed}) and (\ref{e:Gamma_ABresult}) can be approximated for GW capture and collision events as
\begin{align}  \label{APd71}
  \left\langle \frac{ \partial^3 \Gamma }{\partial r \partial m_A \partial m_B } \right\rangle
  \approx 4 \pi r^2 N_{A}N_{B}
 C_{rA} C_{rB}  n_A(r) n_B(r) \zeta(r)\xi(2^{-1/2}) \, ,
\end{align}
 and
\begin{align}  \label{APd7}
  \left\langle \frac{\partial^2 \Gamma }{ \partial m_A \partial m_B } \right\rangle & \approx
  N_A N_B \, C_{rA} \, C_{rB} \, \xi(2^{-1/2}) \,
 \nonumber \\
 & \times \int_{r_\mathrm{min}^{A,B}}^{r_\mathrm{max}} dr \, 4 \pi r^2 n_A(r)
 n_B(r)  \zeta(r) \, ,
\end{align}
 where $r_\mathrm{min}^{A,B} = \max(r_\mathrm{min}^{A},r_\mathrm{min}^{B})$, $C_{rA}$ and $C_{rB}$ are given by Equation (\ref{eq:Cr}), and $\zeta(r)$ and $\xi(\wbar)$ are given by Equations (\ref{eq:captzeta}) and (\ref{eq:collzeta}).

\subsection{Merger Rate Distribution in the $\epsilon_0 - \rho_\mathrm{p0}$ Plane for Fixed Component Masses}
\label{TWORPEPS}

 To derive the merger rate distribution as a function of the initial orbital elements, we first calculate the merger rate distribution as a function of $b$, $w$, and $r$. The differential cross section for a fixed $r$, $w$, and impact parameter between $b$ and $b + db$ is $d \sigma = 2 \pi b \, db$. Using Equation (\ref{e:Gamma_ABrfixed}), we get
\begin{align}  \label{APe1}
 \left\langle \frac{\partial^5 \Gamma }{\partial r \partial b \partial w \partial m_A
 \partial m_B} \, dr \, db \, dw  \right\rangle & = 4 \pi  r^2 C_A  C_B \, n_A(r) \, n_B(r)\,
 \nonumber
 \\&
 \times 2 \pi b w P_{AB}(w| r) \, dr  \,db  \, dw \,
\end{align}
 if $b_\mathrm{min}(w) < b < b_\mathrm{max}(w)$, and zero otherwise. Here, we have introduced the conditional PDF of $w$ at radius $r$,
\begin{equation}  \label{e:P(w|r)}
 P_{AB}( w | r) = \frac{F_{AB}[w/v_\mathrm{max}(r)]}{ v_\mathrm{max}(r)} \, ,
\end{equation}
 where $F_{AB}[w/v_\mathrm{max}(r)]$ is given by Equation (\ref{e:FAB2}).

 Next, we change variables in Equation~(\ref{APe1}) from $(b,w)$ to $(r_\mathrm{p0}, \epsilon_0)$ using $ db\, dw = J \, dr_\mathrm{p0}\, d\epsilon_0$, where $J = |\partial(b, w)/ \partial(r_\mathrm{p0}, \epsilon_0)|$ is the Jacobian. From Equations (\ref{eq:dEGW}), (\ref{eq:Efin}), (\ref{eq:42}), and (\ref{eq:46}), we find that, to leading order,
\begin{align} \label{e:eps(b,w)}
  r_\mathrm{p0}(b,w) = \frac{ b^2 w^2 }{ 2 M_\mathrm{tot} } \, ,\quad
\epsilon_0 (b,w) = \frac{ 340 \pi\, \eta M_{\rm tot}^5 }{ 3\, b^5 w^5 } -
  \frac{ b^2  w^4}{ M_{\rm tot}^2 } \, .
\end{align}
 To understand how the boundaries of the regions transform, we note that $b > b_\mathrm{min}(w)$ corresponds to $r_\mathrm{p0}>r_{\mathrm{p0},\min} = 8 M_\mathrm{tot}$, and $b<b_\mathrm{max}( w )$ corresponds to $\epsilon_0>0$; see Section \ref{sec:PIOP} for both conditions. Substituting in Equation (\ref{APe1}), the merger rate distribution as a function of $r$, $r_{p0}$, and $\epsilon_0$ is
\begin{align}    \label{e:Gamma-r-rp-eps}
  \left\langle \frac{\partial^5 \Gamma }{\partial r \partial r_\mathrm{p0} \partial \epsilon_0
  \partial m_A \partial m_B} \right\rangle & =  2 \pi^2 M_\mathrm{tot}^2 C_A C_B n_A(r) n_B(r)
  \nonumber \\
  & \times \frac{ r^2 F_{AB}( \wbar) }{ r_{\mathrm{p0}} \wbar^2 v_\mathrm{max}^3(r) }
\end{align}
 if $r_\mathrm{p0} > 8 M_\mathrm{tot}$ and $\epsilon_0 > 0$ , and the merger rate is zero otherwise. Here, \mbox{$\wbar \equiv w(r_\mathrm{p0},\epsilon_0) /v_\mathrm{max}(r)$} (see Equation \ref{eq:w_eps_rhop0}), $v_{\max}(r)$ is given by Equation (\ref{eq:vmax}), $C_{A}$ and $C_{B}$ are given by Equations (\ref{eq:Cfact})--(\ref{eq:Cp}), $n_{A}(r)$ and $n_{B}(r)$ are given by Equations (\ref{e:nphi}) and (\ref{eq:alpha_m}), and $F_{AB}(\wbar)$ is given by Equation (\ref{e:FAB2}).

 The distribution in terms of the dimensionless pericenter distance $\rho_{\mathrm{p}0} = r_{\mathrm{p}0}/M_{\mathrm{tot}}$ is
\begin{align}    \label{e:Gamma-r-rhop-eps}
  \left\langle \frac{\partial^5 \Gamma }{\partial r \partial \rho_\mathrm{p0} \partial \epsilon_0
  \partial m_A \partial m_B} \right\rangle =  M_{\rm tot} \left\langle \frac{\partial^5
  \Gamma }{\partial r \partial r_\mathrm{p0} \partial \epsilon_0 \partial m_A \partial m_B}
  \right\rangle \,.
\end{align}
 The merger rate distribution in the $\epsilon_0$--$\rho_{\mathrm{p}0}$ plane can be given by integrating Equation (\ref{e:Gamma-r-rp-eps}) over $r$:
\begin{equation}  \label{APe5}
  \left\langle \frac{\partial^4 \Gamma}{\partial \rho_\mathrm{p0} \partial \epsilon_0
  \partial m_A \partial m_B } \right \rangle = \int_{r_\mathrm{min}^{A,B}}^{r_\mathrm{max}} dr \,
  \left\langle \frac{\partial^5 \Gamma}{\partial r \partial \rho_\mathrm{p0} \partial \epsilon_0
  \partial m_A \partial m_B}  \right\rangle \, .
\end{equation}

\subsection{Merger Rate Distribution as a Function of $\rho_\mathrm{p0}$ for Fixed Component Masses}
 \label{PERICDIST}

 The merger rate distribution as a function of $r_\mathrm{p0}$  can be obtained by integrating Equation (\ref{e:Gamma-r-rp-eps}) over $r$ and $\epsilon_0$. An algebraically simpler result can be derived directly from Equation (\ref{APe1}) by changing variables from $(b,w)$ to $(r_ \mathrm{p0},w)$ and integrating over $r$ and $w$. In this way, we avoid the variable change to $\epsilon_0$. Equation (\ref{eq:42}) implies that \mbox{$dr_\mathrm{p0} \, dw = w^2 b \, db \, dw / M_\mathrm{tot}$}. Thus,
\begin{align}  \label{e:Gamma-r-rp-w}
 \frac{\partial^5 \Gamma}{\partial r \partial r_\mathrm{p0} \partial w \partial m_A \partial m_B}
 \,dr \, dr_\mathrm{p} dw &= 8 \pi^2 C_A C_B n_A(r) n_B(r) r^2 \,dr\,
 \nonumber \\
 & \times \frac{ dr_\mathrm{p0}}{w }  M_\mathrm{tot}  P_{AB}( w | r) \, dw
\end{align}
 if $b < b_\mathrm{max}$ , where $P_{AB}( w | r)$ is given by Equations (\ref{e:FAB2}) and (\ref{e:P(w|r)}), and $\partial^5 \Gamma /\partial r \partial r_\mathrm{p0} \partial w \partial m_A \partial m_B = 0$ otherwise. The condition $b<b_\mathrm{max}$ may be equivalently written as
\begin{equation}  \label{e:rpmax}
 r_\mathrm{p0} < r_{\mathrm{p0},\max}(r,w) =
 \frac{b_\mathrm{max}^2(w) w^2}{2\,M_\mathrm{tot}} \, ,
\end{equation}
 where $b_{\max}(w)$ is given by Equation (\ref{eq:impactlim}), which gives
 \begin{equation}\label{eq:wbarmax}
 \wbar < \wbar_\mathrm{max} (r_\mathrm{p0},r) \equiv (2
 r_\mathrm{p0} )^{-7/4} c_\eta^{1/2} v_\mathrm{max}(r) ^{-1} M_\mathrm{tot}^{7/4} \, .
 \end{equation}
 Now let us substitute Equation (\ref{e:P(w|r)}), change the integration variable to $\wbar = w/v_\mathrm{max}(r)$, and integrate over the allowed regions:
\begin{align}  \label{e:Gamma-rp}
 \left\langle\frac{\partial^3 \Gamma}{\partial r_\mathrm{p0} \partial m_A \partial m_B }\right\rangle
 & = \int_{r_\mathrm{min}^{A^3,B}}^{r_\mathrm{max}} dr \, \frac{C_A C_B n_A(r) n_B(r)
 M_\mathrm{tot} 8 \pi^2  r^2 }{ v_\mathrm{max} (r) }
 \nonumber \\
 & \times \int_{ 0 }^{ \min[2, \wbar_\mathrm{max} (r_\mathrm{p0}, \, r)] } d \wbar \,
 \frac{F_{AB}(\wbar)}{\wbar} \, .
\end{align}
 Equation (\ref{e:Gamma-r-rp-w}) is independent of $r_\mathrm{p0}$, which shows that the merger rates are uniformly distributed in pericenter distance for a fixed $r$ and $w$. The merger rate distribution integrated over $r$ is uniform in $r_\mathrm{p0}$ for
\begin{align}
 r_\mathrm{p0} &\leq r_{\mathrm{p0},\rm uni}\equiv \frac{ 2 \, b_\mathrm{max}^2
 [2 \, v_\mathrm{max}  (r_\mathrm{min})]\, v_\mathrm{max}^2 \,
 (r_\mathrm{min}) }{ M_\mathrm{tot} }
 \nonumber\\ &= \left(\frac{85\pi}{24\sqrt{2}}\right)^{2/7} M_{\mathrm{tot}}
 \frac{\eta^{2/7}}{v_{\max}^{4/7}(r_{\min})}
 \, ,
\end{align}
 where $b_{\max}(\cdot)$ and $v_{\max}(\cdot)$ are given by Equations (\ref{eq:impactlim}) and (\ref{eq:vmax}), respectively. For larger pericenter distances, \mbox{$r_\mathrm{p0}>r_{\mathrm{p0}, \rm uni}$}, the rates $\left\langle \partial^3 \Gamma/ \partial r_\mathrm{p0} \partial m_A \partial m_B \right\rangle$ decrease continuously with $r_\mathrm{p0}$ because only relative velocities less than $\wbar_\mathrm{max}(r_\mathrm{p0},r)$ contribute to the integral over $\wbar$ in Equation (\ref{e:Gamma-rp}).

 The merger rate distribution as a function of the dimensionless pericenter distance is
\begin{equation}  \label{eq:dGammadrhop0}
  \left\langle \frac{ \partial^2 \Gamma }{\partial \rho_\mathrm{p0} \partial m_A \partial m_B} \right \rangle
  = M_{\rm tot} \left\langle \frac{ \partial^2 \Gamma }{\partial r_\mathrm{p0} \partial m_A \partial m_B}
  \right \rangle \, .
\end{equation}

\subsection{Merger Rate Distribution as a Function of $\epsilon_0$ for Fixed Component Masses}
\label{ED}

 Next, we derive the merger rate distribution as a function of $\epsilon_0$ at fixed component masses. To do so, we may integrate $\partial^5 \Gamma /\partial r\,\partial r_\mathrm{p0}\, \partial \epsilon_0 \, \partial m_A \, \partial m_B$ over $r_\mathrm{min}^{A,B} < r <r_\mathrm{max}$ and \mbox{$8M_\mathrm{tot} <r_\mathrm{p0}<r_{\mathrm{p0}_\mathrm{max}(r)}$}. Alternatively, we can change variables from $r_\mathrm{p0}$ to $\wbar$ using the relationship $r_\mathrm{p0}(\wbar, \epsilon_0)$, e.g. by inverting the relative velocity $\wbar(r_\mathrm{p0},\epsilon_0)$ in Equation (\ref{eq:w_eps_rhop0}) and then integrating $\partial^5 \Gamma /\partial r\,\partial \wbar\, \partial \epsilon_0 \, \partial m_A \, \partial m_B$ over $r$ and $\wbar$. The result is
\begin{align}   \label{e:Gamma-eps}
 \left\langle \frac{\partial^3 \Gamma}{\partial \epsilon_0 \partial m_A \partial m_B} \right\rangle
 & = \int_{r_\mathrm{min}^{A,B}}^{r_\mathrm{max}} \, dr \, \frac{ C_A  C_B  n_A(r) n_B(r)
 M_\mathrm{tot} 8 \pi^2  r^2 }{ v_\mathrm{max}(r) }
\nonumber \\
 & \times \int_0^{\min[2, \wbar_\mathrm{max} (r_\mathrm{p0}(\wbar,\epsilon_0), r)]}
 d \wbar \, \frac{F_{AB}(\wbar)}{\wbar}
 \nonumber \\
 & \times \left( \frac{ 5 }{ 2 } \frac{ \epsilon_0 }{ r_\mathrm{p0} (\wbar,\epsilon_0) }
 + \frac{7 \wbar^2 \, v_\mathrm{max} (r)^2 }{ M_\mathrm{tot} } \right)^{-1} \, ,
\end{align}
 where $\wbar_{\max}$ is given by Equation (\ref{eq:wbarmax}), $v_{\max}(r)$ is given by Equation (\ref{eq:vmax}), $C_{A}$ and $C_{B}$ are given by Equations (\ref{eq:Cfact})--(\ref{eq:Cp}), $n_{A}(r)$ and $n_{B}(r)$ are given by Equations (\ref{e:nphi}) and (\ref{eq:alpha_m}), and $F_{AB}(\wbar)$ is given by Equation (\ref{e:FAB2}).

\subsection{Merger Rate Distribution as a Function of Total Mass and Mass Ratio}
\label{sec:MMDGN}

 Up to now, we have considered the merger rate distributions for two components $A$ and $B$, which may have different numbers $N_A$ and $N_B$ and normalization constants $C_A$ and $C_B$; see Equation (\ref{eq:Cfact}). To derive the mass-dependent merger rate distributions, assuming that both components are drawn from a BH mass function $\mathcal{F}(m)$, we set $N_A = N_{\rm BH}\mathcal{F}(m_A) dm_A$ and did similarly for $N_B$, where $N_{\rm BH}$ is the total number of BHs in the GN. This defines the mass distribution of the event rates for the two components. Note that the mass dependence is also implicit in $\alpha( m_{A,B})$, which affects $C_A$, $C_B$, $n_A(r)$, $n_B(r)$, and $F_{AB}(r)$ in Equations (\ref{e:f(r,vbar)})--(\ref{eq:Cp}) and Equation (\ref{e:FAB2}). To spell out the mass dependence more explicitly, we rewrite Equation (\ref{e:Gamma-r-rp-eps}) as
\begin{align}  \label{e:Gamma-r_rp-eps-mA-mB} \left\langle \frac{\partial^5 \Gamma}{\partial r \partial r_\mathrm{p0} \partial \epsilon_0 \partial m_A \partial m_B} \right\rangle & = \frac{ 2 \pi^2  M_\mathrm{tot}^2 r^2 n(r,m_A) n(r,m_B)}{\wbar^2 v_\mathrm{max}^3(r) r_\mathrm{p0} }
 \nonumber \\
 &  \times C(m_A)C(m_B) F[ \wbar, p(m_A), p(m_B) ] \, ,
\end{align}
 where $\wbar=w(r_\mathrm{p0},\epsilon_0,m_A,m_B)/v_\mathrm{max}(r)$ is given by Equation (\ref{eq:w_eps_rhop0}), $C(m_{A,B})$ are given by Equation (\ref{eq:Cfact}) and $n(r, m_{A,B})$ are given by Equations (\ref{e:nphi})-(\ref{eq:Cp}), where \mbox{$F[ w, p(m_A), p(m_B) ] \equiv F_{AB}(\wbar)$} in Equation (\ref{e:FAB2}). This shows that the five-dimensional distribution in Equation (\ref{e:Gamma-r_rp-eps-mA-mB}) is generally non-separable, and mass segregation causes the different mass components to follow different radial and velocity profiles.

 Merger rate distributions with respect to the component masses follows directly from Equation (\ref{e:Gamma_ABresult}). Using the approximation in Equation (\ref{APd5}), we get from Equation (\ref{e:Gamma_ABresult}) that
\begin{align}  \label{APg2}
 \left\langle \frac{\partial^2\Gamma}{\partial m_A\partial m_B}\right\rangle
 & \approx  N_{\rm BH}^2 C_{rA} C_{rB} \mathcal{F}(m_A) \mathcal{F}(m_B)
 \nonumber \\
 & \times \int_{r_\mathrm{min}^{A,B}}^{r_\mathrm{max}} d r \, 4 \pi r^2 v_{\rm circ}(r)
 n_A(r) n_B(r)
\nonumber \\
 & \times \pi \left\{ b_{\rm max}^2[v_{\rm circ}(r)] - b_\mathrm{min}^2[v_{\rm circ}(r)]\right\}
\end{align}
 if $m_\mathrm{BH,min} \leqslant m_{A,B} \leqslant m_\mathrm{BH,max}$, and zero otherwise. Here, $b_{\max}(w)$ and $b_{\min}(w)$ are given by Equation (\ref{eq:impactlim}),  where we substitute \mbox{$w = v_{\mathrm{circ}}(r) = \sqrt{M_{\mathrm{SMBH}}/r}$}.

 For a power-law mass function with exponent $-\beta$ (see Equation \ref{e:f(m)}), we get
\begin{equation}  \label{eq:Ratemulti-mass}
 \left\langle \frac{\partial^2\Gamma}{\partial m_A\partial m_B} \right\rangle = \left\langle
 \frac{\partial^2\Gamma_{\rm capt}}{\partial m_A\partial m_B}\right\rangle - \left\langle
 \frac{\partial^2\Gamma_{\rm coll}}{\partial m_A\partial m_B}\right\rangle \,,
\end{equation}
 where
\begin{align}   \label{eq:56}
\left\langle \frac{\partial^2\Gamma_{\rm capt}}{\partial m_A \partial m_B}
\right\rangle & \approx  \frac{ 9 m_\mathrm{BH,max}^2 - 6 p_0 m_\mathrm{BH,max} M_\mathrm{tot}
 + 4 p_0^2 \mu M_\mathrm{tot} }{ m_\mathrm{BH,max}^2 \,
r_\mathrm{max} ^{ 3 - p_0 \frac{ M_\mathrm{tot} }{ m_\mathrm{BH,max} } } }
 \nonumber \\
 & \times \frac{ M_\mathrm{tot}^{2-\beta} \mu^{-\beta} (1-\beta)^2 }{  (m_\mathrm{BH,max}
 ^{1-\beta}
 - m_\mathrm{BH,min}^{1-\beta} )^2} \frac{N_{\rm BH}^2 c_\eta^{2/7}}{16 \,\Msmbh^{11/14}}
  \nonumber \\
  & \times \frac{ r_\mathrm{max}^{11/14 - p_0 \frac{ M_\mathrm{tot} }{ m_\mathrm{BH,max} }
  } - (r_\mathrm{min}^{A,B})^{ 11/14 - p_0 \frac{ M_\mathrm{tot}}{ m_\mathrm{BH,max} }
  } }{\frac{11}{14} - p_0 \frac{ M_\mathrm{tot} }{ m_\mathrm{BH,max} } }
\end{align}
 if $m_\mathrm{BH,min} \leqslant m_{A,B} \leqslant m_\mathrm{BH,max}$, and
 \begin{align}   \label{eq:561}
\left\langle \frac{\partial^2\Gamma_{\rm coll}}{\partial m_A\partial m_B} \right\rangle
 & \approx \frac{ 9 m_\mathrm{BH,max}^2 - 6 p_0 m_\mathrm{BH,max} M_\mathrm{tot} + 4 p_0^2 \mu
 M_\mathrm{tot}   }{ m_\mathrm{BH,max}^2 r_\mathrm{max}^{ 3 - p_0\frac{ M_\mathrm{tot}
 }{ m_\mathrm{BH,max}  } } }
 \nonumber \\
 & \times \frac{ M_\mathrm{tot}^{2-\beta} \mu^{-\beta} (1-\beta)^2 }{
 (m_\mathrm{BH,max}^{1-\beta}  -  m_\mathrm{BH,min}^{1-\beta} )^2 } \frac{N_{\rm BH}^2 }{
 \Msmbh^{1/2} }
 \nonumber \\
 & \times \frac{ r_\mathrm{max} ^{ 1/2 - p_0  \frac{ M_\mathrm{tot} }{ m_\mathrm{BH,max} } }
 - (r_\mathrm{min}^{A,B})^{ 1/2 - p_0 \frac{ M_\mathrm{tot} }{ m_\mathrm{BH,max} } } }{ \frac12
 - p_0 \frac{ M_\mathrm{tot} }{ m_\mathrm{BH,max} } }
\end{align}
 if $m_\mathrm{BH,min} \leqslant m_{A,B} \leqslant m_\mathrm{BH,max}$. Otherwise, the rates vanish. Here, $c_\eta$ depends on mass ratio as given by Equation (\ref{e:C0}).

 We find that $\left\langle \partial^2 \Gamma_{\rm coll} / \partial m_A \, \partial m_B \right\rangle$ has a small \mbox{($\la 10 \%$)} contribution to $\left\langle \partial^2 \Gamma / \partial m_A \, \partial m_B \right\rangle$. Using the expressions of $N_{\rm BH}$ defined by Equation (\ref{eq:encBH}), we find for that, both multi-mass BH populations\footnote{That is, for a multimass-BH population with a power-law PMF and the \citet{Belczynskietal2014_2} PMF.}, 
\begin{equation}  \label{eq:mergrate_Multi}
  \left\langle \frac{\partial^2\Gamma}{\partial m_A\partial m_B} \right\rangle \propto
  M_{\rm SMBH}^{3/28} \left[1 - \left(\frac{ r_{\rm min}^{A,B} }{ r_{\rm max} } \right)
  ^{\frac{11}{14} - p_0 \frac{ M_\mathrm{tot} }{ m_\mathrm{BH,max} }} \right] \, ,
\end{equation}
 and for a single-mass BH population,
\begin{equation}  \label{eq:mergrate_Single}
  \left\langle \frac{\partial^2\Gamma}{\partial m_A\partial m_B} \right\rangle \propto
  M_{\rm SMBH}^{3/28} \left(\frac{ r_{\rm min}^{A,B} }{ r_{\rm max} } \right)
  ^{- \frac{3}{14}}  \, .
\end{equation}
 Finally, using the expression of $r_{\rm max}$ defined by Equation (\ref{eq:rmax}), along with the conditions that $r_{\rm min}^{A,B} \ll r_{\rm max}$ and that $r_{\rm min}^{A,B} \propto M_{\rm SMBH}^{B_{\rm SMBH}} \approx M_{\rm SMBH}^{0.82}$ (see Appendix \ref{sec:CalRmin} for the BH population dependent
 values of $B_{\rm SMBH}$), implies that
\begin{equation}  \label{eq:mergrate_lowmass}
  \left\langle \frac{\partial^2\Gamma}{\partial m_A\partial m_B} \right\rangle \propto
  M_{\rm SMBH}^{3/28}
\end{equation}
 if $M_\mathrm{tot} \la 11 m_\mathrm{BH,max}/ 14 p_0$,
\begin{align}  \label{eq:mergrate_highmass}
  \left\langle \frac{\partial^2\Gamma}{\partial m_A\partial m_B} \right\rangle & \propto
  M_{\rm SMBH}^{\frac{3}{28} + \left(B_{\rm SMBH} - \sqrt{2}\right) \left( \frac{11}{14}
  - p_0 \frac{ M_\mathrm{tot} }{ m_\mathrm{BH,max} }\right)}
 \nonumber \\
  & \propto  M_{\rm SMBH}^{\frac{3}{28} - 0.6 \left( \frac{11}{14}
  - p_0 \frac{ M_\mathrm{tot} }{ m_\mathrm{BH,max} }\right)}
\end{align}
 for both multi-mass BH populations if $M_\mathrm{tot} > 11 m_\mathrm{BH,max}/ 14 p_0$, and
\begin{align}  \label{eq:mergrate_SingleRed}
  \left\langle \frac{\partial^2\Gamma}{\partial m_A\partial m_B} \right\rangle & \propto
  M_{\rm SMBH}^{\frac{3}{28} + \frac{3}{14} \left(\sqrt{2} - B_{\rm SMBH} \right)}
 \nonumber \\
  & \propto M_{\rm SMBH}^{0.235}
\end{align}
 for a single-mass BH population. Equation (\ref{eq:mergrate_lowmass}) is consistent with the approximate result of \citet{OLearyetal2009}. Moreover, Equations (\ref{eq:mergrate_highmass}) and (\ref{eq:mergrate_SingleRed}) reduce to $\left\langle \partial ^2 \Gamma / \partial m_A \, \partial m_B \right\rangle \propto M_{\rm SMBH} ^{3.5/28}$ for the most massive EBBHs forming in a multi-mass BH population with \mbox{$p_0=0.5$} and for EBBHs forming in a single-mass BH population, which is also consistent with the result of \citet{OLearyetal2009}. However, by setting \mbox{$p_0=0.6$}, we find that $\left\langle \partial ^2 \Gamma / \partial m_A \, \partial m_B \right\rangle \propto M_{\rm SMBH} ^{10/28}$. As a consequence of Equations (\ref{eq:mergrate_lowmass})-(\ref{eq:mergrate_SingleRed}), $\left\langle \partial^2 \Gamma / \partial m_A \, \partial m_B \right\rangle$ is somewhat sensitive to $M_{\rm SMBH}$, but only for the most massive EBBHs in the multi-mass BH population, which implies that $\left\langle \partial ^2 \Gamma / \partial m_A \, \partial m_B \right\rangle$ and any other distribution derived from it will be weakly sensitive to $M_{\rm SMBH}$ (i.e. $\left\langle \partial^2\Gamma / \partial M_{\rm tot} \partial q \right\rangle$, $\left\langle \partial^2\Gamma / \partial M_{\rm tot} \partial \mu \right \rangle$, $\left\langle \partial^2\Gamma /\partial \mathcal{M} \partial \eta \right\rangle$, and the marginalized distributions, see below).

 Given $\left\langle \partial^2 \Gamma / \partial m_A \, \partial m_B \right\rangle$, $\left\langle \partial^2\Gamma / \partial M_{\rm tot} \partial q \right\rangle$ can be obtained by changing variables from $m_A$ and $m_B$ to $M_{\rm tot}=m_A+m_B$ and $q=m_A/m_B$, or conversely:
\begin{equation}  \label{eq:q_Mtot}
 m_A = \frac{q M_{\rm tot}}{1 + q}  \, , \quad m_B = \frac{M_{\rm tot}}{1 + q} \, .
\end{equation}
 The Jacobian is $J = [\partial(m_A,m_B)/\partial(q,M_{\rm tot})]=M_{\rm tot} (1+q)^{-2}$ so $dm_A \, d m_B = M_{\rm tot} (1+q)^{-2} \, dq \, dM_{\rm tot}$, and thus
\begin{equation}  \label{eq:mergrate_q_Mtot}
  \left\langle \frac{\partial^2\Gamma}{\partial M_{\rm tot} \partial q} \right\rangle =
  \frac{ M_{\rm tot} }{ (1+q)^2} \left\langle \frac{\partial^2\Gamma}{\partial m_A \,
  \partial m_B}\right\rangle
\end{equation}
 if
\begin{equation}
  \frac{q \, m_\mathrm{BH,min}}{1+q} < M_{\rm tot} < \frac{ m_\mathrm{BH,max}}{1+q} \, ,
\end{equation}
 and zero otherwise. The merger rate distribution as a function of only $M_{\rm tot}$ or $q$ respectively can be given by marginalizing Equation (\ref{eq:mergrate_q_Mtot}) over $q$ or $M_{\rm tot}$.

 Additionally, we note that $\left\langle \partial^2\Gamma / \partial M_{\rm tot} \partial \mu \right \rangle$ and $\left\langle \partial^2\Gamma /\partial \mathcal{M} \partial \eta \right\rangle$ can be given as
\begin{equation}  \label{eq:rate_Mtot_Mu}
 \left\langle \frac{\partial^2\Gamma}{\partial M_\mathrm{tot} \, \partial \mu}\right\rangle
  = \frac{ 2 M_\mathrm{tot} }{ \sqrt{ M_\mathrm{tot}^2 - 4 \mu \, M_\mathrm{tot} } }
  \left\langle \frac{\partial^2\Gamma}{\partial m_A \, \partial m_B}\right\rangle \, ,
\end{equation}
 and as
\begin{equation}  \label{eq:rate_Mchp_eta}
 \left\langle \frac{\partial^2\Gamma}{\partial \mathcal{M} \partial \eta} \right\rangle =
 \mathcal{M} \eta^{-6/5}(1-4 \eta)^{-1/2} \left\langle \frac{\partial^2\Gamma}{\partial m_A \,
 \partial m_B}\right\rangle 
\end{equation}
 in the range $m_{\rm BH,min} \leqslant m_{A,B} \leqslant m_{\rm BH,max}$, and zero otherwise. It is straightforward to calculate the marginalized $\mu$, $\eta$, and $\mathcal{M}$ merger rate distributions from these equations.

 We use these results to generate panels in Figures \ref{fig:NoPMFDist}-\ref{fig:MassFuncs_PowerLaw_Belcz}.

\subsection{Average Relative Velocity in a Single Galactic Nucleus}
\label{App:AvrRelvelNucleus}

 Here, we derive the average relative velocity $\langle w_{AB}\rangle_{\bm{v}_A,\bm{v}_B,r}$ with which objects with component masses $m_A$ and $m_B$ form EBBHs in GNs in order to clarify the $M_{\rm SMBH}$ dependence of both $e_0$ and $\rho_{\mathrm{p}0}$ discussed in Section \ref{subsec:DistIniParam}.

 Using the distribution of relative velocities at radius $r$ from the SMBH, $P_{AB}(w|r)$, and the radial distribution of mergers $P_{AB}(r)$, Equations (\ref{e:P(w|r)}) and (\ref{APd71}), we have
\begin{equation}
  \langle w_{AB}\rangle_{\bm{v}_A,\bm{v}_B,r} = \int_{r_{\rm min}^{A,B}}^{r_{\rm max}}
  \langle w_{AB}(r)\rangle_{\bm{v}_A,\bm{v}_B} P_{AB}(r) \, dr \, ,
\end{equation}
 where
\begin{equation}
  \langle w_{AB}(r)\rangle_{\bm{v}_A,\bm{v}_B} = \int_{0}^{2 v_{\rm max}(r)} w P_{AB}(w|r) \, dw 
\end{equation}
 is the average relative velocity with which BHs $A$ and $B$ form an EBBH at radius $r$ from the central SMBH. Here, $r \in [r_{\rm min}^{A,B}, r_{\rm max}]$ is defined by Equations (\ref{eq_rminAB}) and (\ref{eq:rmax}), and $\langle w_{AB}(r)\rangle_{\bm{v}_A,\bm{v}_B}$ depends on both $r$ and $M_{\rm SMBH}$ implicitly through $v_{\rm max}(r)$ (Equation \ref{eq:vmax}). Here, $P_{AB}(r)$ is given by Equations (\ref{eq:formrateBHBH_multi}) and (\ref{eq:formrateBHBH_single}) for multi-mass and single-mass BH populations, respectively.

 \begin{figure}
    \centering
    \includegraphics[width=85mm]{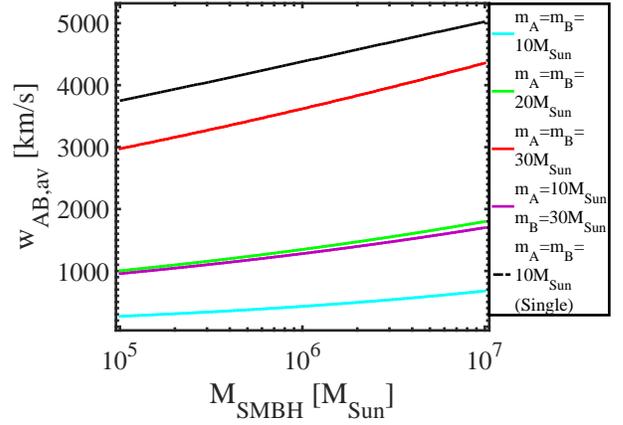}
\caption{ \label{fig:w_Avr_mSMBH} The average relative velocity ($\langle w_{AB} \rangle_{\bm{v}_A, \bm{v}_B}$) with which BHs with component masses $m_A$ and $m_B$ form EBBHs in GNs. Different lines correspond to different component masses in a multi-mass BH population and in a single-mass BH population as in Figure \ref{fig:Distrhop0}. }
\end{figure}

 Examples of the $M_{\rm SMBH}$ dependence of $ \langle w_{AB}\rangle_{\bm{v}_A,\bm{v}_B,r}$ are displayed in Figure \ref{fig:w_Avr_mSMBH}. In the case of a multi-mass BH population, we find that $\langle w_{AB}\rangle_{\bm{v}_A,\bm{v}_B,r}$ depends only on $m_\mathrm{BH,max}$, which is due to the fact that both $P_{AB}(w|r)$ and $P_{AB}(r)$ depend only on $m_\mathrm{max,BH}$ (see Appendices \ref{App:AppGN} and \ref{App:HandRelvel} and Section \ref{sec:NumRes}), and $r_{\rm min}^{A,B}$ is weakly sensitive to the parameters of the multi-mass BH population (Appendix \ref{sec:CalRmin}). These properties of $\langle w_{AB}(r)\rangle_{\bm{v}_A,\bm{v}_B}$ also determine how $\rho_{{\rm p}0}$ and $e_0$ depend on $M_{\rm SMBH}$ (Figure \ref{fig:SMBHdeprhopaeps0}) and the BH population parameters (Figure \ref{fig:Distrho_Param}) through Equation (\ref{e:Gamma-r-rp-eps}).

\section{aLIGO Detection Rate Distribution}
\label{sec:aLIGOeventrate}

 In this appendix, we determine the binary component mass distribution of the total detection rate for GW capture sources in GNs corresponding to the planned aLIGO design sensitivity. We add up the merger rates for all galaxies with nuclear star clusters which host an SMBH within the detectable universe, and we express the results as a function of the component masses $m_A-m_B$, $\partial^2 \mathcal{R}_{\rm aLIGO} / \partial m_A \partial m_B $ and the total mass and mass ratio, $\partial^2 \mathcal{R}_{\rm aLIGO} / \partial M_{\rm tot} \partial q$. This is used to plot the right panels of Figures \ref{fig:NoPMFDist} and \ref{fig:MergRate_BelczModel}.

 The detection rate distribution of EBBHs can be given using Equation (59) in \citet{OLearyetal2009} by omitting the integral over the component masses $m_A$ and $m_B$ as
\begin{equation}  \label{eq:aLIGOevrate_pcoms}
  \frac{ \partial^2 \mathcal{R}_{\rm aLIGO} }{ \partial m_A \partial m_B } =
  \int _{ \rho_\mathrm{p0,min} } ^{ \rho_\mathrm{p0,max} } d \rho_\mathrm{p0} \int_0 ^{z_{\rm max}}
  dz \frac{ dV_{\rm C} }{ dz } \frac{ 1 }{ 1 + z} \frac{ \partial^3 \mathcal{R} }{ \partial
  \rho_\mathrm{p0} \partial m_A \partial m_B }  \, .
\end{equation}
 Here, $\partial^3 \mathcal{R} / \partial \rho_\mathrm{p0} \partial m_A \partial m_B$ is the comoving partial binary formation rate density between masses $(m_A, m_B)$ for fixed $\rho_\mathrm{p0}$, averaged over the number density distribution of $M_{\rm SMBH}$ in the universe. Because $e_0 \sim 1$ for EBBHs forming through GW capture in GNs (Section \ref{subsec:DistIniParam}), it is not necessary to average over the $e_0$ dependence in the comoving partial binary formation rate density, but the results may be well-approximated by setting a single $e_0$ value (see below). In Equation (\ref{eq:aLIGOevrate_pcoms}), $z_{\rm max}$ is the redshift corresponding to the maximum detectable distance of EBBHs with $\rho_\mathrm{p0}$ and component masses $m_A$ and $m_B$, and $dV_{\rm C} / dz$ is the comoving volume density corresponding to the given cosmology. Generally, this is defined as $4\pi d_L^3 (1+z)^{-2}$ \citep{Eisenstein1997}.
 \begin{equation}
 \frac{dV_{\rm C}}{dz} = \frac{c}{(1+z)^2H_0} \frac{d_{\rm L}^2(z)}{\sqrt{\Omega_\mathrm{M} (1+z)^3 +
 \Omega_\Lambda}}\,,
 \end{equation}
 where the luminosity distance for a flat $\Lambda$CDM cosmology, $d_\mathrm{L}$, can be expressed as a function of $z$
\begin{equation}  \label{eq:CovDLz}
  d_\mathrm{L}(z) = \frac{ (1+z)c }{ H_0 } \int_0 ^z \frac{dz'}{\sqrt{ \Omega_\mathrm{M}
  (1+z')^3 + \Omega_{\Lambda} }} \, ,
\end{equation}
 where $H_0 = 68 \, \mathrm{km} \, \mathrm{s}^{-1} \mathrm{Mpc}^{-1}$ is the Hubble constant, and $\Omega _\mathrm{M} = 0.304$ and $\Omega_{\Lambda} = 1-\Omega_\mathrm{M}= 0.696$ are the density parameters for matter and dark energy, respectively \citep{PlanckCollaboration2014a,PlanckCollaboration2014b}.\footnote{ If cosmological effects are ignored, Equation (\ref{eq:aLIGOevrate_pcoms}) reduces to
\begin{equation}  \label{eq:aLIGOevrate}
   \frac{ \partial^2 \mathcal{R}_{\rm aLIGO} }{ \partial m_A \partial m_B } =
   \int _{ \rho_\mathrm{p0,min} } ^{ \rho_\mathrm{p0,max} } d \rho_\mathrm{p0}
   \frac{ \partial^3 \mathcal{R} }{ \partial \rho_\mathrm{p0} \partial m_A \partial m_B }
   \frac{ 4 \pi  d_{\rm L,\max}^3 }{ 3 }  \, ,
\end{equation}
 where $d_{\rm L,\max}$ is the maximum distance of aLIGO detection at a fixed signal-to-noise ratio, for EBBHs with $\rho_\mathrm{p0}$ and component masses $m_A$ and $m_B$.} We determine $\partial^2 \mathcal{R}_{\rm aLIGO} / \partial m_A \partial m_B $ for one aLIGO detector at design sensitivity \citep{Abbottetal2016a}.

 In Equation (\ref{eq:aLIGOevrate_pcoms}), we calculate $z_{\max}$ from the maximum luminosity distance $d_\mathrm{L}(z_{\max})$ of detection with a given signal-to-noise ratio for aLIGO. For binaries on eccentric orbits, the sky position and binary orientation averaged rms signal-to-noise ratio for a single orthogonal arm interferometric GW instrument can be given as \citep{OLearyetal2009}
\begin{equation}  \label{eq:SNR2}
 \left\langle \frac{ S^2 }{ N^2 } \right\rangle = \frac{48}{95} \frac{ \eta M_{\mathrm{tot},z}^3
 \rho_\mathrm{p0} ^2 }{ d_{\rm L}^2 } \int _{e_{\rm LSO}} ^{e_0} \sum _{n=1} ^{ n_{\rm max}(e_0) }
 \frac{ g(n,e) s(e,e_0) }{ n^2 S_h (f_n) } \frac{ de }{ e }\,,
\end{equation}
 where $S_h$ is the one-sided noise spectral density, $e_{\rm LSO}$ is given by Equations (\ref{eq:rhoe}) and (\ref{eq:ELSO}), \mbox{$M_{\mathrm{tot},z} = M_{\rm tot} (1 + z)$} is the redshifted total mass, and $g(n,e)$ and $s(e,e_0)$ are given by Equations (52) and (56) in \citet{OLearyetal2009}. In Equation (\ref{eq:SNR2}), $f_n$ is the observed orbital harmonic at redshift $z$, i.e.,
\begin{equation}  \label{eq:fnz}
  f_n = \frac{n (1-e^{3/2})}{ 2 \pi \rho_{\rm p}^{3/2} M_{\mathrm{tot},z} } \, ,
\end{equation}
 and we truncate $n$ at $n_\mathrm{max}(e_0)$ \citep{OLearyetal2009,Mikoczietal2012}
\begin{equation}  \label{eq:nmax}
  n_\mathrm{max}(e_0) = \left\{ 5 \frac{ (1+e_0)^{1/2} }{ (1-e_0)^{3/2} }
  \right\} \, ,
\end{equation}
 which accounts for $99 \%$ of the signal power \citep{Turner1977}. Here, the bracket $\{ \}$ denotes the floor function. By setting the detection threshold to be $\langle S^2 / N^2 \rangle = \mathrm{SNR}^2 _\mathrm{lim} = 8^2$ in Equation (\ref{eq:SNR2}), the maximum luminosity distance of detection as a function of $M_{\mathrm{tot},z}$, $\eta$, $e_0$, and $\rho_\mathrm{p0}$ can be given as
\begin{equation}  \label{eq:dLmax}
 d_{\rm L}^{\max} = \sqrt{ \frac{48}{95} \frac{ \eta M_{\mathrm{tot},z}^3 \rho_\mathrm{p0} ^2 }{
 \mathrm{SNR}^2_\mathrm{lim} } \int _{e_{\rm LSO}} ^{e_0} \sum _{n=1} ^{ n_{\rm max}(e_0) }
 \frac{ g(n,e) s(e,e_0) }{ n^2 S_h (f_n) } \frac{ de }{ e } }
\end{equation}
 \citep{OLearyetal2009}. EBBHs form with $0.9 < e_0 < 0.9999$ in GNs (Section \ref{subsec:DistIniParam}), therefore the integrand in Equation (\ref{eq:dLmax}) has a small scatter ($\la 5 \%$) over the range of $e_0$ for any $\rho_\mathrm{p0}$ occuring in MC simulations ($8 \leqslant \rho_\mathrm{p0} < 1000$; see Section \ref{subsec:DistIniParam} for details). Here, $d^{\rm max}_{\rm L}\equiv d_{\rm L} (z_{\max})$ given by Equation (\ref{eq:CovDLz}). Furthermore, $d_{\rm L}^{\rm max}$ depends on the redshifted masses as $M_{\mathrm{tot},z}$. Thus, for given $m_{\rm A}$ and $m_{\rm B}$, we calculate $z_{\max}$ from Equations (\ref{eq:CovDLz}) and (\ref{eq:dLmax}), which then yields $d^{\rm max}_{\rm L}$. Here, we adopt the simplifying approximation to set $e_0 = 0.99$.

 In Equation (\ref{eq:aLIGOevrate_pcoms}), $\partial^3 \mathcal{R} / \partial \rho_\mathrm{p0} \partial m_A \partial m_B$ can be given by convolving $\left\langle \partial^3 \Gamma / \partial \rho_\mathrm{p0} \partial m_A \partial m_B \right \rangle$ (Equations \ref{e:Gamma-rp} and \ref{eq:dGammadrhop0}) with the number density of galaxies with nuclear star clusters that host an SMBH of mass $M_{\rm SMBH}$, $d n_{\rm gal} / d M_{\rm SMBH}$:\footnote{Bear in mind here, when using Equations (\ref{e:Gamma-rp}) and (\ref{eq:dGammadrhop0}) and averaging over all galaxies within the aLIGO range, the detection rate is proportional to $\langle \partial^3 \Gamma/ \partial \rho_\mathrm{p0} \partial m_A \partial m_B\rangle \propto \langle n_A n_B \rangle \propto \langle C_A C_B n_{\rm MS}^2\rangle$, which may be much larger than $\langle C_A\rangle \langle C_B\rangle \langle n_{\rm MS}\rangle^2$ due to the Cauchy-Schwarz inequality. This enhancement factor $\xi=\langle n_{\rm MS}^2\rangle / \langle n_{\rm MS}\rangle^2$ was introduced in \citet{OLearyetal2009}. However, this quantity shifts the detection rates uniformly for different BH masses. Such constant normalization factors drop out in Figure
 \ref{fig:NoPMFDist}.}
\begin{align}
   \frac{ \partial^3 \mathcal{R} }{ \partial \rho_\mathrm{p0} \partial m_A \partial m_B } & =
  \int_{10^5 \, \Msun} ^{10^7 \, \Msun} d M_{\rm SMBH} \frac{ d n_{\rm gal} }{ d M_{\rm SMBH} } \xi
  \nonumber \\
  & \times \left\langle \frac{\partial^3 \Gamma }{ \partial \rho_\mathrm{p0} \partial m_A \partial m_B }
  \right \rangle \, .
\end{align}
 Several studies have provided fitting functions for the number density distribution of massive SMBHs \citep{Saluccietal1999,AllerRichstone2002,Marconietal2004,Shankaretal2004,Bensonetal2007,Grahametal2007,Shankaretal2009,Vikaetal2009,Sijackietal2015}, but most of them are only valid for $M_{\rm SMBH} \ga 10^6 \, \Msun $. In this paper, the SMBH mass range of interest is $10^5 \, \Msun - 10^7 \, \Msun$ (Section \ref{subsec:GNModels}); therefore, we interpolate the results of \citet{AllerRichstone2002}, who found the best-fit number density distribution of massive SMBHs to be
\begin{equation}  \label{eq:SMBHdens}
  \frac{ d n_{\rm gal} }{ d M_{\rm SMBH} } = c_0 \left( \frac{ M_{\rm SMBH} }{ M_* } \right)^{-1.25}
  \mathrm{exp} (- M_{\rm SMBH} / M_*)
\end{equation}
 in the SMBH mass range of $10^4 \, \Msun - 10^9 \, \Msun$. In Equation (\ref{eq:SMBHdens}), $M_* = 1.3 \times 10^8 \, \Msun$, and $c_0 = 3.2 \times 10^{-11} \, \Msun ^{-1} \, \mathrm{Mpc} ^{-3}$.

 In Equation (\ref{eq:aLIGOevrate_pcoms}), we set \mbox{$\rho_\mathrm{p0,min} = 8$} and \mbox{$\rho_\mathrm{p0,max} = 1000$}, which captures most of the mergers for the considered ranges of SMBH mass, BH mass, and BH population parameters (Section \ref{subsec:DistIniParam}).

 Using Equation (\ref{eq:mergrate_q_Mtot}), the total-mass- and mass-ratio-dependent rates may be calculated as
\begin{equation}  \label{eq:evrate_q_Mtot}
  \frac{ \partial^2 \mathcal{R}_{\rm aLIGO} }{ \partial M_{\rm tot} \partial q } = \frac{ M_{\rm tot} }{ (1+q)^2 }
   \frac{ \partial^2 \mathcal{R}_{\rm aLIGO} }{ \partial m_A \partial m_B }
\end{equation}
 if
\begin{equation}
 \frac{q \, m_\mathrm{BH,min}}{1+q} < M_{\rm tot} < \frac{ m_\mathrm{BH,max}}{1+q}  \, ,
\end{equation}
 and zero otherwise. The 1D distribution of aLIGO detection rate as a function of either $M_{\rm tot}$ or $q$ can be given by marginalizing Equation (\ref{eq:evrate_q_Mtot}) over the other parameter.

 Similarly to Equations (\ref{eq:rate_Mtot_Mu}) and (\ref{eq:rate_Mchp_eta}), the functions $ \partial^2 \mathcal{R}_{\rm aLIGO} / \partial M_{\rm tot} \partial \mu $ and $ \partial^2 \mathcal{R}_{\rm aLIGO} /\partial \mathcal{M} \partial \eta $ can be given as
\begin{equation}  \label{eq:evrate_Mtot_Mu}
 \frac{ \partial^2 \mathcal{R}_{\rm aLIGO} }{\partial M_\mathrm{tot} \, \partial \mu} =
  \frac{ 2 M_\mathrm{tot} }{ \sqrt{ M_\mathrm{tot}^2 - 4 \mu \, M_\mathrm{tot} } }
  \frac{ \partial^2 \mathcal{R}_{\rm aLIGO} }{\partial m_A \, \partial m_B} \, ,
\end{equation}
 and as
\begin{equation}  \label{eq:evrate_Mchp_eta}
 \frac{ \partial^2 \mathcal{R}_{\rm aLIGO} }{\partial \mathcal{M} \partial \eta} =
 \mathcal{M} \eta^{-6/5}(1-4 \eta)^{-1/2} \frac{ \partial^2 \mathcal{R}_{\rm aLIGO} }{\partial m_A \,
 \partial m_B} 
\end{equation}
 in the range $m_{\rm BH,min} \leqslant m_{A,B} \leqslant m_{\rm BH,max}$, and zero otherwise. It is straightforward to calculate the marginalized $\mu$, $\eta$, and $\mathcal{M}$ detection rate distributions from these equations.

 We use these results to generate Figures \ref{fig:NoPMFDist}, \ref{fig:MergRate_BelczModel}, and \ref{fig:MassDists_DetRate}.

\bibliographystyle{yahapj}
\bibliography{refs}

\end{document}